\documentclass[]{jfm}

\usepackage{graphicx}
\usepackage{newtxtext}
\usepackage{newtxmath}
\usepackage{natbib}

\usepackage{hyperref}
\hypersetup{
    colorlinks = true,
    urlcolor   = blue,
    citecolor  = blue,
}

\newcommand{\RomanNumeralCaps}[1]
\linenumbers

\usepackage{floatrow}
\usepackage[caption=false]{subfig}
\usepackage{threeparttable}
\usepackage[scientific-notation=true]{siunitx}
\usepackage{makecell}
\usepackage{commath}
\usepackage{chemformula}
\usepackage[scientific-notation=true]{siunitx}
\usepackage{booktabs}
\usepackage[bold]{hhtensor}
\usepackage{cancel}

\usepackage{tikz}
\DeclareRobustCommand\full  {\tikz[baseline=-0.6ex]\draw[thick] (0,0)--(0.5,0);}

\DeclareRobustCommand\dashed{\tikz[baseline=-0.6ex]\draw[thick,dashed] (0,0)--(0.54,0);}
\DeclareRobustCommand\longdashed{\tikz[baseline=-0.6ex]\draw[thick, dash pattern= on 6pt off 3pt on 6pt] (0,0)--(0.54,0);}
\DeclareRobustCommand\chain {\tikz[baseline=-0.6ex]\draw[thick,dash dot] (0,0)--(0.5,0);}
\DeclareRobustCommand\dashdoubledot {\tikz[baseline=-0.6ex]\draw[thick,dash dot dot] (0,0)--(0.5,0);}

\title{Kinetic modelling of three-dimensional shock/laminar separation bubble instabilities in hypersonic flows over a double wedge}

\author{Saurabh S. Sawant\aff{1}
  \corresp{\email{sssawan2@illinois.edu}},
  V. Theofilis\aff{2,3}
 \and  D. A. Levin\aff{1}}

\affiliation{\aff{1}Department of Aerospace, University of Illinois at Urbana-Champaign, 104 S. Wright St, Champaign, Illinois, USA
\aff{2} School of Engineering, University of Liverpool, The Quadrangle, Brownlow Hill, L69 3GH, UK
\aff{3} Escola Politecnica, Universidade S\~{a}o Paulo, Av. Prof. Mello Moraes 2231, CEP 5508-900, S\~{a}o Paulo-SP, Brasil}

\begin{document}

\maketitle

\begin{abstract}
Linear global instability of the three-dimensional (3-D), spanwise-homogeneous laminar separation bubble (LSB) induced by shock-wave/boundary-layer interaction (SBLI) in a Mach 7 flow of nitrogen over a $30^{\circ}-55^{\circ}$ double wedge is studied.
At these conditions corresponding to a freestream unit Reynolds number, $Re_1=\num{5.2e4}$~m$^{-1}$, the flow exhibits  rarefaction effects and comparable shock-thicknesses to the size of the boundary-layer at separation.  
This, in turn, requires the use of the high-fidelity Direct Simulation Monte Carlo (DSMC) method to accurately resolve unsteady flow features.

We show for the first time that the LSB sustains self-excited, small-amplitude, 3-D perturbations that lead to spanwise-periodic flow structures not only in and downstream of the separated region, as seen in a multitude of experiments and numerical simulations, but also in the internal structure of the separation and detached shock layers.
The spanwise-periodicity length and growth rate of the structures in the two zones are found to be identical.
It is shown that the linear global instability leads to low-frequency unsteadiness of the triple point formed by the intersection of separation and detached shocks, corresponding to a Strouhal number of $St\sim0.02$.
Linear superposition of the spanwise-homogeneous base flow and the leading 3-D flow eigenmode provides further evidence of the strong coupling between linear instability in the LSB and the shock layer.
\end{abstract}

\begin{keywords}
Authors should not enter keywords on the manuscript, as these must be chosen by the author during the online submission process and will then be added during the typesetting process (see \href{https://www.cambridge.org/core/journals/journal-of-fluid-mechanics/information/list-of-keywords}{Keyword PDF} for the full list).  Other classifications will be added at the same time.
\end{keywords}

\newpage
\section{Introduction}\label{Intro}
Laminar SBLI has been a topic of extensive study since the best part of last century.
The early experimental and theoretical work primarily focused on the upstream influence of disturbances in  boundary layers, as can be found in seminal contributions such as \citet{czarnecki1950investigation, liepmann1951reflection,lighthill1953boundary1, lighthill1953boundary2, lighthill2000upstream,chapman1958investigation,stewartson1964theory}. In subsequent research, triple deck theory~\citep{stewartson1969self, smith1986steady, neyland2008asymptotic} was developed and used to understand  boundary layer instability mechanisms that lead to separation in supersonic and hypersonic flows over compression ramps at moderate to high Reynolds numbers~\citep{rizzetta_burggraf_jenson_1978,cowley_hall_1990, smith_khorrami_1991,cassel_ruban_walker_1995,korolev_gajjar_ruban_2002,fletcher_ruban_walker_2004}.
More recent topics of study on shock-induced LSB include  3-D effects~\citep{lusher_sandham_2020}, unsteadiness and underlying instability mechanisms~\citep{sansica_sandham_hu_2016}, and the coupling between LSB and shock structure~\citep{tumuklu2018POF2, sawantIUTAM2019}.
\vspace{\baselineskip}

Experimental investigations of hypersonic SBLI primarily exist on compression ramps in flow regimes from laminar to turbulent and are at large $Re_1\sim O(10^6-10^7)$~m$^{-1}$~\citep{holdenThesis, holdenRamp1978, needhamThesis,needham1965heat,elifstromThesis,elfstrom1972turbulent, hankey1975two, simeonides1995experimental, schneider2004hypersonic, roghelia2017experimental,chuvakhov2017effect}.
Experiments on flows over double wedges can be found at moderate Reynolds number, $Re_1\sim O(10^5-10^6)$~m$^{-1}$, but are limited, exhibit far more complicated SBLI than compression ramps, and suffer from test times significantly lower than the characteristic times involved in the development of instabilities and unsteadiness.
\citet{schrijer2006quantitative,schrijer2009three} performed experiments at $Re_1\sim O(10^7)$~m$^{-1}$ and $M=7$ in a Ludwieg tube facility and observed an unsteady flow exhibiting Edney type-$VI$ and $V$ interactions on 15$^{\circ}$-30$^{\circ}$ and 15$^{\circ}$-45$^{\circ}$ double wedge configurations, respectively.
In a moderate Reynolds number regime, $Re_1\sim O(10^5)$~m$^{-1}$, \citet{hashimoto2009experimental} performed experiments in a free piston shock tunnel, where the flow of air on a double wedge was tested for 300~$\mu$s at $Re_1=0.18-\num{3.5e6}$~m$^{-1}$ and $M=7$, and was found to exhibit Edney type-$V$ interactions.
Recently, \citet{swantek2015flowfield} have performed experiments on a 30$^{\circ}$-25$^{\circ}$ double wedge in Hypervelocity Expansion Tube (HET) facility to test air and nitrogen at $Re_1=0.44-\num{4.6e6}$~m$^{-1}$ and $M=4.01-7.14$, where the test times ranged from 361-562~$\mu$s.
Experiments of \citet{knisely2016geometry} include nitrogen flow over the 30$^{\circ}$-55$^{\circ}$ double wedge geometry considered in this work at higher $Re_1=0.435-\num{1.1e6}$~m$^{-1}$ and $M=6.64-7.14$, where, in addition to HET facility, the T5 reflected shock tunnel was used that allows for a longer test time of 1~ms.
\vspace{\baselineskip}

On the numerical side, existing studies in laminar and transitional regimes primarily use compressible Navier-Stokes equations to understand different aspects of SBLI in hypersonic flows over compression ramps such as three-dimensionality of an LSB~\citep{rudy1991computation}, stability of hypersonic boundary layers~\citep{balakumar2005stability}, development of 3-D instability in the form of spanwise periodic striations of the LSB \citep{egorov2011three,dwivedi_2019}, and formation of secondary vortices and their fragmentation inside an LSB~\citep{gai2019hypersonic}.
Efforts such as these have been extended to simulate laminar SBLI over double wedge geometries~\citep{knight2017assessment}.
\citet{durna2016shock} simulated a 2-D Mach 7 flow of nitrogen over a double wedge at the 2~MJ low enthalpy conditions of Swantek and Austin ($Re_1=\num{1.1e6}$) to study the effect of the aft wedge angle on the flow characteristics with additional, recently included 3-D effects~\citep{durna2020effects}.
\citet{sidharth2018onset} carried out global stability analysis and Direct Numerical Simulation (DNS) of a Mach 5 perfect gas flow at $Re_1=\num{1.36e6}$ over double ramps with forward and aft angles of 12$^{\circ}$ and 12$^{\circ}$-22$^{\circ}$, respectively.
For aft angle of 20$^{\circ}$ and greater, they observed a linear instability of the 2-D separation bubble in the absence of upstream perturbations and associated that with streamwise streaks in wall temperature near the reattachment region.
Recently, \citet{reinert2020simulations} simulated 3-D flows at Mach 7 over a 30$^{\circ}$-55$^{\circ}$ double wedge at the experimental conditions of \citet{swantek2015flowfield} and \citet{knisely2016geometry} for much longer flow times than the duration of the experiments and reported unsteady asymmetric 3-D separation bubble.
\vspace{\baselineskip}

However, despite  extensive numerical and experimental work, the physics of complicated SBLI formed in a hypersonic flow over double-wedge configurations 
is not well-understood.  In this work,
 we investigate questions about the instability mechanism of a 3-D LSB, the coupling between the  shock structure and LSB, and the low-frequency unsteadiness of the shock structure.  We focus on the  linear instability of a shock-induced, 3-D LSB formed in a Mach 7 nitrogen flow over a 30$^{\circ}$-55$^{\circ}$ double wedge configuration at a freestream unit Reynolds number of $Re_1=\num{5.2e4}$ corresponding to an altitude of about 60~km.  We will show that even for this lower density free stream condition, that is typically not studied, our fully-resolved, kinetic DSMC simulations of this complex flow allow us to study the strong coupling between the separation shock and the LSB.
 In our previous work, a
 2-D (spanwise independent) flow over the same configuration and freestream conditions was simulated by \citet{tumuklu2018PhysRevF}, who demonstrated that the flow reaches a steady-state in $\sim 0.9~ms$ after the leading damped global modes, recovered by the residuals algorithm~\citep{theofilis2000origins} and proper orthogonal decomposition (POD), have decayed. Yet we will show  in our 3-D treatment using this  2-D, steady base flow of \citet{tumuklu2018PhysRevF}, that indeed the flow is linearly unstable to self-excited, small-amplitude, spanwise-homogeneous perturbations and will ultimately transition to turbulence.
\vspace{\baselineskip}

An important goal of  the research  discussed in this paper is to understand the
 effect of three-dimensionality on the coupling between the LSB and shock structure.
The accurate modelling of the internal shock structure is an essential feature of relevance to the study of linear instability in high-speed boundary layer flow, where coupling between the separation bubble and the {shock has been demonstrated through the amplitude function of the underlying global modes} in a number of studies~\citep[e.g.][]{crouch2007predicting}.
In the hypersonic regime, \citet{tumuklu2018POF1,tumuklu2018POF2} used DSMC to study the effect of unit Reynolds number, $Re_1=0.935-\num{3.74e5}~m^{-1}$, on laminar SBLI in a Mach 16 nitrogen flow over an axisymmetric double cone configuration.
The authors demonstrated a strong coupling between oscillations of the shock structure and instability of the laminar separated flow region through the spatial structure of the amplitude functions as well as Kelvin-Helmholtz waves formed at the contact surface downstream of the triple point.
In this work, we  focus on the coupling mechanism between a fully 3-D LSB and shock, and show that the instability in the LSB as well as {\em{inside}} the strong gradient region of shocks are intimately related.
\vspace{\baselineskip}

To capture the complex physics of SBLI with highest fidelity, we use the DSMC method.
Continuous developments spanning the past fifty decades have resulted in this method being well-suited for the study of the physics of unsteady laminar SBLI to deliver accurate results in five critical aspects: (a) computations of molecular thermal fluctuations~\citep{garcia1986nonequilibrium, bruno2019direct,sawant2020Kinetic}, (b) calculation of anisotropic stresses and heat fluxes in strong shock layers (M $>>$ 1.6)~\citep{bird1970aspects,cercignani1999structure}, (c) prediction of rarefaction effects such as velocity slip and temperature jump~\citep{moss2001dsmc,moss2005direct,tumuklu2018POF1}, (d) quantification of translational, rotational, and vibrational nonequilibrium~\citep{sawant2018application}, and (e) time-accurate evolution of self-excited perturbations~\citep{tumuklu2018POF1,tumuklu2018POF2, tumuklu2018PhysRevF}.
As a result, the method is gaining momentum in the study of hydrodynamic instabilities~\citep{bird1998recent,stefanov_part1,stefanov_part2,stefanov_part3,kadau2004nanohydrodynamics, kadau2010atomistic, gallis2015direct,gallis2016direct}.
In our flow, even though the freestream Knudsen number of $\num{3.e-3}$ is continuum, the local Knudsen number in the shock-LSB region is much higher due to the steep gradients in macroscopic flow parameters.  
These non-continuum features, also known as local rarefaction zones, are crucial to understanding the coupling between the shock and LSB, as this work will demonstrate. 
Furthermore, the high-fidelity kinetic modelling of these regions is crucial because the thicknesses of shocks and the boundary-layer in the separation region are comparable.
\vspace{\baselineskip}

The time-accuracy of the DSMC method in modelling unsteady evolution of 3-D perturbations allows for quantification of the low-frequency unsteadiness of the shock structure.
This phenomenon has been extensively investigated in turbulent SBLI at $Re_1\sim O(10^{6}-10^{8})$ using DNS and large eddy simulation (LES)~\citep[see][]{pirozzoli2006direct, touber2009large, piponniau2009simple,grilli_schmid_hickel_adams_2012, priebe_martin_2012,clemens2014low,gaitonde2015progress, priebe2016low, pasquariello2017unsteady}, where numerical studies report a Strouhal number associated with unsteadiness within a range of 0.01 to 0.05, consistent with findings of many experiments~\citep{dussauge2006unsteadiness}.
However, in hypersonic, 3-D laminar SBLI, such investigations are sparse.
\citet{tumuklu2018POF2} observed a similar Strouhal number of $\sim$0.08, corresponding to the bow shock oscillation in their axisymmetric flow over a double cone simulated using DSMC.
In the 3-D, Mach 7, finite-span double-wedge flow simulation of \citet{reinert2020simulations}, however, such unsteadiness was not observed for conditions at a  freestream flow enthalpy of 8~MJ, although, the Reynolds number of their case was a factor of 8 higher.
In this work, we show that our 8~MJ enthalpy, Mach 7, spanwise-periodic flow over the same configuration (at eight-times lower density) exhibits low-frequency unsteadiness after the onset of linear instability.
\vspace{\baselineskip}

Finally, topology analysis of separated flows is an important way to characterize and compare complex 3-D flows for different input conditions and shapes
~\citep[see, e.g.][]{lighthill, tobak1982topology,hornungAndPerry, perryAndHornung,perryAndChong,dallmann1983topological,dallmann1985structural}.  
Topologies of 3-D flow constructed from the linear superposition of the leading stationary eigenmode due to a linear instability of an LSB and 2-D base flow were analyzed by \citet{rodriguez_theofilis_2010} in the incompressible regime and \citet{robinet_2007,boin20063d} in the compressible regime involving oblique SBLI.
In the analysis presented here, we estimate the changes in the 3-D wall-streamline topology for an increasing amplitude of the 3-D perturbations based on a linear combination of the 2-D base flow and 3-D perturbations.  
In this simplified approach, we will demonstrate that the wall-streamline signature is very different depending on whether the coupling is considered.

\vspace{\baselineskip}
The paper is organized as follows: section~\ref{sec:Method} describes the methodology, which includes a brief description of the DSMC method in section~\ref{sec:DSMC} and details about numerical models, the DSMC solver, the input conditions, and flow initialization in section~\ref{sec:Numerical}.
The features of 2-D base flow are described in section~\ref{sec:2DBase}.
Section~\ref{sec:3DMechanisms} is devoted to the key findings of this paper.
Section~\ref{sec:LinearInstab} describes the linear instability mechanism and its spatial origin through a detailed discussion of boundary layer profiles.
The correlation between the shock and the separation bubble is explained in section~\ref{sec:Correlation}.
The surface rarefaction effects are described in section~\ref{sec:Slip}, whereas the isocontours of spanwise periodic flow structures are discussed in section~\ref{sec:SpanPeriodicFlowStructures}.
The topology of the LSB is discussed in section~\ref{sec:Topology}, first without taking into account the coupling between the bubble and the shock in section~\ref{sec:withoutCoupling} and then the effect of their coupling in section~\ref{sec:withCoupling}.
The important findings are summarized in section~\ref{sec:Conclusion}.
\vspace{\baselineskip}

\section{Methodology}\label{sec:Method}
\subsection{The DSMC algorithm}\label{sec:DSMC}
The equation for the evolution of velocity distribution function of molecules, $f(t,\vec{r},\vec{v})$ with respect to time $t$, position vector $\vec{r}$, and instantaneous velocity vector $\vec{v}$, is written as,
\begin{equation} 
\begin{split}
 \frac{\partial f}{\partial t} + 
 (\vec{v} \cdot \nabla) f + 
 \left( \frac{\vec{F}}{m} \cdot \nabla_v \right)f =  \left[\frac{df}{dt}\right]_{coll}\\
\end{split}
\label{Boltzmann}
\end{equation}
where $\nabla$ and $\nabla_v$ are gradient operators in physical and velocity spaces, respectively.
The first, second, and third terms on the left-hand side describe the change of $f$ with time, change due to convection of molecules in physical space, and that in the velocity space, respectively.
The latter can happen due to the action of external conservative force per unit mass $\vec{F}/m$, such as gravity or electric field, which are ignored in this work.
The right-hand side (RHS) term accounts for changes in $f$ in an element of space-velocity phase-space due to intermolecular collisions.
For a thorough description, see \citet{vincenti1965introduction}.
\vspace{\baselineskip}

The DSMC method \citep[see][]{bird:94mgd} decouples the advection of molecules and their intermolecular collisions. 
Each simulated particle represents $F_{n}$ amount of actual gas molecules and is advected for a discrete timestep.
Based on the choice of boundary conditions, particles are introduced, removed, or reflected from the domain boundaries and interacted with the embedded surfaces using gas-surface collision models for the duration of the timestep.
They are then mapped to an adaptively refined collision mesh ($C$-mesh) that encompasses the flow domain and ensures the spatial proximity of particles that are potential candidates for binary collisions.
Next follows a collision scheme, which selects particle pairs that are collided based on the appropriate (elastic/inelastic) collision cross-section and are assigned with post-collisional instantaneous velocities and internal energies.
Macroscopic flow parameters of interest such as pressure, velocities, etc., can be derived from the microscopic properties of simulated particles using statistical relations of kinetic theory.
These parameters are represented on the sampling mesh ($S$-mesh), which has coarser cells than the $C$-mesh.
Note that the unique characteristic of the DSMC method, the advection-collision decoupling, is justified if the local cell size of $C$-Mesh, $\Delta x$, is smaller than the local mean-free-path of molecules, $\lambda$, and the timestep, $\Delta t$, is lower than the mean-collision-time, $\tau$.
A sufficient number of instantaneous particles in the smallest collision cells must also be ensured for unbiased collisions.
Conveniently, the satisfaction of only these three numerical criteria leads to accurate modelling of internal structure of shocks, their mutual interaction, and surface rarefaction effects.
This warrants the use of DSMC for detailed modelling of SBLIs at high altitudes, compared to ad-hoc techniques of modelling shocks in computational fluid dynamics (CFD) simulations that fall short of accurately capturing the internal structure of shocks.

\subsection{The numerical implementation and flow initialization}\label{sec:Numerical}
The fulfillment of the numerical criteria demands a large number of computational particles and collision cells.
To overcome this challenge, we have previously developed an octree-based, 3-D DSMC solver known as Scalable Unstructured Gas-dynamic Adaptive mesh-Refinement (SUGAR-3D).
See~\citet{sawant2018application} for a comprehensive account of the implementation strategies, validation, and performance studies of the solver.
In summary, the code takes advantage of message-passing-interface (MPI) for parallel communication between processors, adaptive mesh refinement (AMR) of coarser Cartesian octree cells to achieve spatial fidelity at regions of strong gradients, a cutcell algorithm to correctly capture physics in the vicinity of embedded surfaces, a domain decomposition strategy based on Morton-Z space-filling-curves, capability of parallel input/output, inclusion of thermal nonequilibrium models, and numerous run-time memory reduction strategies.
In the octree-based AMR framework, the $C$-mesh is formed from a user-defined, uniform Cartesian grid.
The cells of this grid are known as `root' cells, which are recursively subdivided into eight parts until the local cell-size is smaller than the local mean-free-path.
Note that a subdivision based on the above criterion is performed only if there are at least 32 particles in a collision cell.
The satisfaction of both of these criteria in the presented flow over a double wedge requires $\sim$60~billion computational particles and $\sim$4.5~billion collision cells of an adaptively refined octree grid.
See the appendix of \citet{sawantIUTAM2019} for the details of convergence study.

\begin{table}[H]
\centering
\normalsize
\begin{threeparttable}
\caption{Freestream and numerical parameters for the Mach 7 nitrogen flow.}
\label{tab:input}       
\begin{tabular}{cccc}
\hline\noalign{\smallskip}
\textbf{Parameters}                   & \textbf{Values}          \\\hline
Unit Reynolds number, $Re_{1}$                 & \num{5.22e4}           \\
Knudsen number$^a$                             & \num{3.2e-3}     \\
Number density, $n_1$/(m$^{3}$)                & 10$^{22}$        \\
Streamwise velocity, $u_{x,1}$/(m.s$^{-1}$)    & 3812             \\
Equilibrium translational temperature, $T_{tr,1}$/(K)                & 710              \\
Surface$^{b}$ temperature, $T_s$/(K)           & 298.5            \\
Species mass, $m$/(kg)                         & \num{4.65e-26}              \\
Species diameter, $d$/(m)                      & \num{4.17e-10}              \\
Viscosity index, $\omega$                      & 0.745            \\
Reference temperature, T$_{r}$/(K)             & 273              \\
Parker model parameters, Z$_{r,\infty}$ and T$^{*}$/(K) & 18.5 and 91             \\
Vibrational characteristic temperature, $\theta$/(K) & 3371             \\
Domain size, ($L_x$,$L_y$,$L_z$)/(mm)             & (80, 28.8, 80) \\
Number of octree and sampling cells along ($X$,$Y$,$Z$)      & (400, 144, 400) \\
Number of gas-surface interaction cells along ($X$,$Y$,$Z$)        & (25, 10, 25) \\
F$_{n}$$^c$                                              & \num{6.1e7}                \\
Timestep, $\Delta t$/(ns)                         & 5                \\
Adaptive mesh refinement interval /($\mu$s)        & 5                \\
Relaxation probability computation interval /($\mu$s)& 1                \\
\noalign{\smallskip}\hline\noalign{\smallskip}
\end{tabular}
\begin{tablenotes}
\item $^a$ Based on the length of the lower wedge, 50.8 mm.
\item $^b$ Surface is fully accommodated~\citep{bird:94mgd}, i.e., isothermal.
\item $^c$ Number of actual molecules represented by a computational particle.
\end{tablenotes}
\end{threeparttable}
\end{table}
The DSMC specific input and numerical parameters used in this work are listed in table~\ref{tab:input}.
Note that the Cartesian coordinates are used with streamwise, spanwise, streamwise-normal directions as $X$, $Y$, and $Z$, respectively.
The code uses the majorant frequency scheme (MFS) of \citet{ivanov1988analysis} derived using the Kac stochastic model for the selection of collision pair and the variable hard sphere (VHS) model for elastic collisions.
Appendix~\ref{A:MCCS} describes an essential modification to the computation of maximum collision cross-section used in the MFS scheme for accurate spectral analysis of unsteady flows simulated on adaptively refined grids.
For rotational relaxation, the \citet{borgnakke:75BLContinuous} model is employed with rates by \citet{parker:59rotational} and DSMC correction factors~\citep{lumpkin:91Zcorrection,gimelshein2002vibrational} to account for the temperature dependence of the rotational probability.
For vibrational relaxation, the semi-empirical expression of \citet{millikan1963systematics} is used with the high-temperature correction of \citet{park1984problems}.
\vspace{\baselineskip}

For this work, the SUGAR solver is also employed with spanwise-periodic boundary conditions as follows.
Suppose a particle, during its discrete movement, intersects the spanwise domain boundary, $Y=0$ or $Y=L_y$, within a period, $\delta t$, smaller than the timestep.
In that case, its spanwise position index is changed to the periodically opposite $Y$ boundary index, i.e., $Y=L_y$ or $Y=0$, respectively.
After this translation, the particle continues its movement for the remaining period, $\Delta t - \delta t$.
This simple algorithm is implemented in SUGAR's parallel framework by ensuring that the processors containing a portion of the flow domain adjacent to any $Y$-boundary must also store the information of processors containing the periodically opposite portion of the domain.
Such information includes the `location code array' and the triangulated panels of the embedded surface.
The location code arrays are special arrays used in the efficient particle mapping strategy based on the Morton-based space-filling-curve approach.
See~\citep{sawant2018application} for details of these arrays and optimized gas-surface interaction strategies employed in the SUGAR code.
\vspace{\baselineskip}

The 2-D, steady-state solution of the flow over a double wedge, previously simulated by \citet{tumuklu2018PhysRevF}, is extruded in the spanwise direction ($Y$) with as many replicas as the number of spanwise octrees. 
See figure~\ref{f:SBLI_Features} for understanding the simulation domain setup in the $X-Z$ plane, where $X$ and $Z$ are streamwise and streamwise-normal directions. 
The spanwise $Y$ boundaries are periodic.
From the inlet boundary, $X=0$, inward-directed ($X>0$) local Maxwellian flow is introduced at an average number density, bulk velocity, and temperature of $n_1$, $u_{x,1}$, and $T_{tr,1}$, respectively.
Particles with the same properties are also introduced within one mean-free-path distance from the $Z$ boundaries, such that the streamlines of the flow are parallel to the $Z$-boundaries.
If particles move out of the domain from either $X$ or $Z$ boundaries, they are deleted.
The chosen spanwise extent of $L_y$=28.8~mm was estimated from a preliminary simulation with a span length of 72~mm for 30 flow times and was expected to contain four spanwise periodic structures.
However, this turned out to be an underestimate, because when linear instability was detected after 50 flow times, the flow was found to exhibit a much larger spanwise wavelength.
The spanwise extent of the current simulation is long enough to capture one linearly growing periodic structure. 
Contours and isocontours detailing spanwise periodic structures are shown with two periodic wavelengths for clarity.
Note that a flow time, $T$, is defined as the time it takes for the flow to traverse a length of the separation bubble in the base (or mean) flow, $L_s=40$~mm, at a freestream velocity of $u_{x,1}$ where 
$L_s$ is defined as a straight-line distance from the separation point, $P_s$, to the reattachment point, $P_R$.
{\color{black}Note that the spanwise-periodic simulation takes $\sim$5 hours per flow time using 19.2k Intel Xeon Platinum 8280 (``Cascade Lake") processors of the \citet{Frontera}.}

\section{Features of Two-dimensional Base flow}~\label{sec:2DBase}
\begin{figure}[H]
    \centering
      \sidesubfloat[]{\label{f:SBLI_Features}{\includegraphics[width=0.46\textwidth]{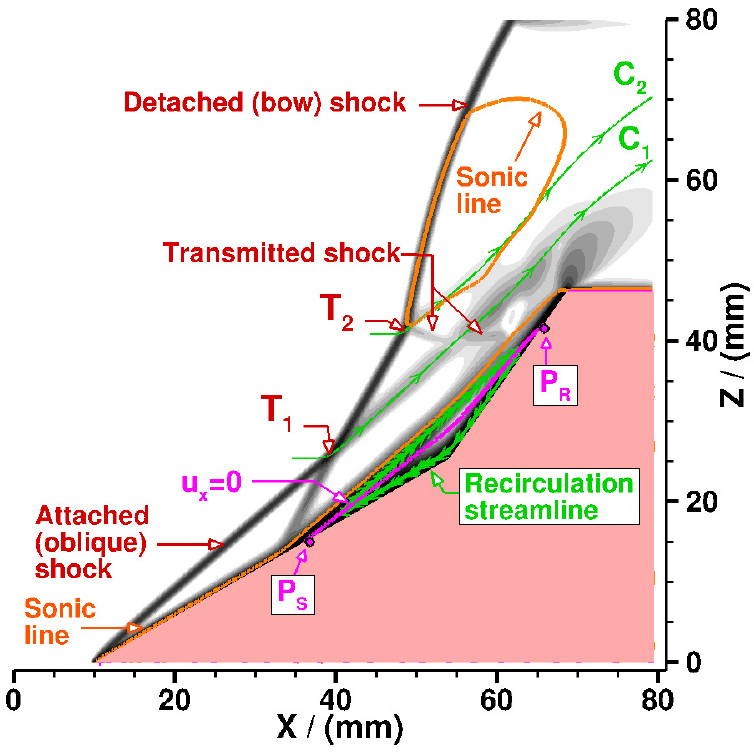}}}\,
      \sidesubfloat[]{\label{f:SBLI_ProbesAndPlanes}{\includegraphics[width=0.46\textwidth]{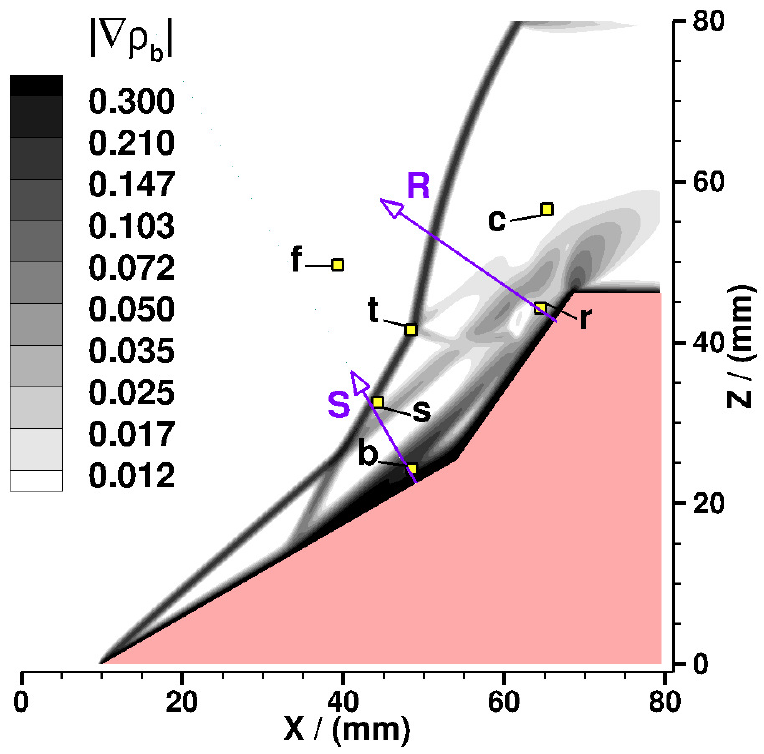}}}\,
    \caption{(\textit{a}) SBLI features shown in the magnitude of mass density gradient of the base flow, $|\nabla\rho_b|$, normalized by $\rho_1L_s^{-1}$, where $\rho_1$ is freestream mass density. Contour levels are shown in (\textit{b}).
(\textit{b}) On same flowfield, overlay of wall-normal directions $S$ and $R$, and numerical probes $b$ inside separation bubble ($X$=48.496~mm, $Z$=24.270~mm), $r$ at reattachment (64.396, 44.358), $s$ in the separation shock (44.165, 32.597), $c$ near contact surface (65.191, 56.593), $t$ at the triple point $T_2$ (48.347, 41.624), $f$ in the freestream (39.212, 49.722). $S$ and $R$ directions intersect the $X$-axis at 62 and 127~mm, respectively.
}
\label{f:SBLI}
\end{figure}

Figure~\ref{f:SBLI_Features} shows the typical features of an Edney-IV type SBLI~\citep{edney1968effects} in the base (or mean) flow, which is similar to that observed on double cones~\citep{druguet2005effects,babinsky_harvey_2011}.
The base flow macroscopic parameters are denoted by the subscript `$b$'.
For details of the time evolution of 2-D SBLI interaction over the double wedge, see the work of \citet{tumuklu2018PhysRevF}.
In summary, these features are formed by the interaction of a leading-edge attached (oblique) and detached (bow) shocks generated by the lower and upper wedge surfaces, respectively.
This interaction generates a transmitted shock that impinges on the upper wedge surface and increases the pressure and heat flux at the reattachment (or impingement) location, $P_R$.
The induced adverse pressure gradient results in the separation of the supersonic boundary layer on the lower wedge surface at $P_S$ and the formation of flow recirculation zone in the vicinity of the intersection of two surfaces, also known as the hinge.
Inside the separation bubble, a shear layer is represented by the line contour of $u_{x}=0$ from $P_S$ to $P_R$.
The separation zone significantly alters the SBLI system, such that the compression waves generated at the separation coalesce into a separation shock that interacts with the attached and the detached shocks at triple points.
Two contact surfaces, $C_1$ and $C_2$, are formed downstream of triple points $T_1$ and $T_2$, respectively. The former is between two supersonic streams formed downstream of the separation shock, and the latter is between the lower supersonic and upper, hotter subsonic flow formed downstream of the detached shock.
The transmitted shock is also affected by the contact surface $C_1$ and causes the reattachment point to move downstream, and the separation bubble to increase in size.
A reflected shock is formed downstream of the transmitted shock to guide the supersonic stream along the upper wedge surface.
If the upper wedge surface were longer, such interaction would have resulted in a $ \lambda $-shock pattern, which was observed on the double cone by \citet{tumuklu2018POF1,tumuklu2018POF2}.
Instead, the flow encounters the corner of the upper wedge and goes through the Prandtl-Meyer expansion.

The initial 2-D SBLI system moves slightly downstream within the first 30 flow times because of low spanwise relaxation that leads to a decrease in pressure downstream of the primary shocks.
This spanwise relaxation is induced by the thermal fluctuations of spanwise velocity about zero in the spanwise periodic simulation.
This is consistent with the fact that all macroscopic quantities fluctuate about their mean~\citep[][chapter XII]{landauLifshitz}.
A strictly imposed zero bulk velocity in the purely 2-D solution is unrealistic in that it does not account for such thermal fluctuations.
The new 2-D flow state is defined by spanwise and temporally averaging the solution between 48 to 60 flow times.
This is referred to as the base state, which fosters the growth of linear instability, detectable after 50 flow times.
Note that the DSMC-derived instantaneous data at 90.5 flow times, shown in this work, \emph{i.e.}, the boundary-layer profiles shown in section~\ref{sec:LinearInstab}, the perturbation flow field contours shown in section~\ref{sec:Correlation}, isocontours shown in section~\ref{sec:SpanPeriodicFlowStructures}, and the perturbation field used for superposition in section~\ref{sec:Topology}, are noise-filtered using the POD method (see Appendix~\ref{A:PODCompare}).
\vspace{\baselineskip}

In spite of molecular fluctuations, DSMC allows for the detection of the onset of instability.
Statistical mechanics predicts the standard deviation in the fluctuations of the directed bulk velocity such as $u_x$ in a gas at local equilibrium as, $\sqrt{R\langle T_{tr}\rangle/\langle N\rangle}$, where $R$, $\langle T_{tr}\rangle$, $\langle N\rangle$ are the gas constant, average translational temperature and average number of particles~\citep[][chapter XII]{hadjiconstantinou2003statistical, landauLifshitz}.
Similarly, we can estimate the level of spanwise fluctuations about the spanwise average in a 2-D flow at local equilibrium conditions exhibiting small-amplitude, self-excited fluctuations by calculating $\sqrt{R\langle T_{tr}\rangle_s/\langle N\rangle_s}$.
Subscript `$s$' attached to the averaged quantities denote a spanwise average.
If the DSMC-computed standard deviation is greater than the equilibrium estimate, then the fluctuations are not entirely thermal but are due to self-excited linear instability.
The only exception is the finite thick region of shock layers, where additional fluctuations are present due to strong translational nonequilibrium \citep{sawant2020Kinetic}.
This test was used as a first confirmation of the onset of linear instability at approximately 50 flow times, when the self-excited fluctuations in the separation bubble became slightly but noticeably larger than the thermal fluctuations.

\section{Three-dimensional Instability Mechanisms}\label{sec:3DMechanisms}
\subsection{Linear instability: growth rate and spatial origin}~\label{sec:LinearInstab}

\begin{figure}[H]
    \centering
      \sidesubfloat[]{\label{f:CF_trot_15}{\includegraphics[width=0.6\textwidth]{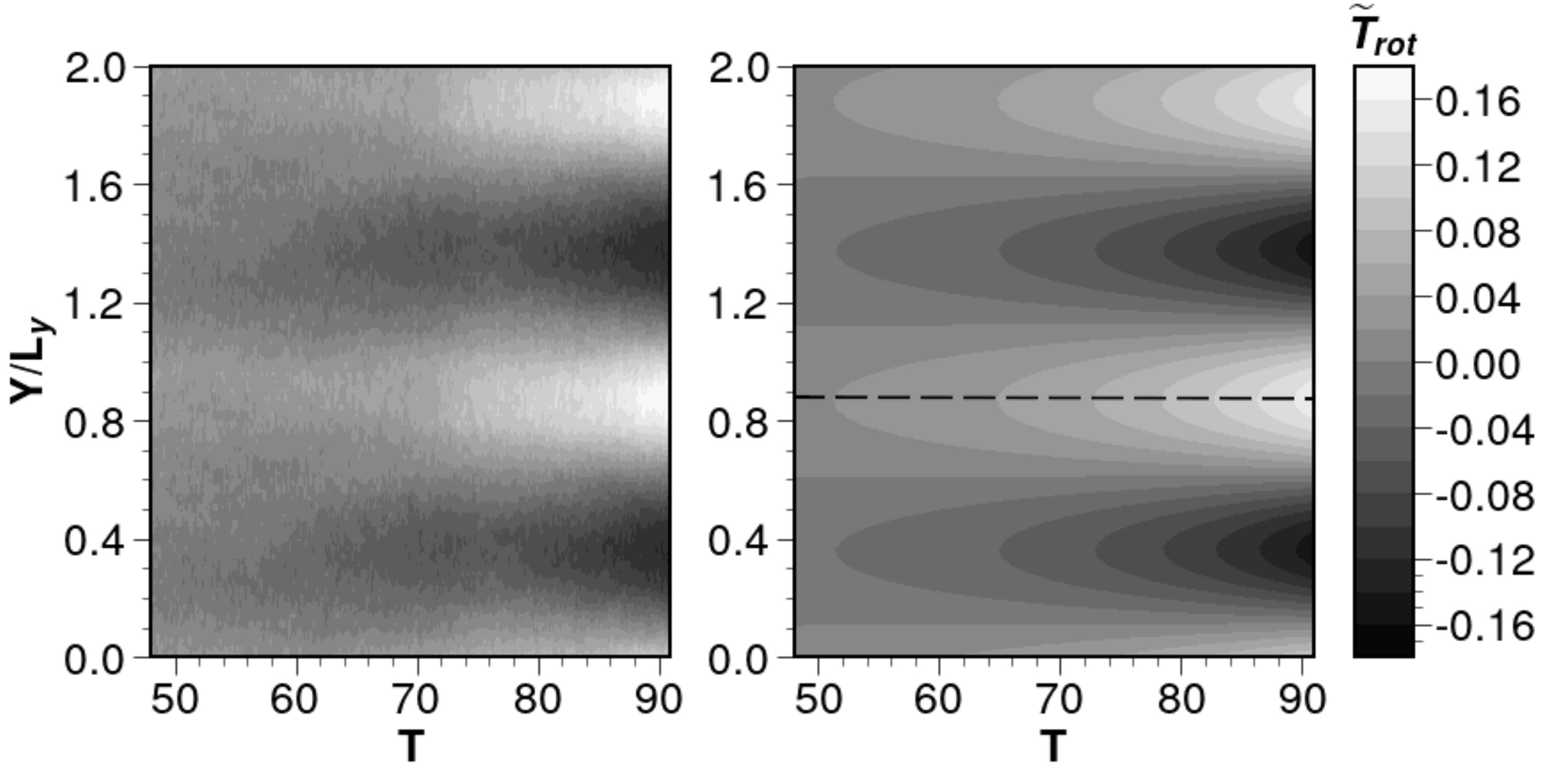}}}\,
      \sidesubfloat[]{\label{f:CF_All_15_1D}{\includegraphics[width=0.46\textwidth]{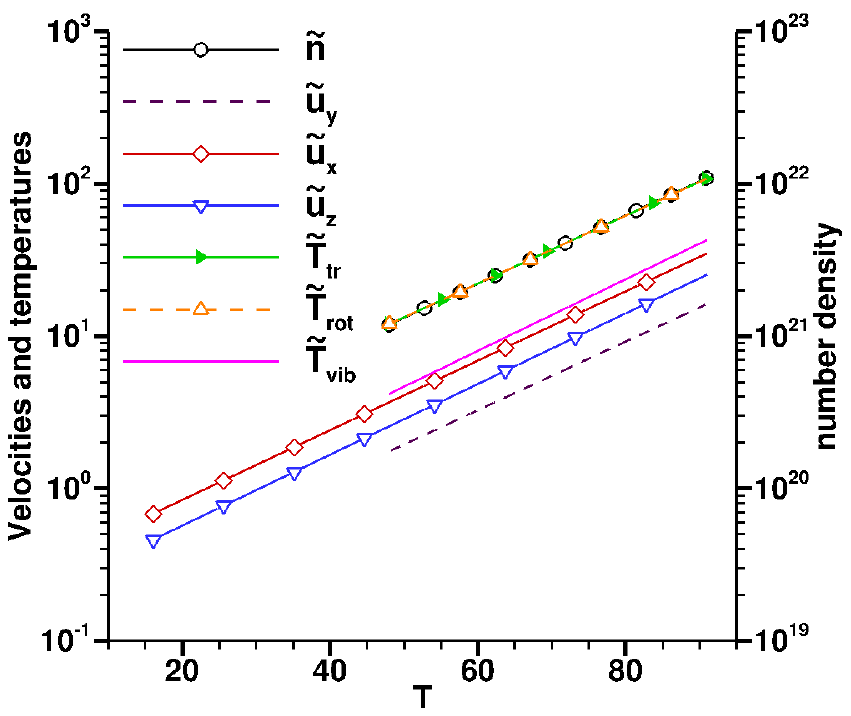}}}\,
      \sidesubfloat[]{\label{f:uyFitsAtOtherProbes}{\includegraphics[width=0.44\textwidth]{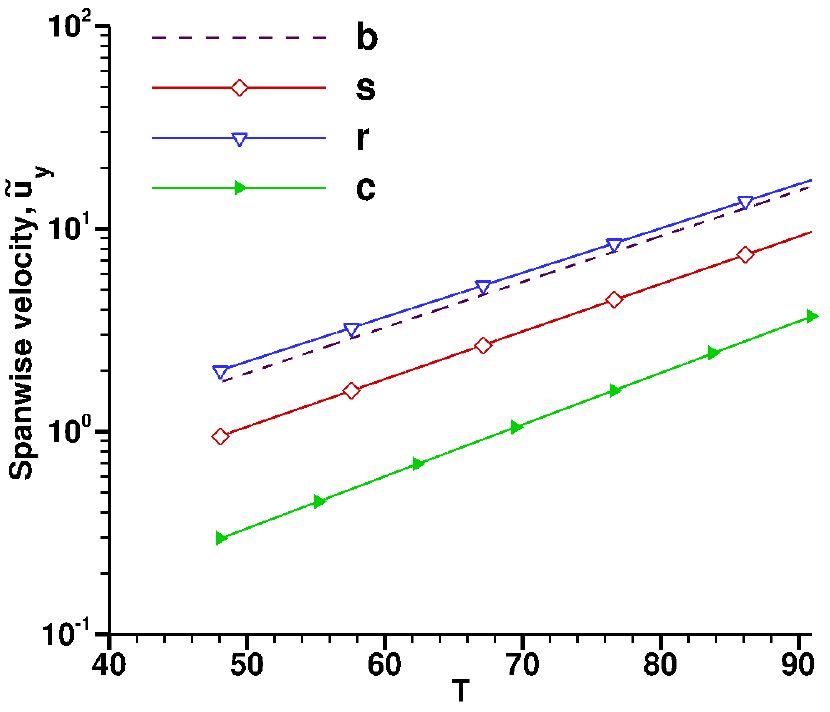}}}\,
    \caption{(\textit{a}) At probe $b$ inside the separation bubble, (left) temporal evolution of DSMC-derived perturbation rotational temperature, $\tilde{T}_{rot}$, normalized by freestream temperature, $T_{tr,1}$, and (right) 2-D linear fit. 
(\textit{b}) Comparison of linear fits of all residuals at a spanwise location that corresponds to the peak. For $\tilde{T}_{rot}$, it is indicated at $Y/L_y$=0.88. Same holds true for $\tilde{T}_{tr}$ and $\tilde{T}_{vib}$. For $\tilde{n}$, $\tilde{u}_x$, and $\tilde{u}_z$, it is at 1.38, whereas for $\tilde{u}_y$, it is at 1.13. (\textit{c}) Comparison of the linear fit of $\tilde{u}_y$ through the peak at probes $b$, $s$, $r$, $c$. For $b$ and $s$, the peak location is at $Y/L_y$=1.13, whereas for $r$ and $c$, it is at 0.63.}
\label{f:2DFits}
\end{figure}

A linear instability responsible for making the 2-D base flow unstable to self-excited spanwise-homogeneous perturbations is verified in figure~\ref{f:2DFits}.
Figure~\ref{f:CF_trot_15} shows the good comparison of the temporal evolution of perturbation rotational temperature $\tilde{T}_{rot}$, obtained from DSMC and a 2-D linear function that fits the DSMC solution.
Note that the perturbation part of a macroscopic flow variable $Q \in (n,u_x,u_y, u_z, T_{tr}, T_{rot}, T_{vib})$ is given by subtracting the 2-D base flow state $Q_b(x,z)$ as,
\begin{equation} 
\epsilon \tilde{Q}(x,y,z,t)=Q(x,y,z,t)-Q_b(x,z)
\label{QperturbFromQandQb}
\end{equation}
Note that $\epsilon << 1$, which indicates the the perturbation is small.
$\tilde{n}$ is the perturbation number density, $\tilde{u}_x,\tilde{u}_y, \tilde{u}_z$ are perturbation velocities in the $X$, $Y$, and $Z$ directions, and $\tilde{T}_{tr}, \tilde{T}_{rot}, \tilde{T}_{vib}$ are perturbation translational, rotational, and vibrational temperatures, respectively.
A DSMC-computed perturbation flow parameter $\tilde{Q} \in (\tilde{n},\tilde{u}_x,\tilde{u}_y, \tilde{u}_z, \tilde{T}_{tr}, \tilde{T}_{rot}, \tilde{T}_{vib})$ is fitted by a linear function written as,
\begin{equation} 
\centering
\begin{split}
\tilde{Q}(x,y,z,t) &= \hat{Q}(x,z)\exp{(i\Theta)} + c.c.\\
\end{split}
\label{Ansatz}
\end{equation}
where $\hat{Q}(x,z)$ is a spanwise homogeneous amplitude function, and $\Theta$ is a phase function of the linear perturbation that has the form,
\begin{equation} 
\centering
\Theta = \beta y - \Omega t
\label{Theta}
\end{equation}
$\beta=2\pi/L_y$ is a real spatial wavenumber indicating spanwise wavelength of the mode, $\Omega=\omega_r + i\omega_i$ is a complex parameter, whose real part indicates frequency and the imaginary part is the growth rate in time $t$, and $c.c.$ indicates complex conjugation so that $\tilde{Q}$ is real.
A 2-D linear fit is performed using the generalized least-squares method using Python's \citet{LMFIT} module, which gives the mean value of unknown fit parameters, $\omega_i$, $\hat{Q}$, $\omega_r$, and $1\sigma$-uncertainty (standard error) in these parameters.
These are listed in table~\ref{tab:AllMacroparamGrowthRates}.
Note that by keeping $\omega_r$ as an unknown resulted in a small number for $\omega_r$ and imposing it as $\omega_r=0$ did not change the value of other three fit parameters, indicating that the linearly growing mode is stationary.

Similar linear fits are performed on other DSMC-computed macroscopic flow parameters, and a 1-D extracted curve passing through the peak spanwise structure, such as that marked in figure~\ref{f:CF_trot_15} by a dashed line, is compared in figure~\ref{f:CF_All_15_1D}.
All curves are parallel to each other, indicating similar growth rates.
Also, figure~\ref{f:uyFitsAtOtherProbes} shows the comparison of curve-fitting functions through the peak structure of $\tilde{u}_y$ of probe $b$ with probes at other important locations, $s$, $r$, and $c$.
Nearly parallel curves are observed, which indicates that linear growth is global.
By comparing the absolute values of the amplitude of $\tilde{u}_y$, it is seen that probes $r$, $b$, $s$, $c$ have largest to lowest amplitude, indicating decreasing magnitude of perturbation.
The average of the mean growth rate for each parameter listed in table~\ref{tab:AllMacroparamGrowthRates} is $\omega_i=5.0$~kHz, with bounds of +0.16\% and -0.16\%.
A maximum deviation of 11.4\% is observed at probe $c$.

\begin{table}[H]
\centering
\normalsize
\begin{threeparttable}
\caption{2-D linear curve fit parameters in equations~\ref{Ansatz} and~\ref{Theta} corresponding to figures~\ref{f:CF_All_15_1D} and~\ref{f:uyFitsAtOtherProbes}.}
\label{tab:AllMacroparamGrowthRates}       
\begin{tabular}{cccccc}
\hline\noalign{\smallskip}
\textbf{\makecell{Perturbation\\parameter$^a$}} & \textbf{\makecell{Growth rate\\$\omega_i$/(kHz)}}   & \textbf{\makecell{Amplitude\\ $\hat{Q}$}} \\\hline
$\tilde{n}       $  & 4.91 $\pm$ 0.06\%     &  -5.013e+19 $\pm$ 0.24\%    \\
$\tilde{u}_x     $  & 4.90 $\pm$ 0.07\%     &  -0.1613 $\pm$ 0.30\%       \\
$\tilde{u}_z     $  & 4.95 $\pm$ 0.08\%     &  -0.1108 $\pm$ 0.33\%       \\
$\tilde{T}_{tr}  $  & 4.88 $\pm$ 0.04\%     &  0.5111 $\pm$ 0.17\%         \\
$\tilde{T}_{rot} $  & 4.88 $\pm$ 0.05\%     &  0.5128 $\pm$ 0.19\%         \\
$\tilde{T}_{vib} $  & 5.15 $\pm$ 0.11\%     &  0.1560 $\pm$ 0.51\%        \\
$\tilde{u}_y$       & 4.89 $\pm$ 0.10\%     &  0.0762 $\pm$ 0.43\%        \\
$\tilde{u}_y$ (at $s$)  & 5.12 $\pm$ 0.26\% &  0.03648 $\pm$ 1.14\%       \\
$\tilde{u}_y$ (at $r$)  & 4.77 $\pm$ 0.11\% &   -0.0914 $\pm$ 0.46\%      \\
$\tilde{u}_y$ (at $c$)  & 5.55 $\pm$ 0.66\% &   -0.0092 $\pm$ 3.20\%      \\
\noalign{\smallskip}\hline\noalign{\smallskip}
\end{tabular}
\begin{tablenotes}
\item $^{a}$ Probe locations other than $b$ are explicitly denoted.
\end{tablenotes}
\end{threeparttable}
\end{table}
In comparison, \citet{tumuklu2018PhysRevF}, using the POD analysis, had found a least damped eigenmode of $-5.88$~kHz that leads the 2-D (spanwise independent) solution to reach steady state, unlike we find here.
Also, our growth rate is larger than that obtained by \citet{sidharth2018onset}, which is consistent with their finding that a larger growth rate is expected for a larger angle difference between the upper and lower wedges.
They performed a Mach 5 hypersonic flow of calorically perfect gas and obtained a nondimensional growth rate of approximately $\num{7.5e-4}$ for a 12$^{\circ}$-20$^{\circ}$ double wedge (angle difference of 8$^{\circ}$).
Following their nondimensionalization, where the growth rate is multiplied by the $\delta_{99}$ boundary-layer thickness at separation equal to 3.35~mm, and divided by the freestream velocity downstream of the leading-edge shock derived from the inviscid shock theory~\citep{andersonModern} for observed shock angle of 41$^{\circ}$, $u_{x,2}=2930.8$, we obtain a value of $\num{0.0057}$.
\vspace{\baselineskip}

\begin{figure}[H]
    \centering
        \sidesubfloat[]{\label{f:d1Tod10Profiles}{\includegraphics[width=0.44\textwidth]{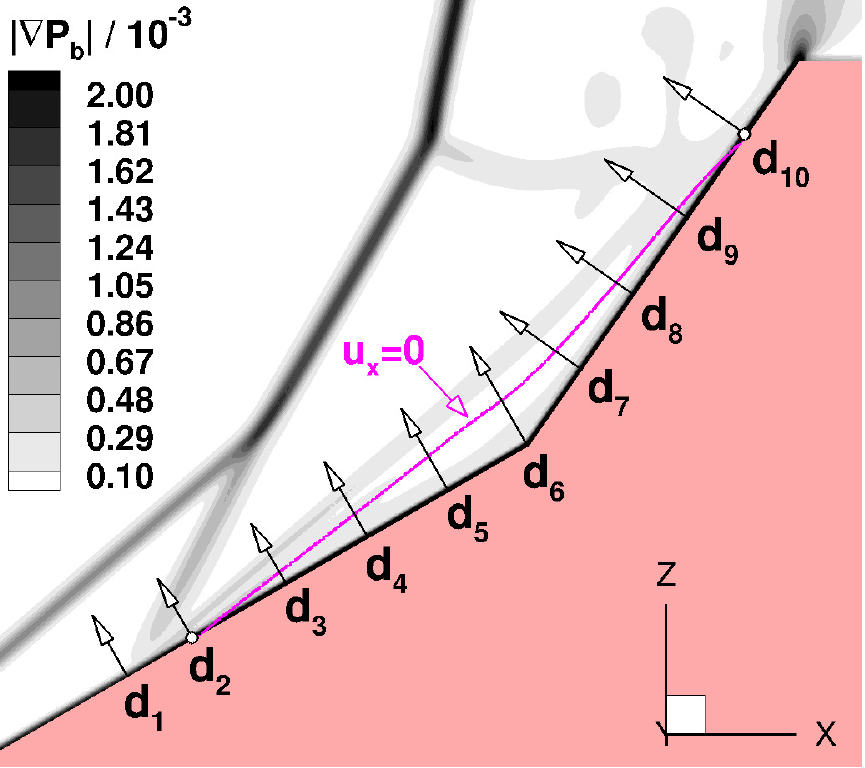}}}\,
        \sidesubfloat[]{\label{f:d1Tod3}{\includegraphics[width=0.47\textwidth]{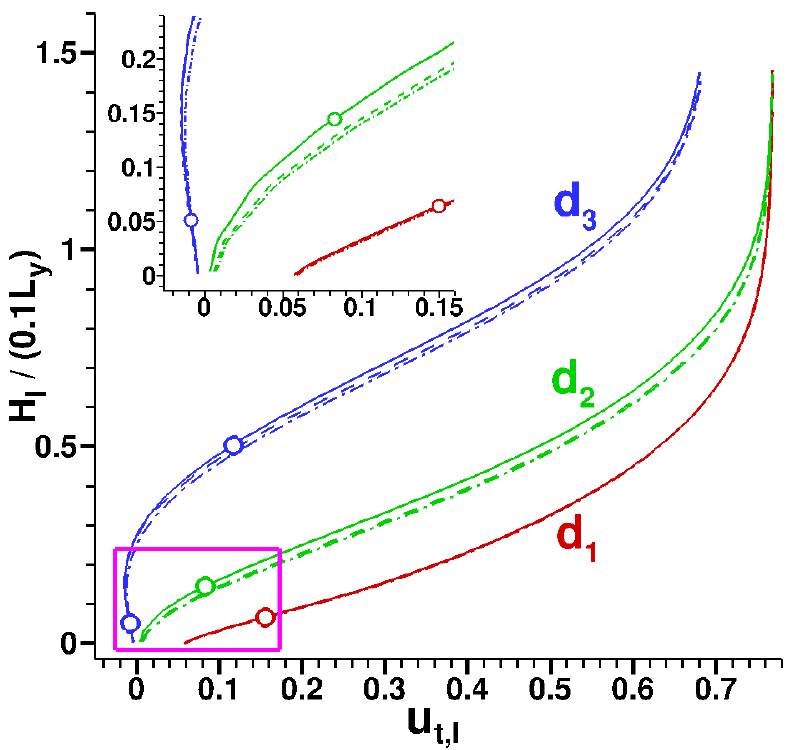}}}\,
        \sidesubfloat[]{\label{f:d4Tod6}{\includegraphics[width=0.47\textwidth]{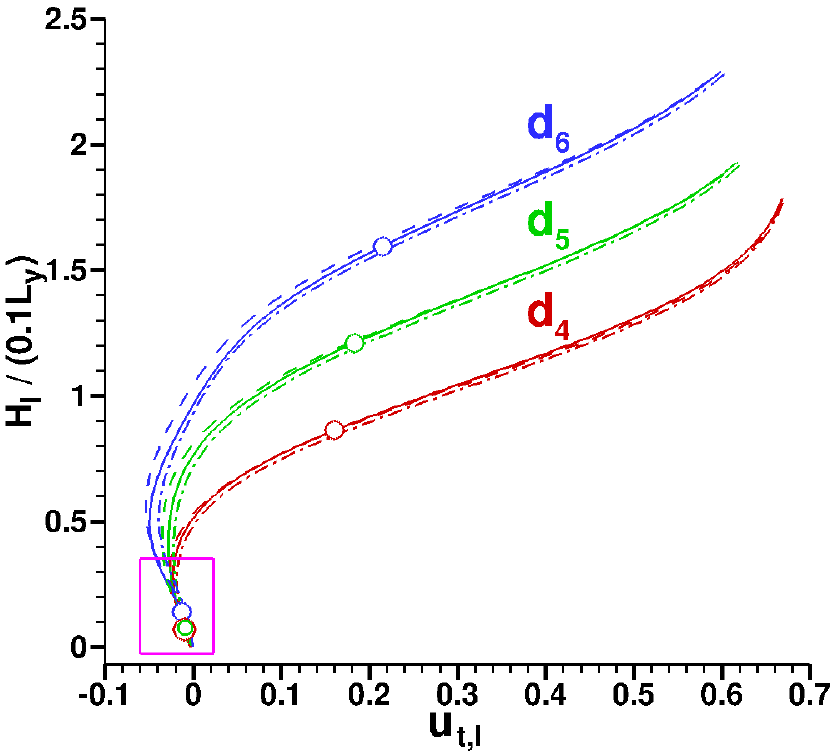}}}\hfill
        \sidesubfloat[]{\label{f:d4Tod6_Zoom}{\includegraphics[width=0.47\textwidth]{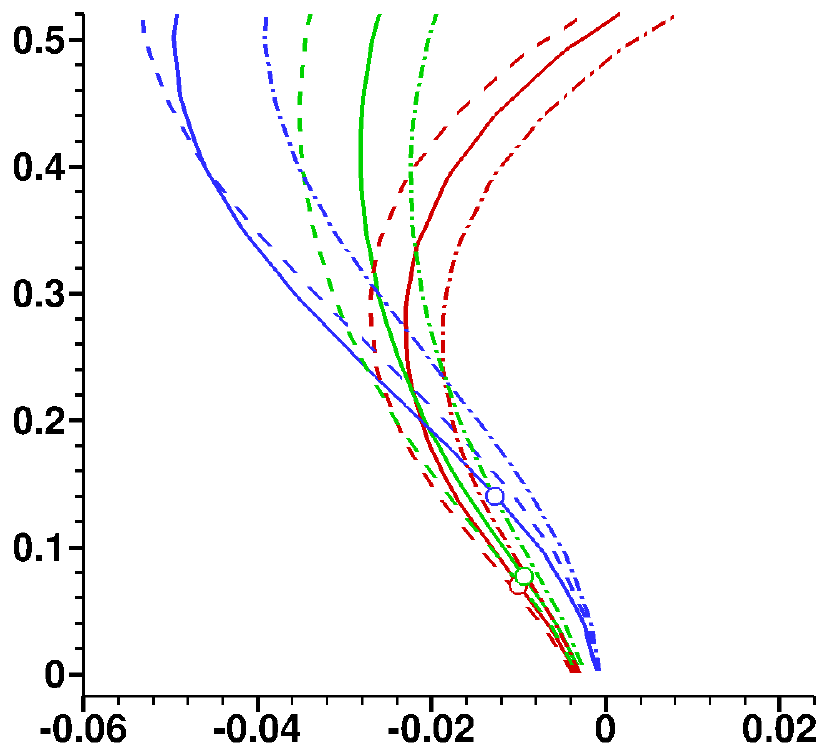}}}\,
        \sidesubfloat[]{\label{f:d7Tod10}{\includegraphics[width=0.47\textwidth]{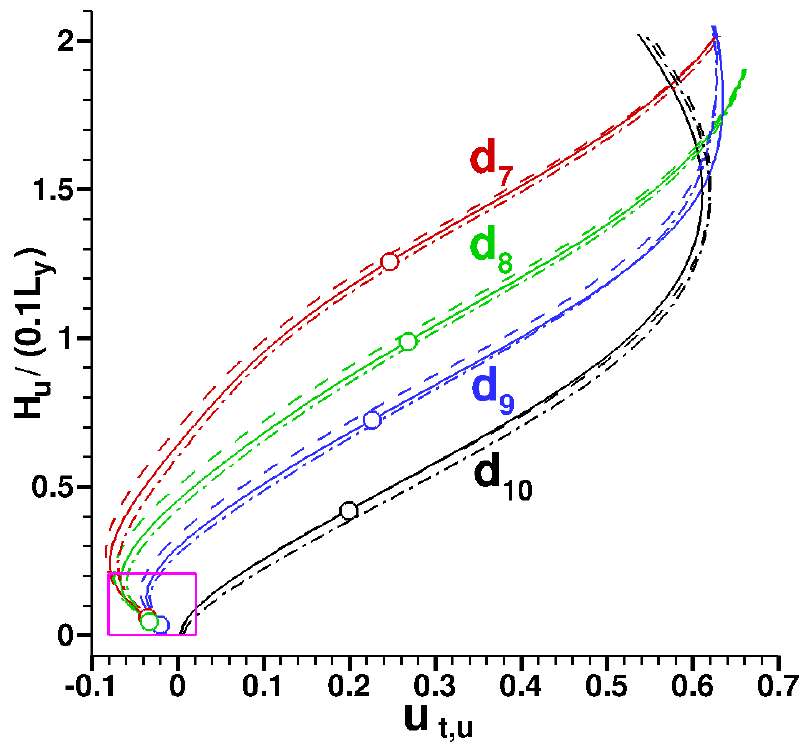}}}\hfill
        \sidesubfloat[]{\label{f:d7Tod10_Zoom}{\includegraphics[width=0.47\textwidth]{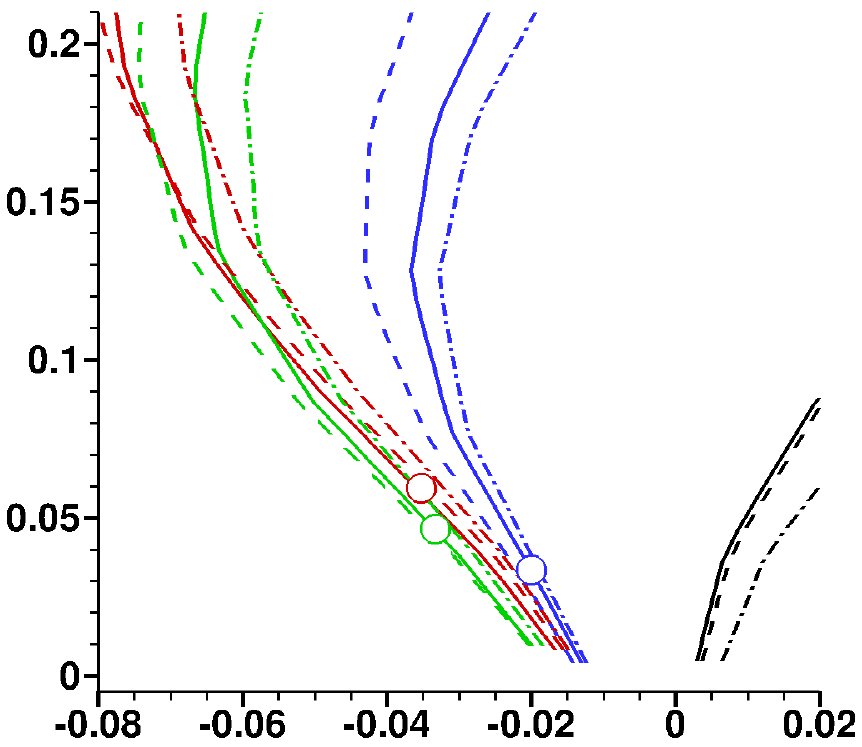}}}\,
        \caption{(\textit{a}) Base flow pressure gradient magnitude, $|\nabla p_b|$, normalized by $p_1L_s^{-1}$, where $p_1$ is freestream pressure. Overlaid wall-normal directions $d_1$ to $d_{10}$ are shown at a local streamwise distance from the hinge normalized by the length of separation $L_s$ as -0.625, -0.5225, -0.375, -0.25, -0.125, 0, 0.125, 0.25, 0.375, 0.51, respectively. (\textit{b}) Local streamwise velocity tangential to lower wedge surface $u_{t,l}$ normalized by freestream velocity, $u_{x,1}$, versus wall-normal height at locations $d_1$, $d_2$, $d_3$. Insert shows zoom of the marked rectangular box. (\textit{c}) Similar profiles at locations $d_4$, $d_5$, $d_6$. (\textit{d}) Zoom of the rectanular region marked in (\textit{c}). (\textit{e}) Local streamwise velocity tangential to upper wedge surface, $u_{t,u}$, normalized by $u_{x,1}$ versus wall-normal height at locations $d_7$, $d_8$, $d_9$, $d_{10}$. (\textit{f}) Zoom of the rectanular region marked in (\textit{e}). \\Legends for (\textit{b}) to (\textit{f}):\protect (\full) base state profile, (\dashed) profile on an $X-Z$ slice passing through location $A$ ($Y/L_y$=0.88) at $T=90.5$, (\chain) profile on an $X-Z$ slice passing through $B$ ($Y/L_y$=1.38) at $T=90.5$.}
\label{f:BLProfiles}
\end{figure}

Now we turn to the question of the spatial origin of the linear instability and answer whether these spanwise structures seen in figure~\ref{f:CF_trot_15} start upstream, at or inside the separation bubble by comparing the boundary layer profiles at wall-normal directions $d_1$ to $d_{10}$ shown in figure~\ref{f:BLProfiles}.
These are denoted in figure~\ref{f:d1Tod10Profiles} on top of the contours of pressure gradient magnitude, $|\nabla p_b|$, in the base flow, which identifies the location of shock structure and the recirculation zone.
The shear layer ($u_x=0$) and the separation and reattachment points are also overlaid.
Along each wall-normal direction, three boundary layer profiles are shown--one in the base flow and two at $T$=90.5 on spanwise locations $Y/L_y$=0.88 (A) and 1.38 (B).
These spanwise locations correspond to a spanwise peak and a trough of the local-streamwise (or wall-tangential) velocity so that the maximum spanwise deviation at $T=90.5$ from the base flow state can be assessed.
For profiles corresponding to the lower wedge, $d_1$ to $d_6$, the local-streamwise velocity, denoted as $u_{t,l}$, is plotted as a function of wall-normal height $H_l$. 
Subscript $`t'$ stands for the wall-tangential (or local-streamwise) component and $`l'$ is associated with the lower wedge surface.
For profiles corresponding to upper wedge, $d_7$ to $d_{10}$, the local-streamwise velocity, denoted as $u_{t,u}$, is plotted as a function of wall-normal height $H_u$. 
Similarly, subscript $`l'$ is associated with the lower wedge surface.
Note that $H_l$ and $H_u$ are zero at the respective surfaces.
\vspace{\baselineskip}

The boundary layer profiles just upstream of separation shock ($d_1$), at the separation ($d_2$), and just downstream of separation ($d_3$) are shown in figure~\ref{f:d1Tod3}.
At $d_1$, all profiles overlap, indicating that the flow is 2-D upstream of the separation.
Along $d_2$, at the separation, the absolute maximum difference of 0.72\% of the freestream velocity, $u_{x,1}$, is seen between $A$ and $B$ profiles at $H_l/(0.1L_y)=0.29$, which indicates spanwise modulation.
The difference decreases above this height but remains nonzero even inside the shock layer, indicating the origin of linear instability inside the interaction region of the separation shock layer with the LSB. 
Profiles $A$ and $B$ also differ from the base state profile, indicating deviation from the base flow.
Along $d_3$, just inside the separation bubble, $A$ and $B$ profiles deviate from each other by a maximum of 1\% at $H_l/(0.1L_y)=0.7$.
Further inside the separation bubble, along directions $d_4$, $d_5$, and $d_6$, similar profiles are shown in figures~\ref{f:d4Tod6} and~\ref{f:d4Tod6_Zoom}, where the latter figure is a zoom of the rectanular boxed region denoted in the former.
The absolute maximum deviation between $A$ and $B$ profiles increases along the local streamwise direction.
At $d_4$, $d_5$ and $d_6$, it is 1.34, 1.92, 2.52\% at locations $H_l/(0.1L_y)=0.88, 1.11, 1.57$, respectively, 
For $d_5$ and $d_6$ directions, these profiles are on either side of their respective base profiles, indicating spanwise modulation about the base flow.
On the upper wedge surface, the boundary layer profiles are shown along $d_7$ to $d_{10}$ in figures~\ref{f:d7Tod10} and~\ref{f:d7Tod10_Zoom}, where the latter figure is a zoom of the rectangular boxed region denoted in the former.
The difference between $A$ and $B$ is even larger on the upper wedge, indicating larger amplitude of spanwise perturbations.
At $d_7$, $d_8$, $d_9$, it is 2.8, 3.33, 3.34\% at locations $H_u/(0.1L_y)=0.6, 0.59, 0.62$, respectively, 
At $d_{10}$ at the reattachment location, the maximum difference decreases to 2.46\% at $H_u/(0.1L_y)=0.44$.

The generalized inflection point (GIP) is also denoted on each boundary layer profile (open circle).
Profiles $d_1$, $d_2$ and $d_{10}$ have only one GIP, whereas profiles $d_3$ to $d_9$, inside the separation bubble, have two GIPs.
The GIP closest to the wall is induced in the recirculating flow between the shear layer and the surface.
The GIP located farthest from the wall is induced between the shear layer and the supersonic flow outside the separation bubble.
From profiles $d_3$ to $d_6$, the upper inflection point moves further away from the wall as the distance between the top enclosure of the bubble and the wall increases.
The lower inflection point, more clearly seen in the respective zooms, also moves away from the surface as the distance between the shear layer and the wall increases.
From profiles $d_7$ to $d_{9}$, both inflection points move closer to the wall.

Additionally, notice that each profile exhibits a non-zero local streamwise velocity at the wall, the magnitude of which is maximum before the separation, lowest inside the separation zone on the lower wedge, and relatively larger on the upper wedge.
This variation is explained by the rarefaction effects at the wall, more details of which are provided in section~\ref{sec:Slip}.

\subsection{Correlation between the shock and separation bubble}\label{sec:Correlation}
\begin{figure}[H]
    \centering
       \sidesubfloat[]{\label{f:uy_R}{\includegraphics[width=0.46\textwidth]{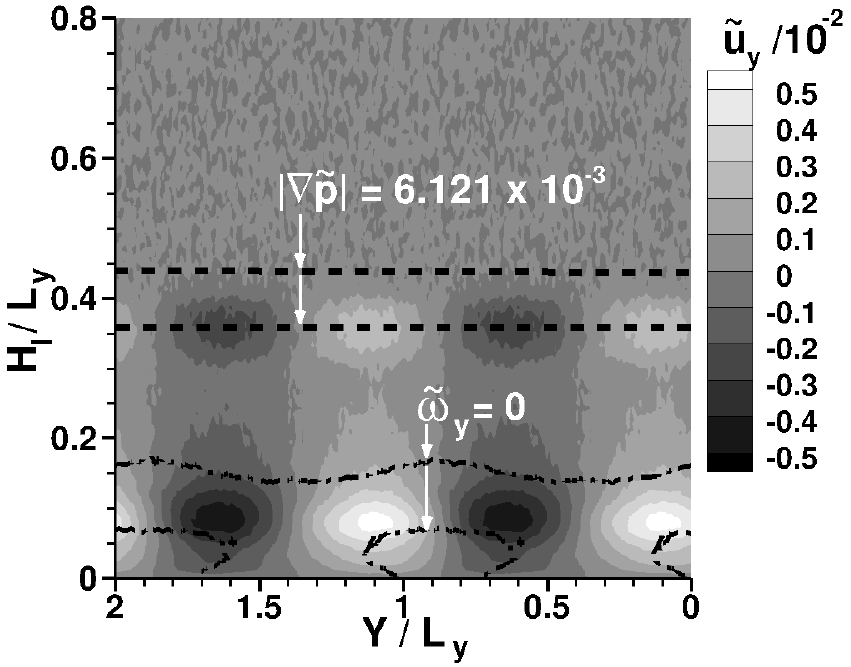}}}\hfill
       \sidesubfloat[]{\label{f:SepShock}{\includegraphics[width=0.46\textwidth]{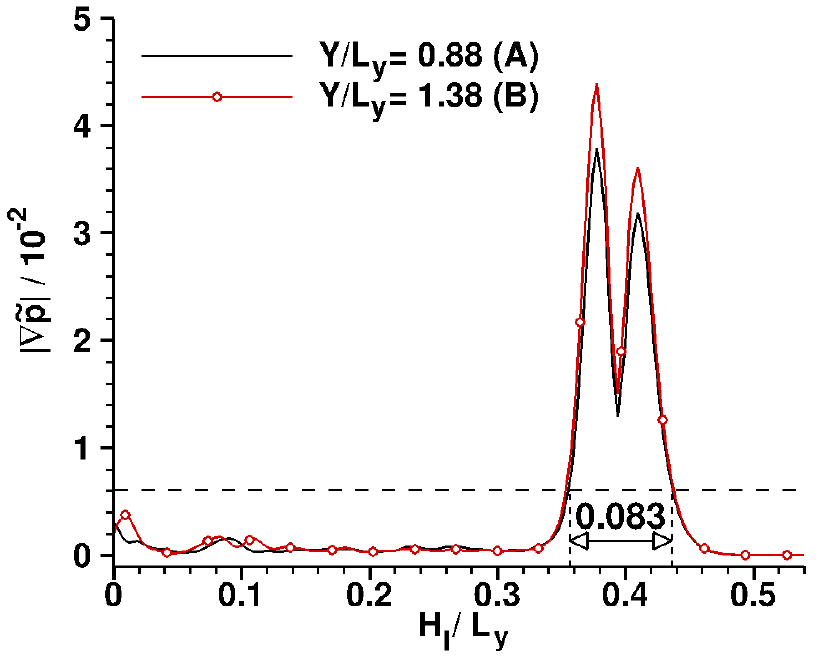}}}\hfill
        \sidesubfloat[]{\label{f:uy_S}{\includegraphics[width=0.46\textwidth]{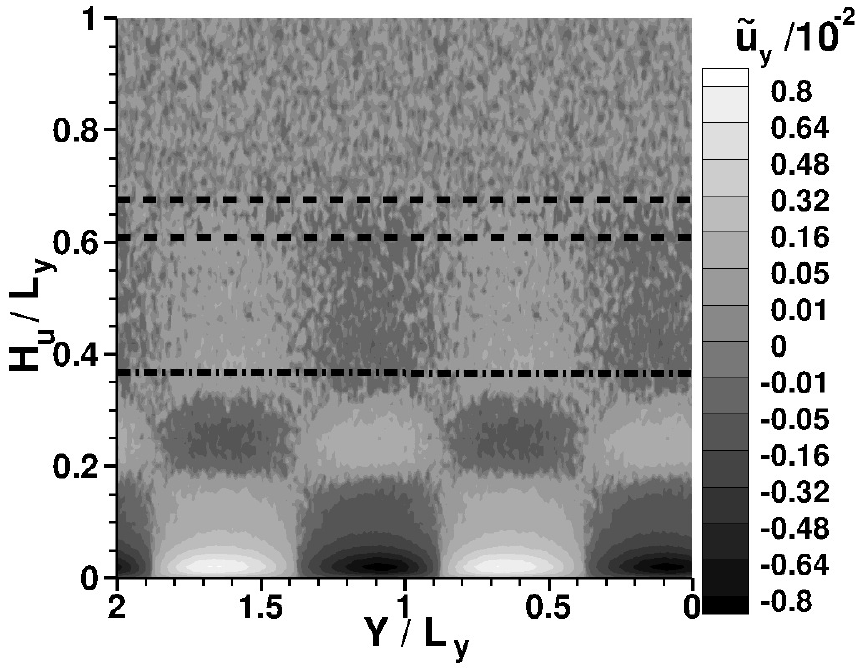}}}\,
    \caption{(\textit{a}) Contours of $\tilde{u}_y$ normalized by $u_{x,1}$ at $T$=90.5 on a plane defined along wall-normal direction $S$, marked in figure~\ref{f:SBLI_ProbesAndPlanes}. $X$ and $Y$ axes are the normalized span and the wall-normal height, respectively. (\textit{b}) Perturbation pressure gradient magnitude, $|\nabla \tilde{p}|$, normalized by $p_1L_s^{-1}$ along direction $S$ as a function of wall-normal height at two spanwise locations $A$ and $B$. (\textit{c}) Contours of $\tilde{u}_y$ normalized by $u_{x,1}$ at $T$=90.5 on plane defined along wall-normal direction $R$, marked in figure~\ref{f:SBLI_ProbesAndPlanes}. Overlaid line contours: ({\color{black}\dashed}) $|\nabla \tilde{p}|=\num{6.121e-3}$ and ({\color{black}\chain}) $\tilde{\omega}_y$=0.}
\label{f:uy_RST}
\end{figure}
The self-excited linear instability leads to the presence of spanwise periodic flow structures in perturbation flow parameters with a spanwise wavelength of $L_y$.
Figure~\ref{f:uy_RST} shows the contours of spanwise perturbation velocity, $\tilde{u}_y$, at $T$=90.5 in the wall-normal planes $S$ and $R$ denoted in figure~\ref{f:SBLI_ProbesAndPlanes}.
On the $S$-plane, the spanwise periodic flow structures inside the separation bubble are seen between the surface ($H_l$=0) and the upper envelope of the separation bubble at $H_l=0.15L_y$ where the spanwise vorticity, $\tilde{\omega}_y$, is zero, as shown in figure~\ref{f:uy_R}.
These structures have elliptical cross-sections with major and minor axes of lengths roughly equal to 0.4$L_y$ and 0.2$L_y$, respectively.
Note that the upper envelope of the bubble also has a spanwise sinusoidal shape.
The overlaid line contours of zero spanwise vorticity between $0 < H_l < 0.1L_y$ that are elliptical in shape shows a 90$^{\circ}$ phase shift in its spanwise mode and that of the spanwise velocity, \emph{i.e.}, the center of the circular structure of $\tilde{\omega}_y$ is at $Y$=0.88$L_y$, inbetween a peak and a trough of $\tilde{u}_y$.
The spanwise vorticity of the flow is negative inside these elliplical shaped contour lines of $\tilde{\omega}_y=0$, i.e. the flow rolls down the surface, and it is positive outside this zone and below the $\tilde{\omega}_y$=0 contour line at $H_l=0.15L_y$, i.e. the flow rolls up the surface.
This shows that the flow moves in the spanwise direction while swirling about the spanwise axis ($Y$).
\vspace{\baselineskip}

Further away from the wall, figure~\ref{f:uy_RST} shows, for the first time, the spanwise periodic flow structures inside the strong gradient region of the separation shock ($0.36 < H_l/L_y < 0.44$).
These structures are in phase with structures inside the separation bubble and they have the same periodicity length. 
This is consistent with the boundary-layer profiles shown in the previous section that showed the origin of linear instability inside the separation shock layer and the linear stability analysis that showed identical growth rate inside the LSB (probe $b$) and the separation shock (probe $s$).
Note that the approximate boundary of the finite shock is marked by dashed horizontal lines corresponding to the isocontour line of normalized perturbation pressure gradient magnitude, $|\nabla \tilde{p}|=\num{0.612e-2}$.
To justify the choice of this value, figure~\ref{f:SepShock} shows the variation of $|\nabla \tilde{p}|$ as a function of wall-normal height, $H_l$, along the $S$-plane at two spanwise locations, $A$ ($Y/L_y$=0.88) and $B$ ($Y/L_y$=1.38).
The rapid increase of $|\nabla \tilde{p}|$ at $H_l=0.36L_y$ is indicative of the separation shock, inside of which the value of $|\nabla \tilde{p}|$ far exceeds that in the vicinity of the surface.
Note that the thickness of the shock layer, $0.083L_y=2.39$~mm, is comparable the boundary-layer thickness at separation, $\delta_{99}=3.35$~mm.
The locations $A$ and $B$ correspond to the peak and trough of the sinusoidal modulation of $|\nabla \tilde{p}|$ inside the separation shock.
The difference between the two profiles also highlights the spanwise changes inside the shock layer.
\vspace{\baselineskip}

In the $R$-plane at the reattachment, a similar contour plot of $\tilde{u}_y$ is shown in figure~\ref{f:uy_S}, which exhibits spanwise periodic structures inside the reattached boundary layer.
Such structures also exist in the vicinity of contour line $\tilde{\omega}_y$=0 at $H_u=0.36L_y$, which indicates the presence of a contact surface $C_2$ downstream of the triple point $T_2$ at the intersection of separation and detached shocks.
Further away from the wall, the contour lines of $|\nabla \tilde{p}|$ at $H_u=0.61L_y$ and $0.677L_y$ indicate the approximate layer of detached shock, which is slightly smaller in thickness than the separation shock because the detached shock strength is higher.
The spanwise structures inside this shock are not as noticeable as the separation shock.
\vspace{\baselineskip}

\begin{figure}[H]
    \centering
        \sidesubfloat[]{\label{f:n_R}{\includegraphics[width=0.46\textwidth]{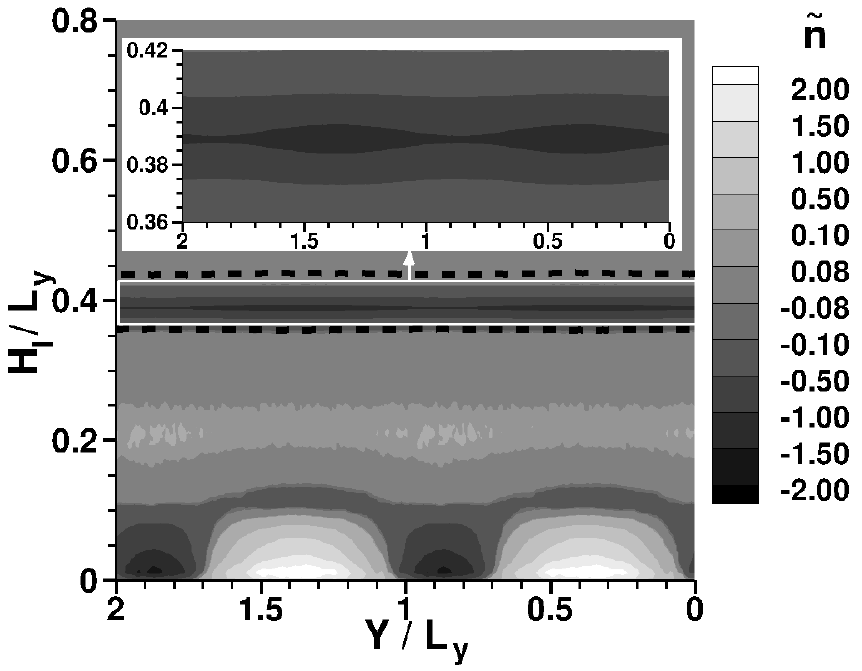}}}\hfill
       \sidesubfloat[]{\label{f:ut_R}{\includegraphics[width=0.46\textwidth]{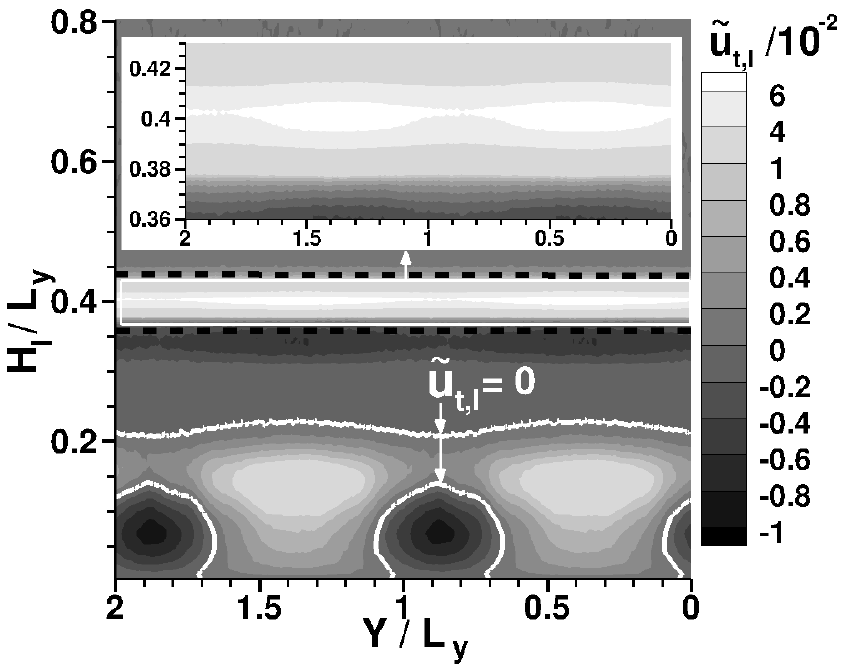}}}\hfill
       \sidesubfloat[]{\label{f:un_R}{\includegraphics[width=0.46\textwidth]{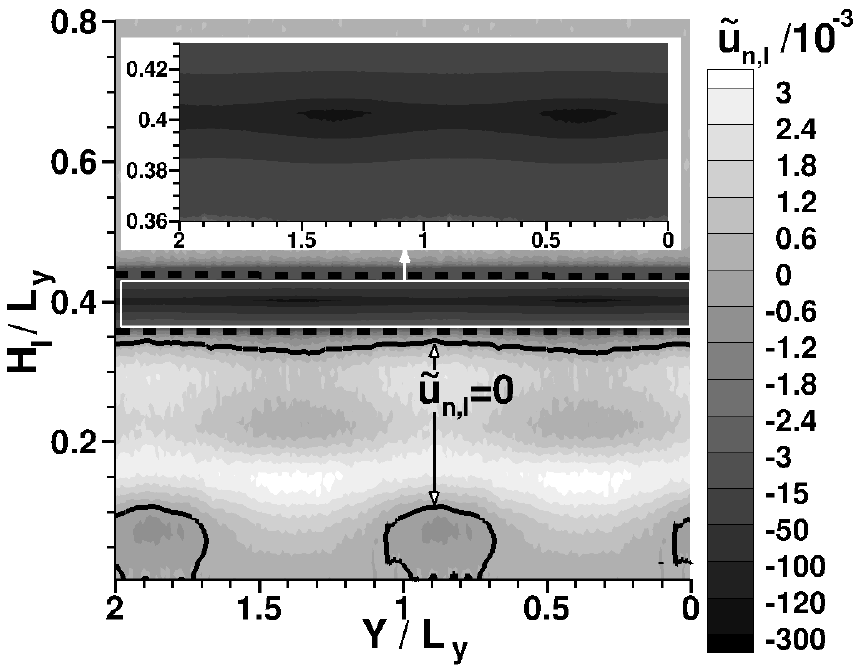}}}\,
     \sidesubfloat[]{\label{f:Ttr_R}{\includegraphics[width=0.46\textwidth] {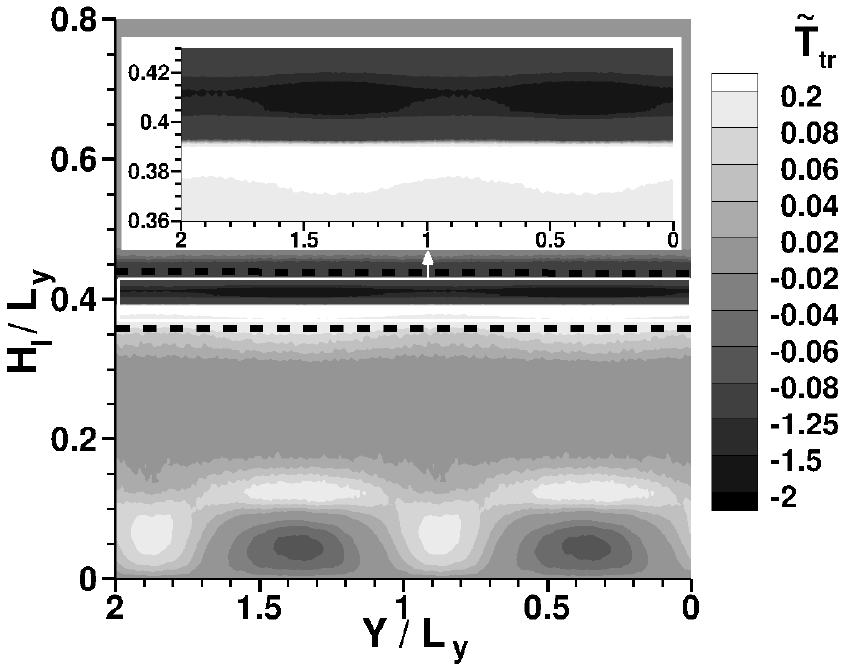}}}\hfill
     \sidesubfloat[]{\label{f:Trot_R}{\includegraphics[width=0.46\textwidth]{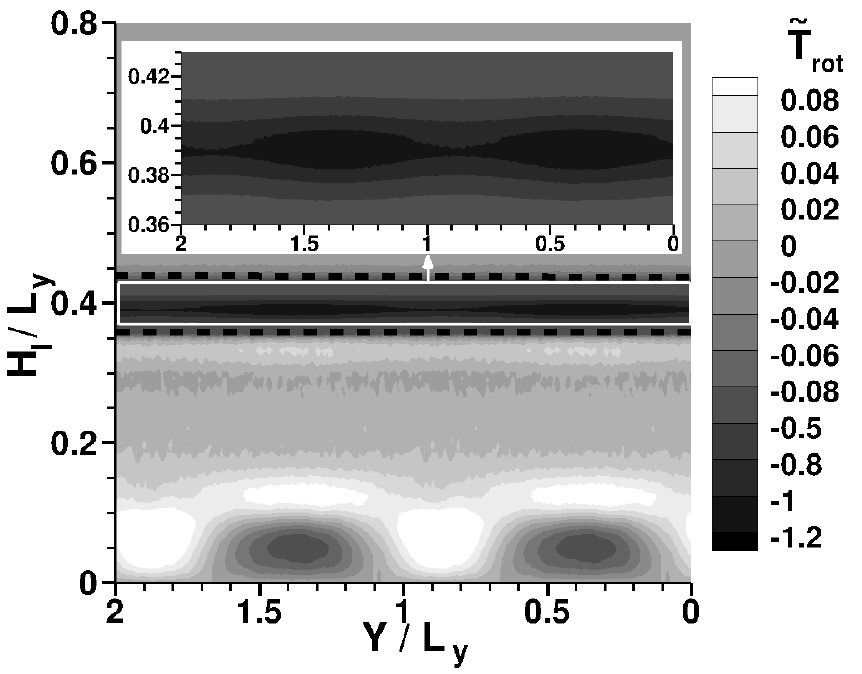}}}\,
      \sidesubfloat[]{\label{f:Tvib_R}{\includegraphics[width=0.46\textwidth] {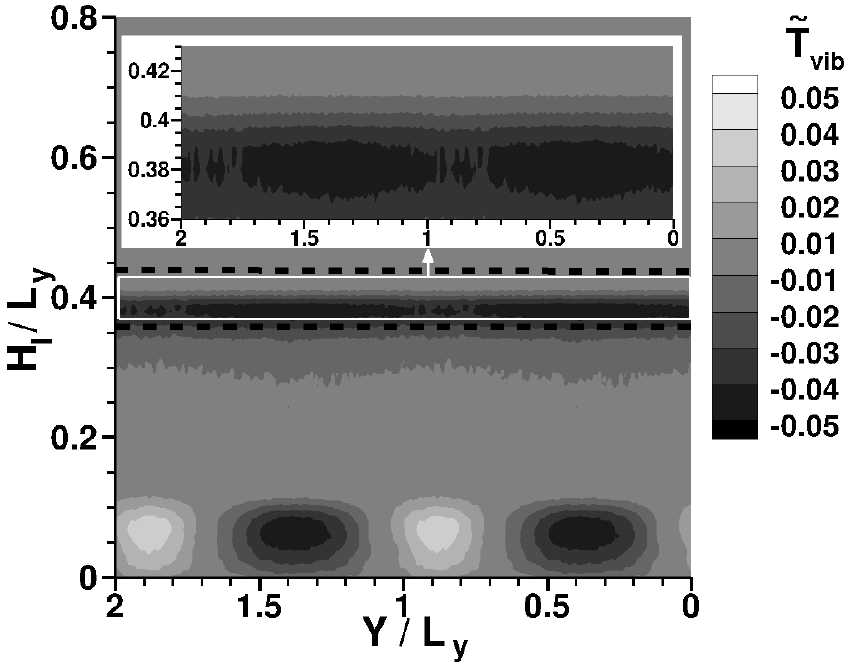}}}\,
    \caption{Contours of perturbation macroscopic flow parameters at $T$=90.5 on a plane defined along $S$, same as figure~\ref{f:uy_R}. (\textit{a}) number density $\tilde{n}$ (\textit{b}) local streamwise velocity (i.e., direction perpendicular to $S$), $u_{t,l}$, (\textit{c}) wall-normal velocity (in the direction of $S$), $u_{n,l}$, (\textit{d}) translational temperature, $\tilde{T}_{tr}$, (\textit{e}) rotational temperture,$\tilde{T}_{rot}$, (\textit{f}) vibrational temperature, $\tilde{T}_{vib}$. All quantities are normalized by freestream values, i.e., number density by $n_1$, velocities by $u_{x,1}$, and temperatures by $T_{tr,1}$.}
\label{f:AllMacro_R}
\end{figure}

Additionally, figure~\ref{f:AllMacro_R} shows that the spanwise structures inside the separation bubble are present in the contours of all other perturbation flow parameters.
Interestingly, inside the separation shock, all flow parameters exhibit spanwise modulations, as shown in the inserts of respective figures.
The minimum (negative) and maximum (positive) values of spanwise structures in $\tilde{u}_{t,l}$, $\tilde{u}_{n,l}$, and $\tilde{n}$ are at spanwise location $Y/L_y$=0.88 ($A$) and 1.38 ($B$), respectively.
All three perturbation temperatures have primary spanwise structures adjacent to the wall having minimum and maximum values at spanwise locations $Y/L_y$=1.38 and 0.88, respectively, i.e., 180$^{\circ}$ out of phase with that of velocities and number density.
$\tilde{T}_{tr}$ and $\tilde{T}_{rot}$ also exhibit secondary structures right above the primary structures within $0.1 < H_l < 0.15$.
Such secondary structures are also seen in $\tilde{u}_{n,l}$ and $\tilde{n}$, but are farther along the height within $0.2 < H_l < 0.35$.
\vspace{\baselineskip}

\begin{figure}[H]
    \centering
        \sidesubfloat[]{\label{f:ShockMovement_2D}{\includegraphics[width=0.48\textwidth]{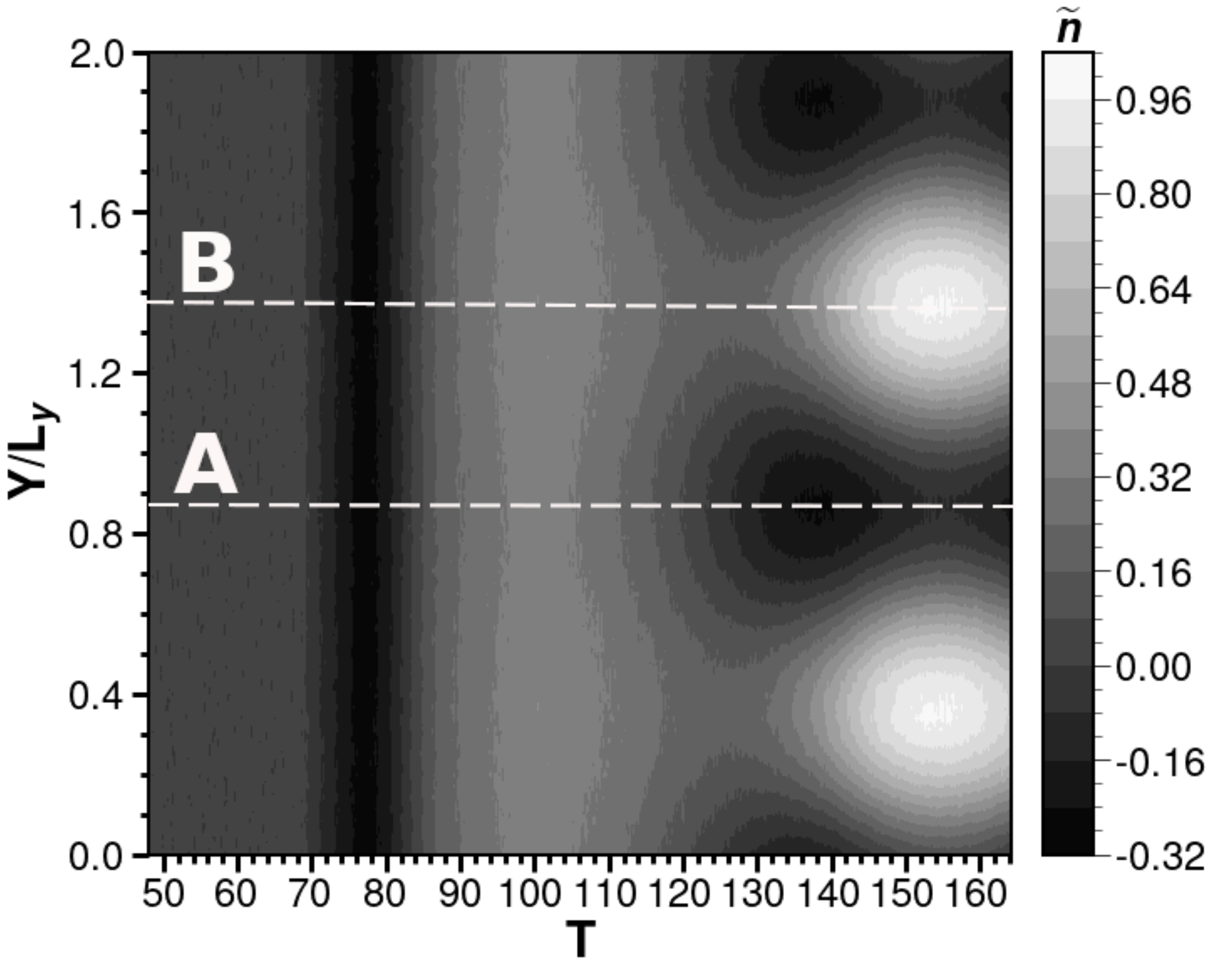}}}\hfill
        \sidesubfloat[]{\label{f:ShockMovement_1D}{\includegraphics[width=0.43\textwidth]{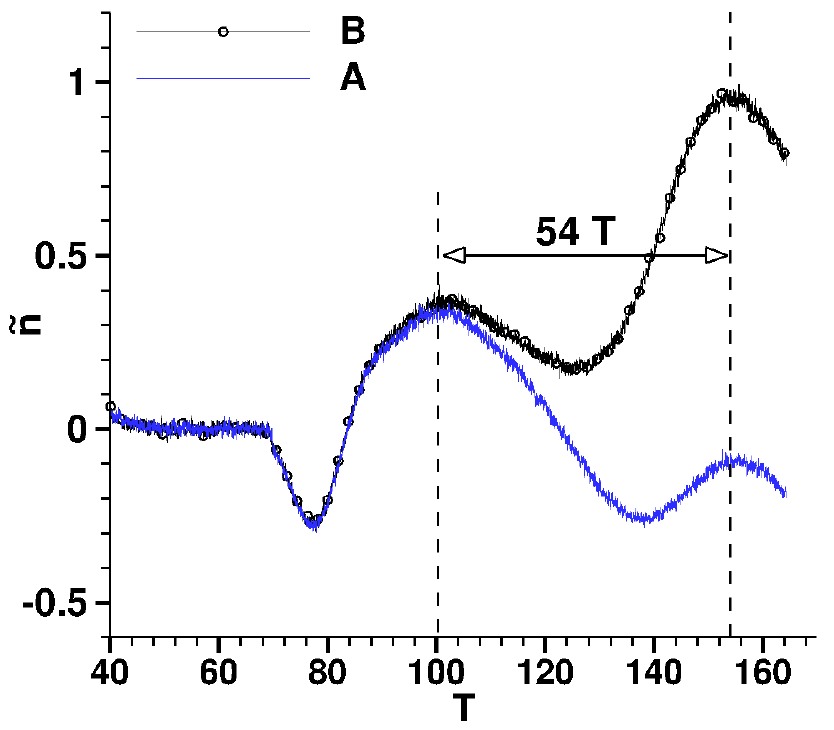}}}\hfill
    \caption{(\textit{a}) At probe $t$ in the vicinity of the triple point $T_2$, denoted in figure~\ref{f:SBLI_ProbesAndPlanes}, the temporal evolution of DSMC-derived perturbation number density, $\tilde{n}$, normalized by $n_1$, indicating low-frequency unsteadiness. (\textit{b}) Normalized $\tilde{n}$ at spanwise locations $A$ and $B$, also marked in (\textit{a}), indicating the period of unsteadiness.}
\label{f:LFU}
\end{figure}

Furthermore, the onset of global linear instability ($T$=50) in the separation bubble is followed by the low-frequency unsteadiness of the shock structure. 
Figure~\ref{f:LFU} shows the spatio-temporal variation of normalized perturbation number density, $\tilde{n}$, at the triple point $T_2$ formed by the intersection of the detached and separation shocks.
To capture one cycle of unsteadiness, the simulation had to be continued much longer up to $T$=165.
Figure~\ref{f:ShockMovement_2D} shows that the triple point starts to oscillate at $T$=70 and its motion remains 2-D up to approximately $T$=85, as there is no variation in $\tilde{n}$ along the spanwise direction within this period.
Afterword, however, linear instability begins at the triple point, which results in spanwise modulation of $\tilde{n}$. 
After $T$=100, we can see the presence of both the linear instability and the low-frequency unsteadiness at the triple point, where we see spanwise structures changing in time.
These features are more clearly seen in figure~\ref{f:ShockMovement_1D} at spanwise locations $A$ and $B$.
The period of oscillation is 54~$T$, which corresponds to the Strouhal number $St$ of 0.0185, defined based on the length of the separation bubble in the base flow, $L_s=$40~mm, and the freestream velocity, $u_{x,1}=$3812~m.s$^{-1}$ as,
\begin{equation} 
St = \frac{f L_{s}}{u_{x,1}}
\label{StrouhalNumber}
\end{equation}
This number is within the low-frequency range, $0.01 \le St \le 0.05$, reported in the literature (see section~\ref{Intro}).

\subsection{Rarefaction effects in the surface parameters}~\label{sec:Slip}

\begin{figure}[H]
    \centering
        \sidesubfloat[]{\label{f:Vslip}{\includegraphics[width=0.46\textwidth]{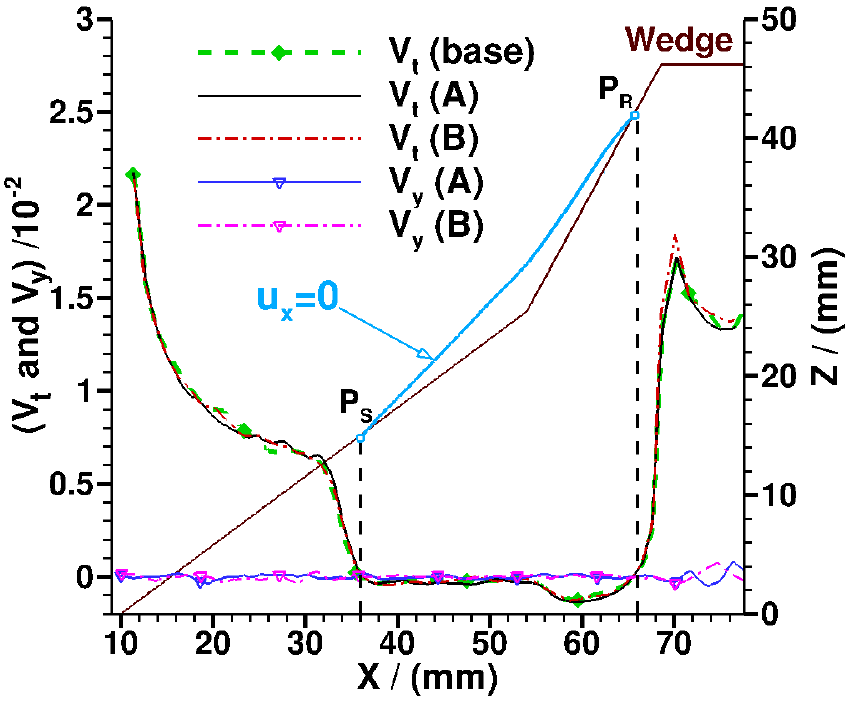}}}\,
        \sidesubfloat[]{\label{f:Tslip}{\includegraphics[width=0.46\textwidth]{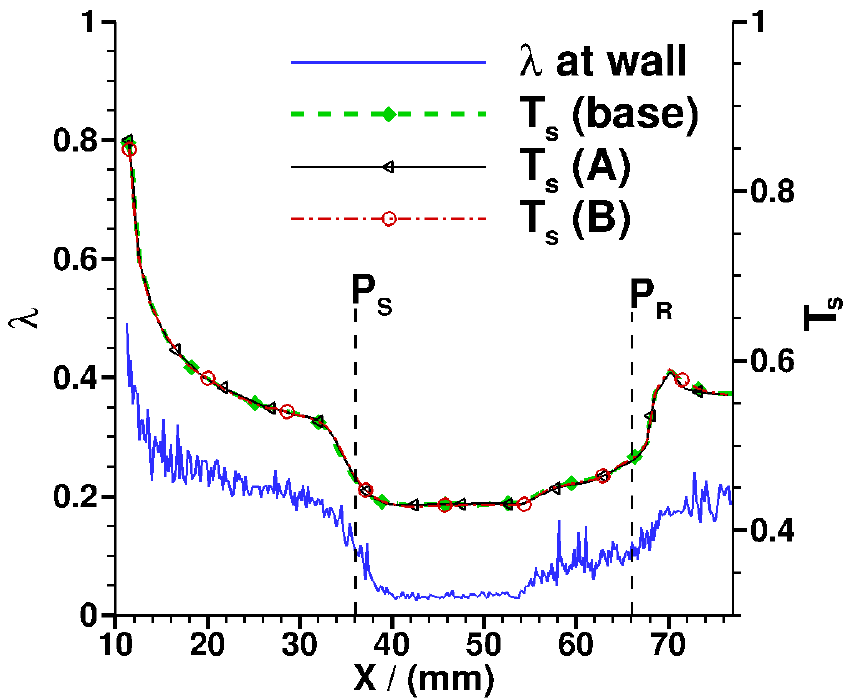}}}\,
        \sidesubfloat[]{\label{f:ChCp}{\includegraphics[width=0.47\textwidth]{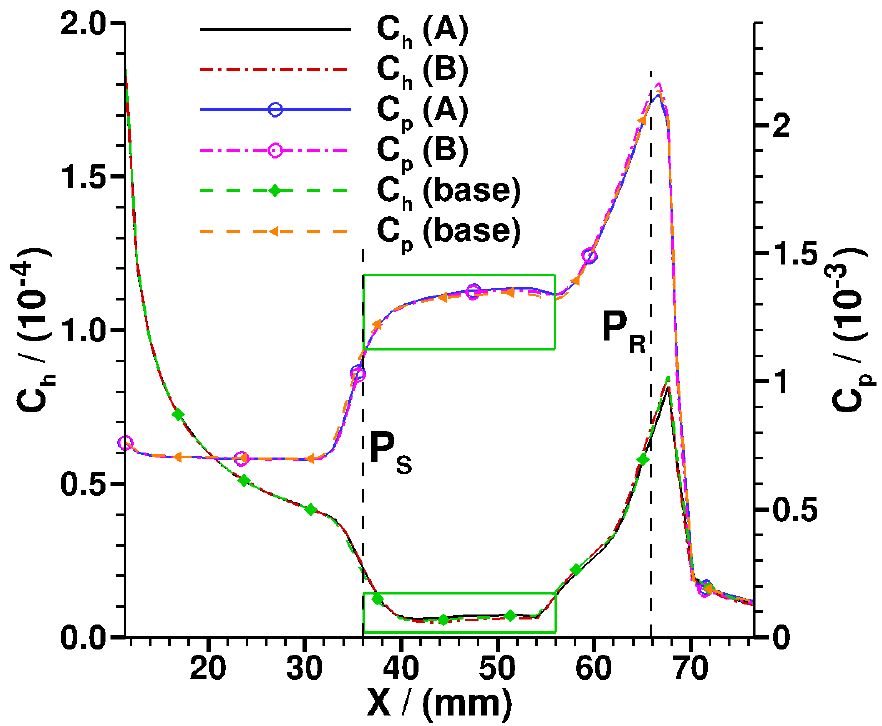}}}\hfill
        \sidesubfloat[]{\label{f:ChCp_zoom}{\includegraphics[width=0.47\textwidth]{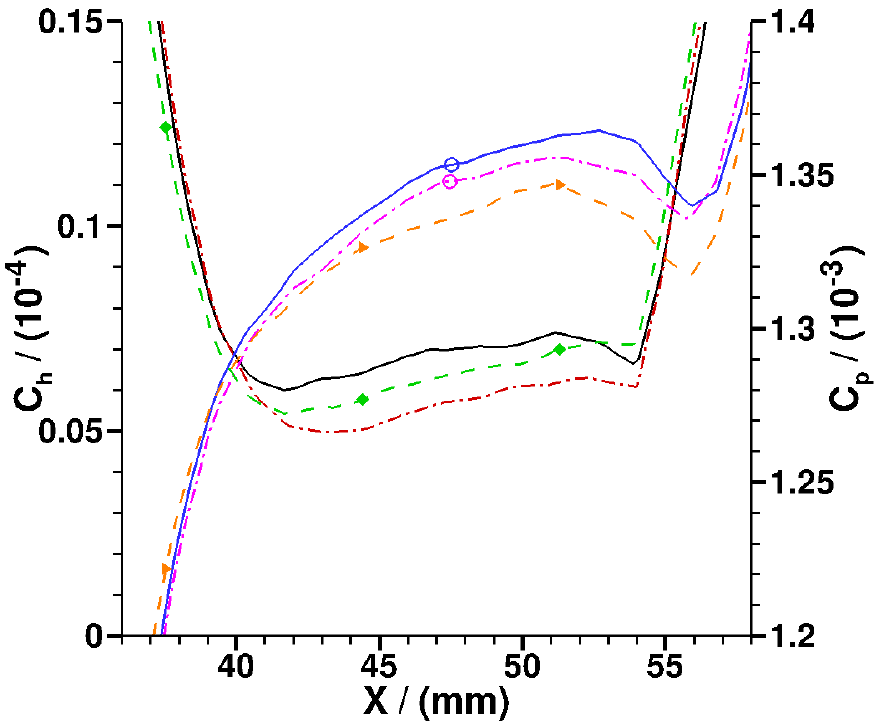}}}\,
        \caption{Surface macroscopic flow parameters in the base state and at the latest time at two spanwise locations $A$ ($Y/L_y=0.88$) and $B$ ($Y/L_y=1.38$). Note that the base state profiles are time-averaged from 48 to 53 flow times and those at $A$ and $B$ from 85 to 90 flow times. (\textit{a}) Surface velocity slips $V_{t}$ and $V_{y}$, normalized by $u_{x,1}$. (\textit{b}) $\lambda$ adjacent to the wall normalized by freestream mean-free-path $\lambda_1$, and temperature jump $T_{s}$, normalized by $T_{tr,1}$. (\textit{c}) The heat transfer and pressure coefficients, $C_h$ and $C_p$, respectively. (\textit{d}) A zoom of the boxed regions marked in (\textit{c}).}
\label{f:Surface}
\end{figure}

To understand the flow behaviour near the wall, figure~\ref{f:Surface} shows surface parameters at two spanwise locations $A$ ($Y/L_y=0.88$) and $B$ ($Y/L_y=1.38$) at the latest timestep $T$=90.5 and in the base state.
Figure~\ref{f:Vslip} shows local-streamwise (tangential) and spanwise velocity slips, $V_{t}$ and $V_{l}$, respectively, and figure~\ref{f:Tslip} shows the local mean-free-path adjacent to the wall, $\lambda$, and the translational temperature jump at the surface, $T_s$.
Velocity slip and temperature jump are rarefaction effects that are proportional to the Knudsen layer in the vicinity of the wall~\citep{koganRGDSpringer,SchaafAndChambre}.
Within this layer, two classes of molecules coexist--those reflected from the wall (in our case, diffusely), and those impinging on the wall which enters this layer from the outside region.
As a result, the average velocity and temperature of the gas are different from the respective velocity and temperature of the wall.
The Knudsen layer is approximately on the order of $\lambda$, the profile of which is noisy because it is obtained on the adaptively refined $C$-mesh.
Note that $\lambda$ is inversely proportional to number density, $n$, and proportional to the translational temperature, $T_{tr}^{\omega-0.5}$, where $\omega=0.745$ is the viscosity index of the gas.
\vspace{\baselineskip}

Figure~\ref{f:Vslip} shows a maximum tangential velocity slip of 2.16\% of the freestream velocity at the leading edge ($X$=10~mm),  which decreases along the local streamwise direction to 0.6\% at $X$=32~mm.
\cite{tumuklu2018PhysRevF} had obtained a maximum velocity slip of 2.45\% at the leading edge in their 2-D flow simulation of nitrogen over a double wedge.
A large slip at the leading edge is due to the increased rarefaction of gas induced by steep gradients of the leading edge shock.
It can be seen from figure~\ref{f:Tslip} that $\lambda$ adjacent to the wall also follows the same behavior as $V_{t}$ in the local streamwise direction, although they are not exactly proportional to each other by a constant factor.
Just upstream of the separation, $P_S$, within a region from $X$=32 to 36~mm, the local streamwise velocity, $u_{t,l}$, as well as $V_{t}$ decrease rapidly and become zero at the separation point, $P_S$ ($X$=36~mm).
$\lambda$ also decreases within this region as there is a rapid increase in number density, $n$, and a decrease in translational temperature, $T_{tr}$, near the wall (not shown).
Inside the recirculation zone, from $P_S$ to $P_R$, the point of reattachment, $V_{t}$ is negative because the flow impinging on the wall is opposite to the local streamwise direction.
$V_{t}$ and $\lambda$ remain constant on the lower wedge, where the latter is about 3.69\% of the freestream mean-free-path, $\lambda_1$.
On the upper wedge, $V_{t}$ increases in magnitude and so does $\lambda$, as $n$ decreases and $T_{tr}$ increases in the local streamwise direction.
From $P_R$ to the upper corner of the wedge, $V_{t}$ continues to increase similar to $\lambda$ as the rates of decrease of $n$ and increase of $T_{tr}$ are larger.
At the location of expansion on the shoulder, $V_{t}$ decreases a bit before it plateaus.
The profiles of $V_{t}$ at $A$, $B$, and the base state, are similar to each other, indicating no significant change so far due to linearly growing mode.
The lateral slip, $V_{l}$, also remains within 0.078\% on the entire surface of the wedge.
The translational temperature jump, $T_s$ follows a similar behavior as $V_{t}$, where it is maximum at the leading-edge of the wedge and decreases up to the recirculation region, in which it remains constant on the lower wedge and increases on the upper wedge.
From $P_R$ to the upper corner of the wedge, the rate of increase of temperature jump is larger, whereas on the shoulder, it plateaus.
Also, no difference is seen in the profiles of $T_s$ at $A$, $B$, and the base state.
Figure~\ref{f:ChCp} shows the surface heat flux and pressure coefficients, $C_h$ and $C_p$, respectively.
Similar to local streamwise velocity and temperature slips, $C_h$ is maximum at the leading edge of the wedge, decreases along the local streamwise direction, and remains at a nearly constant minimum value from the separation to the hinge.
On the upper wedge surface, it increases rapidly up to the upper corner of the wedge, while the rate of increase is larger beyond $X$=61~mm.
The pressure coefficient, $C_p$, is constant on the lower wedge, which increases sharply between $X$=32 to 38~mm, which is the local streamwise region in the vicinity of the separation point.
Inside the recirculation zone on the lower wedge, $C_p$ is nearly constant but increases rapidly on the upper wedge up to the top corner of the wedge, where it is maximum.
On the shoulder of the wedge, both coefficients decrease significantly.
These coefficients are similar in value for profiles $A$, $B$, and the base state, yet figure~\ref{f:ChCp_zoom} shows a zoom of the boxed region marked in figure~\ref{f:ChCp}, to highlight small differences in these profiles on the lower wedge surface inside the recirculation zone.
$C_h$ is at most 11.8\% higher for $A$ and 10.43\% lower for $B$ than the base state, indicating spanwise modulation about the base state.
$C_p$ is at most 0.852\% higher for $A$ than $B$, while both profiles are higher than the base state, indicating a small overall increase in pressure.

\subsection{Spanwise periodic flow structures}\label{sec:SpanPeriodicFlowStructures}
\begin{figure}[H]
    \centering
    \sidesubfloat[]{\label{f:uy_IsoSide}{\includegraphics[width=0.3\textwidth]{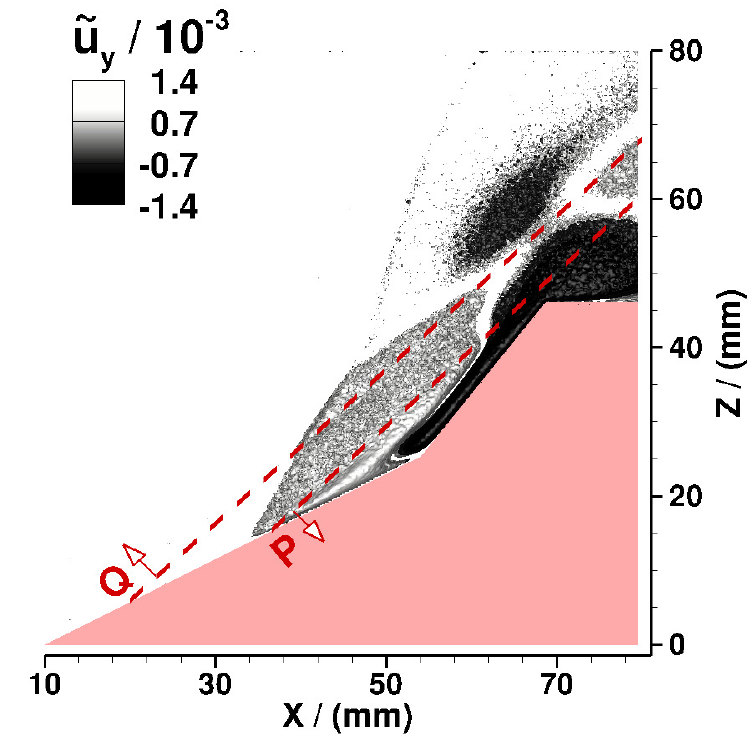}}}\,
    \sidesubfloat[]{\label{f:uy_Iso1}{\includegraphics[width=0.3\textwidth]{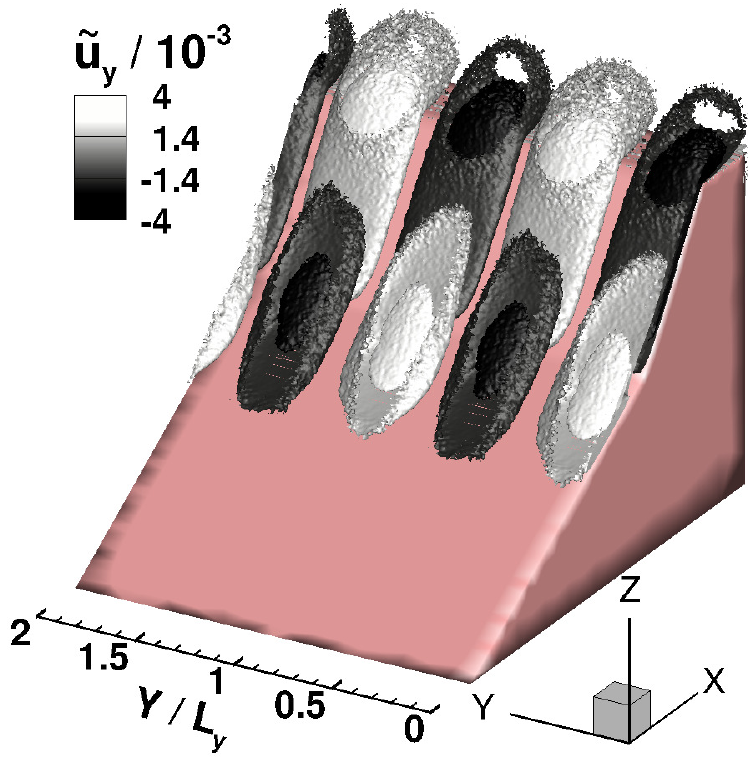}}}\hfill
    \sidesubfloat[]{\label{f:uy_Iso2}{\includegraphics[width=0.3\textwidth]{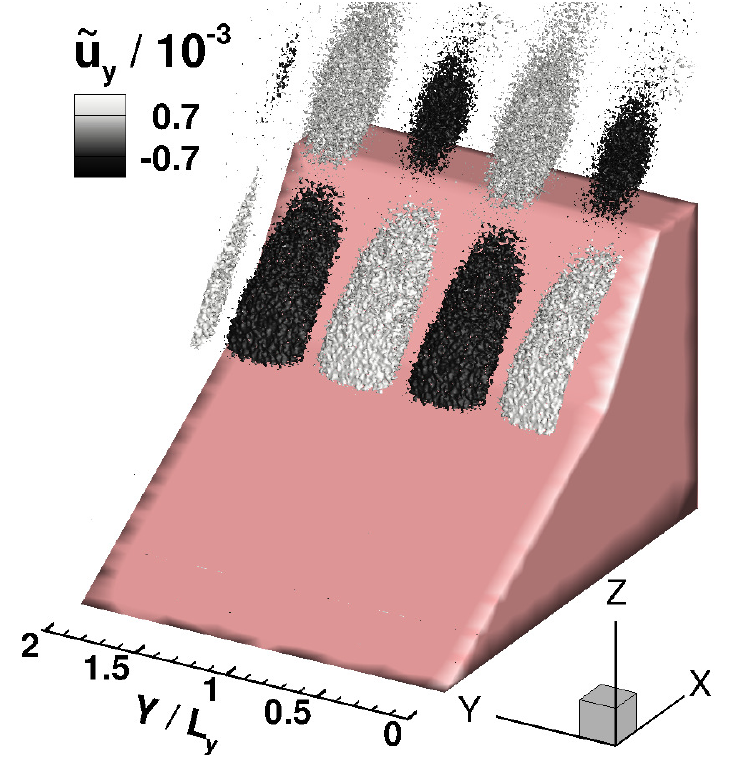}}}\hfill
    \sidesubfloat[]{\label{f:OmegaX_Iso}{\includegraphics[width=0.30\textwidth]{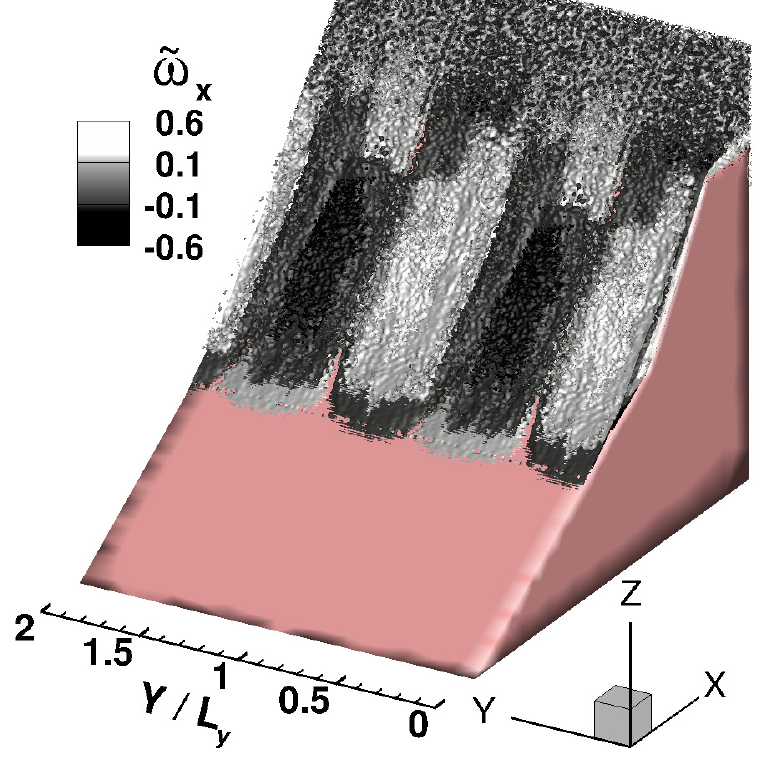}}}\,
    \sidesubfloat[]{\label{f:OmegaY_Iso}{\includegraphics[width=0.30\textwidth]{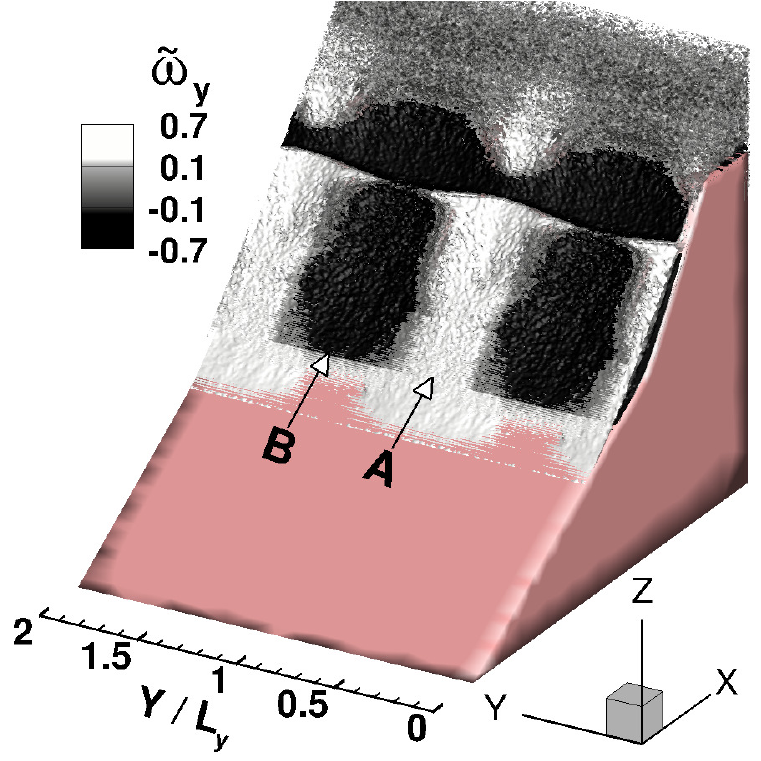}}}\hfill
    \sidesubfloat[]{\label{f:OmegaZ_Iso}{\includegraphics[width=0.30\textwidth]{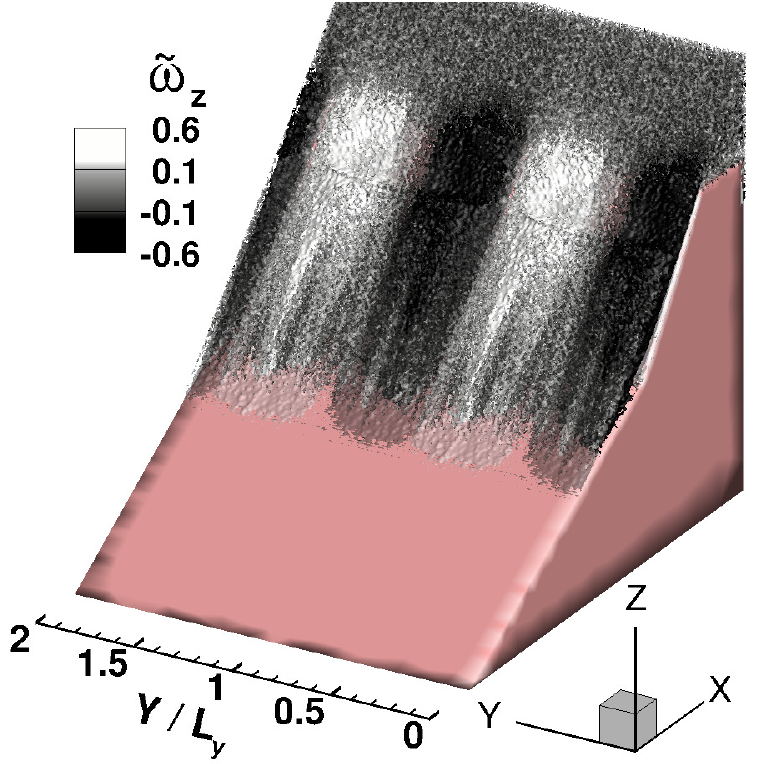}}}\hfill
    \caption{At $T$=90.5, isocontour surfaces of $\tilde{u}_y$ normalized by $u_{x,1}$ and vorticity components $\tilde{\omega}_x$, $\tilde{\omega}_y$, $\tilde{\omega}_z$ normalized by the local vorticity mangitude. (\textit{a}) $\tilde{u}_y$ in side view along with overlaid cut-boundaries $P$ and $Q$ with arrows attached to them that denote normal vectors [-0.7193~$\hat{i}$ + 0.6946~$\hat{k}$] and [0.7193~$\hat{i}$ - 0.6946~$\hat{k}$], respectively. If extended, the cut-boundaries would intersect the $X$ axis at 21.3 and 14~mm, respectively. (\textit{b}) $\tilde{u}_y$ on the normal side of cut-boundary $P$, (\textit{c}) $\tilde{u}_y$ on the normal side of cut-boundary $Q$, (\textit{d}) $\tilde{\omega}_x$, (\textit{e}) $\tilde{\omega}_y$, (\textit{f}) $\tilde{\omega}_z$. Isocontours of all vorticity components are shown on the normal side of cut-boundary $P$.}
\label{f:MacroParam_Iso}
\end{figure}
In summary, the 2-D base flow is unstable to self-excited, small-amplitude, spanwise-homogeneous perturbations, and a linearly growing stationary global mode is observed, which is characterized by spanwise periodic structures in the perturbation flow fields.
The spanwise perturbation velocity, $\tilde{u}_y=u_y$, which was zero at the beginning of the simulation, attains a sinusoidally varying amplitude not only inside the separation bubble but also inside shock layers and downstream of triple points.
This section shows the spanwise periodic sinusoidal flow structures in $\tilde{u}_y$ and vorticity components.
Between the wedge surface and the cut-boundary $P$, marked in figure~\ref{f:uy_IsoSide}, the spanwise periodic structures are shown in figure~\ref{f:uy_Iso1}.
The cut-boundary $P$ cuts through the outer isosurface of $\tilde{u}_y=0.07\%$ of $u_{x,1}$ to reveal core structures having a larger magnitude of $\tilde{u}_y=0.14\%$.
The spanwise structures are seen to extend downstream of the reattachment and on the shoulder of the wedge.
The global mode is also present in the subsonic and supersonic regions downstream of separation and detached shocks, respectively, as seen from figure~\ref{f:uy_Iso2} upstream of the cut-boundary $Q$ marked in figure~\ref{f:uy_IsoSide}.
Such global behavior is expected due to the strong coupling of shocks and the separation bubble.
Finally, the spanwise mode in the isocontours of $X$, $Y$, and $Z$ perturbation vorticity components are also shown in figures~\ref{f:OmegaX_Iso},~\ref{f:OmegaY_Iso}, and~\ref{f:OmegaZ_Iso}, respectively.
The $X$ and $Z$ components are in phase with each other and 90$^{\circ}$ out of phase with the $Y$ component.

\section{Topology of Three-dimensional Laminar Separation Bubble}\label{sec:Topology}
\begin{figure}[H]
\begin{center}
\includegraphics[width=0.55\columnwidth]{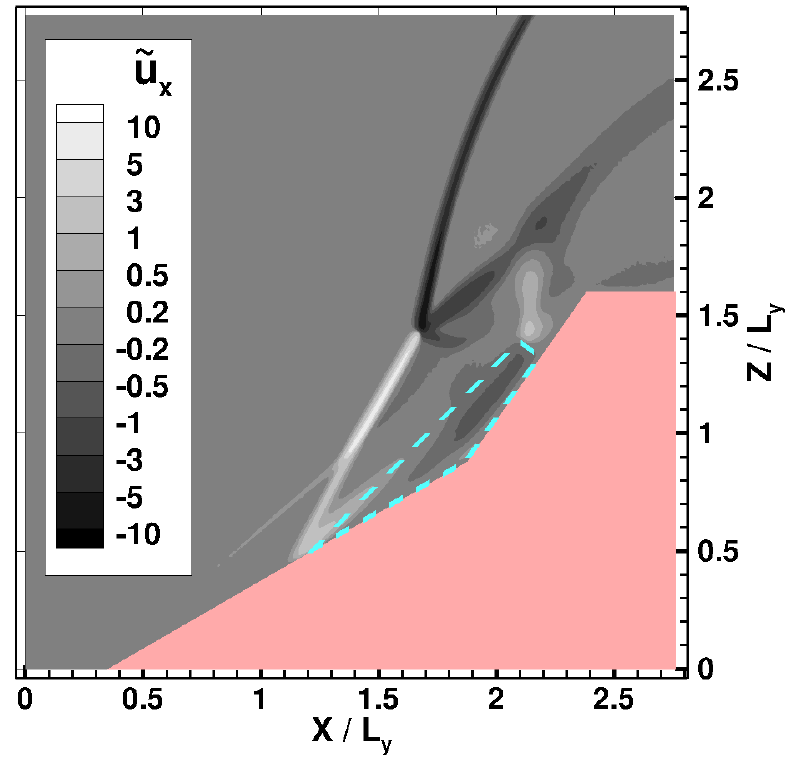}
\caption{At $Y/L_y=0.38$, the $X$-perturbation velocity, $\tilde{u}_x$, normalized by the absolute maximum component of velocity inside the zone marked by a dashed line. Note that the zone is extended in the entire span.}
\label{f:Zone}
\end{center}
\end{figure}

This section investigates the changes in wall-streamlines and three-dimensionality inside the separation bubble by linearly superposing to the 2-D normalized base flow field, $Q_b$, a 3-D normalized perturbation field, $\tilde{Q}$, with a small amplitude, $\epsilon$, ranging from 0.005 to 0.1, using equation~\ref{QperturbFromQandQb}.
Note that the velocity field of the base flow is normalized by the $X$-directional freestream velocity component, $u_{x,1}$.
The perturbation velocity field at $T=90.5$ is normalized in two ways--by the absolute maximum component of velocity inside the separation bubble, i.e., inside the zone marked in figure~\ref{f:Zone} (section~\ref{sec:withoutCoupling}) and by the absolute maximum component of velocity in the entire flow field, which is located in the detached shock near the triple point $T_2$ (section~\ref{sec:withCoupling}).
This distinction will highlight why one cannot draw conclusions about flow topology by decoupling the shock and a separation bubble.
In the former case, the absolute maximum values of normalized $X$, $Y$, and $Z$ perturbation velocities inside the zone marked in figure~\ref{f:Zone} are 0.954, 0.455, and 1, respectively.
In the latter case, these are 1, 0.0565, and 0.518, respectively.

\subsection{Analysis without the coupling of shock and separation bubble}~\label{sec:withoutCoupling}
\begin{figure}[h]
    \centering
    \sidesubfloat[]{\label{f:ws_5em3}{\includegraphics[width=0.46\textwidth]{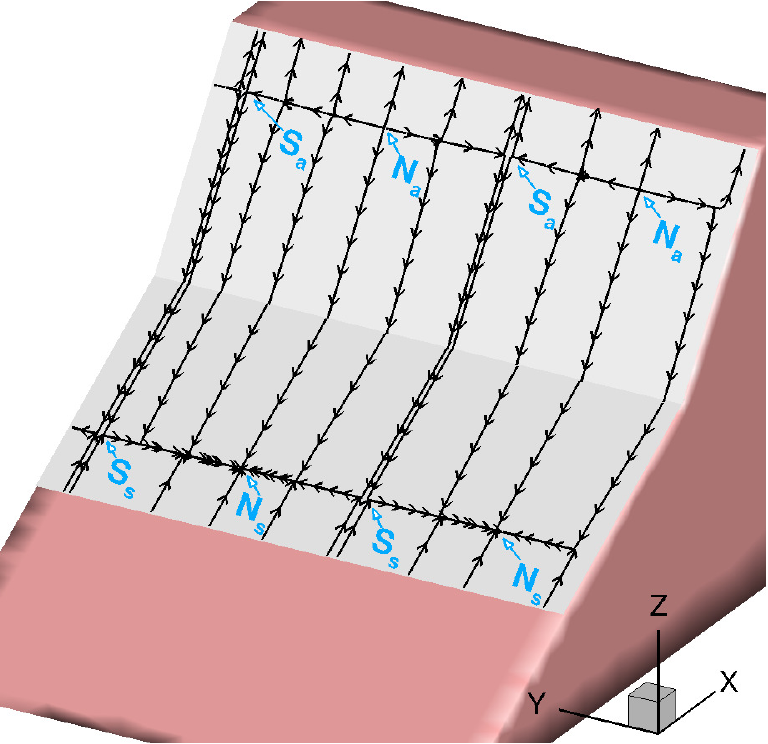}}}\,
    \sidesubfloat[]{\label{f:ws_1em2}{\includegraphics[width=0.46\textwidth]{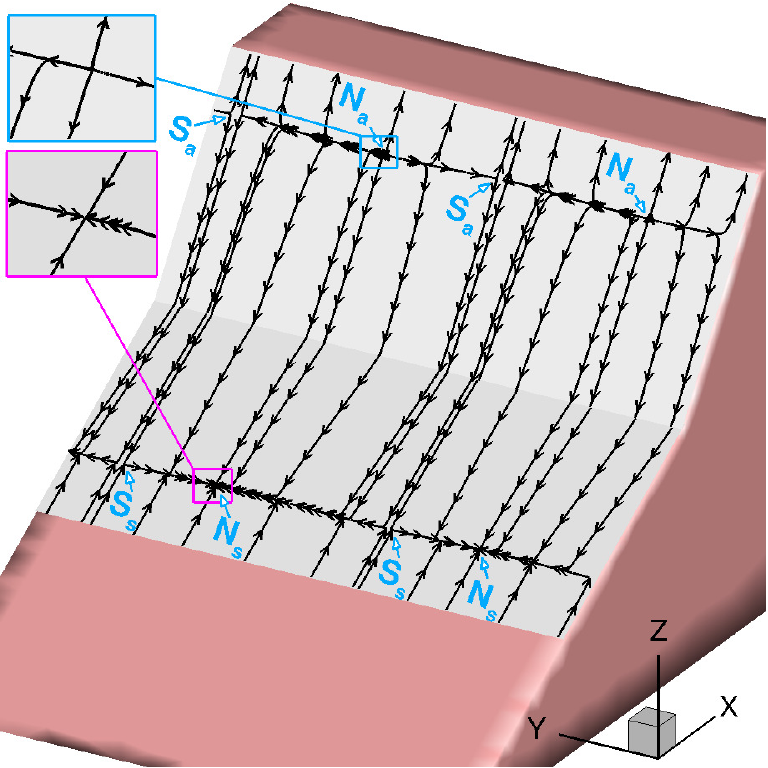}}}\hfill
    \sidesubfloat[]{\label{f:ws_5em2}{\includegraphics[width=0.46\textwidth]{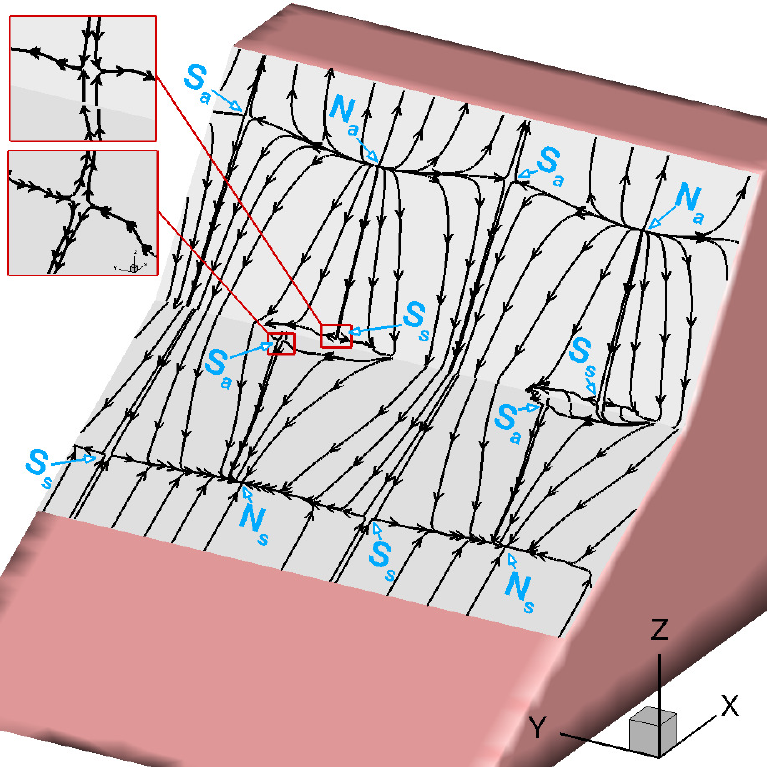}}}\hfill
    \sidesubfloat[]{\label{f:ws_5em1}{\includegraphics[width=0.46\textwidth]{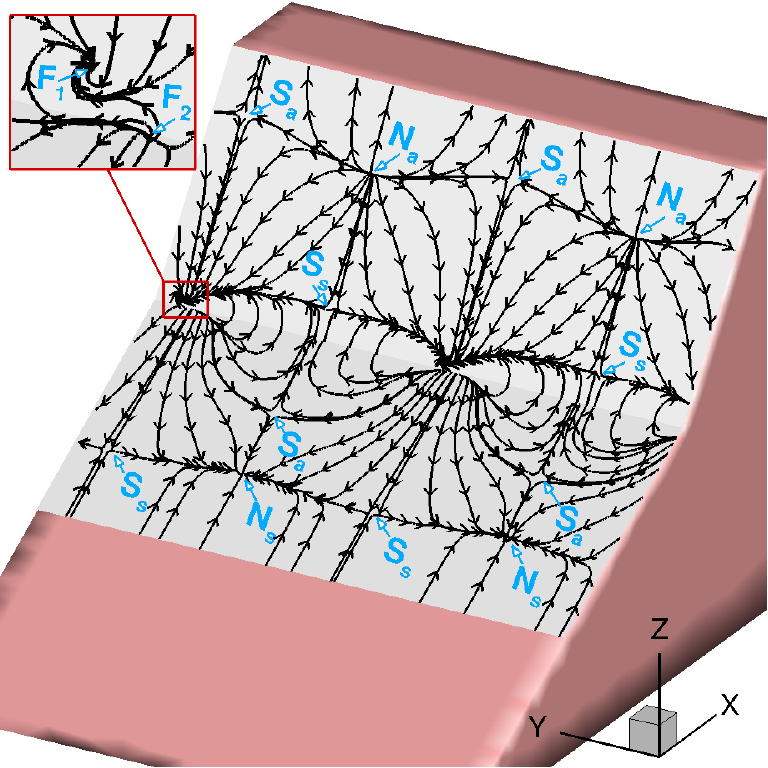}}}\,
    \caption{Wall streamlines in the flow constructed by superposition of scaled 2-D base flow with scaled linear perturbations having amplitude $\epsilon$. (\textit{a}) $\epsilon=0.005$, (\textit{b}) $\epsilon=0.01$, (\textit{c}) $\epsilon=0.05$, (\textit{d}) $\epsilon=0.1$.}
\label{f:ws_topology}
\end{figure}
Figure~\ref{f:ws_topology} shows profiles of wall-streamlines in the superposed flow field for different amplitudes, $\epsilon$.
The signature observed in figure~\ref{f:ws_5em3} typically results from small-amplitude spanwise homogeneous perturbations to the 2-D separation bubble, as was shown by \citet{rodriguez_theofilis_2010} in an incompressible flow.
A series of critical points are formed on the separation and reattachment lines between which the wall-streamlines are slightly bent in the spanwise direction, indicating three-dimensionality of the separated flow.
At a saddle point of separation, $S_s$, on the line of separation, the flow is attracted in the local streamwise direction and is diverted in the spanwise direction.
In the middle of two $S_s$ points on the line of separation, a node point of separation, $N_s$, is formed, where the flow coming from both saddle points meets and leaves in the wall-normal direction.
On the reattachment line, a node point of attachment, $N_a$, is formed, where the flow coming from the wall-normal direction is diverted in the spanwise and local streamwise directions.
Between two $N_a$ points, a saddle point of attachment, $S_a$, is formed where the flow coming from the spanwise direction is diverted in the local streamwise direction.
Figure~\ref{f:ws_1em2} shows a similar pattern for a larger amplitude of $\epsilon=0.01$; however, now the node and saddle points on the separation and reattachment lines are not colinear in the local streamwise direction.
As a result, the flow exhibits two new saddle points near the hinge, as seen in figure~\ref{f:ws_5em2} for $\epsilon=0.05$.
At the larger amplitude of $\epsilon=0.1$, the two saddle points are aligned with the node points of separation and reattachment; however, in the vicinity of the line connecting the saddle point of separation and reattachment, two counter-rotating foci, $F_1$ and $F_2$, are formed.
Further increase in amplitude may lead to the merging of points $S_a$ and $N_s$ on the lower wedge and $S_s$ and $N_a$ on the upper wedge such that the node points on the separation and reattachment lines will disappear.
Such a signature would resemble a simple $U$-shaped separation, first classified by \citet{perryAndHornung}. However, these speculations are beyond the purview of linear analysis.
We will also see in section~\ref{sec:withCoupling} that such topology cannot be studied without accounting for the perturbations in the shock.

\begin{figure}[H]
    \centering
    \sidesubfloat[]{\label{f:ss_5em3}{\includegraphics[width=0.46\textwidth]{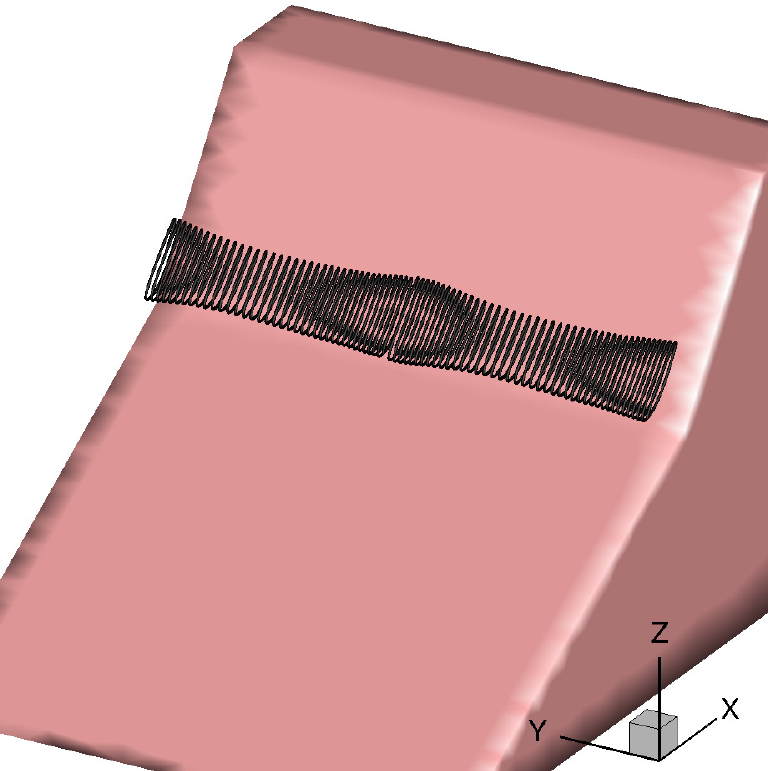}}}\,
    \sidesubfloat[]{\label{f:ss_1em2}{\includegraphics[width=0.46\textwidth]{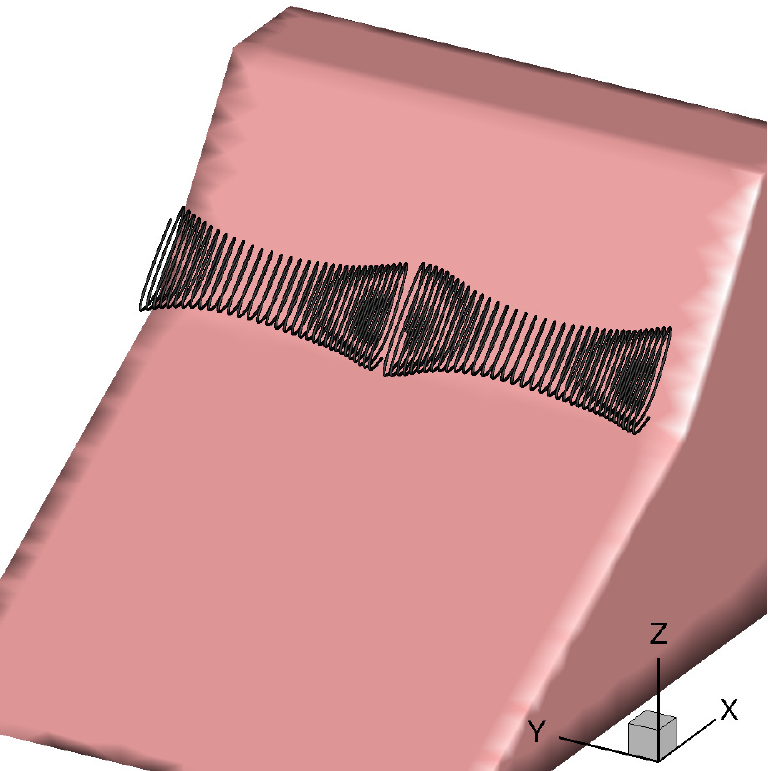}}}\hfill
    \sidesubfloat[]{\label{f:ss_5em2}{\includegraphics[width=0.46\textwidth]{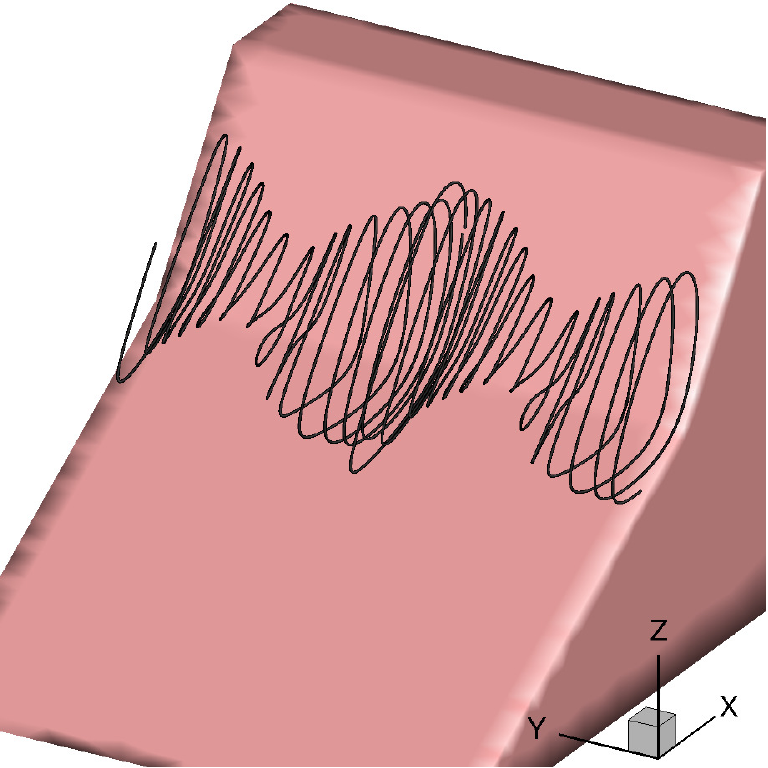}}}\hfill
    \sidesubfloat[]{\label{f:ss_5em1}{\includegraphics[width=0.46\textwidth]{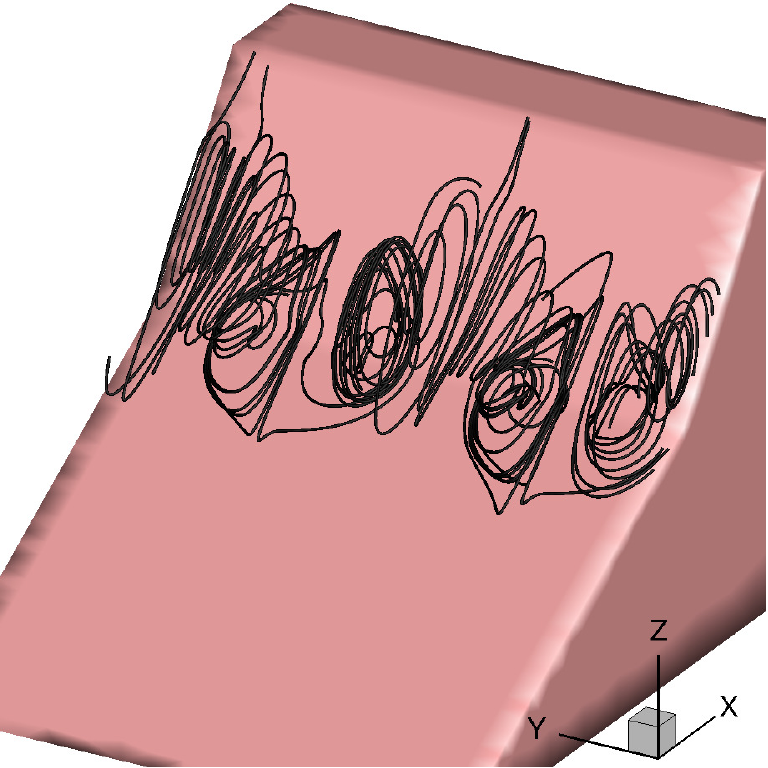}}}\,
    \caption{Three-dimensionality of the flow inside the separation bubble shown using volume streamlines. A volume line is a streamline traveling through 3-D volume data rather than being confined to a surface~\citep{Tecplot}. As in the earlier figure, the flow is constructed by superposition of scaled 2-D base flow with scaled linear perturbations having amplitude $\epsilon$. (\textit{a}) $\epsilon=0.005$, (\textit{b}) $\epsilon=0.01$, (\textit{c}) $\epsilon=0.05$, (\textit{d}) $\epsilon=0.1$.}
\label{f:ss_topology}
\end{figure}

The increasing three-dimensionality of the separation bubble is seen in figure~\ref{f:ss_topology} for superpositions with the above amplitudes.
Comparison of figures~\ref{f:ss_5em3} and~\ref{f:ss_1em2} shows increasing spanwise modulation of recirculating streamlines from $\epsilon=0.005$ to $\epsilon=0.01$, while the axis of rotation remains parallel to the spanwise-direction ($Y$).
For $\epsilon=0.05$ in figure~\ref{f:ss_5em2}, the streamlines become 3-D, where the axes of rotation are seen to deviate from $Y$, and spanwise modulation is increased.
For $\epsilon=0.1$ in figure~\ref{f:ss_5em1}, the streamlines are fully 3-D, where the axes of rotation diverge significantly from $Y$, so much that at some locations it is perpendicular to $Y$.
\subsection{Analysis with the coupling of shock and separation bubble}~\label{sec:withCoupling}
\begin{figure}[H]
    \centering
    \sidesubfloat[]{\label{f:SepShockCorrugation}{\includegraphics[width=0.46\textwidth]{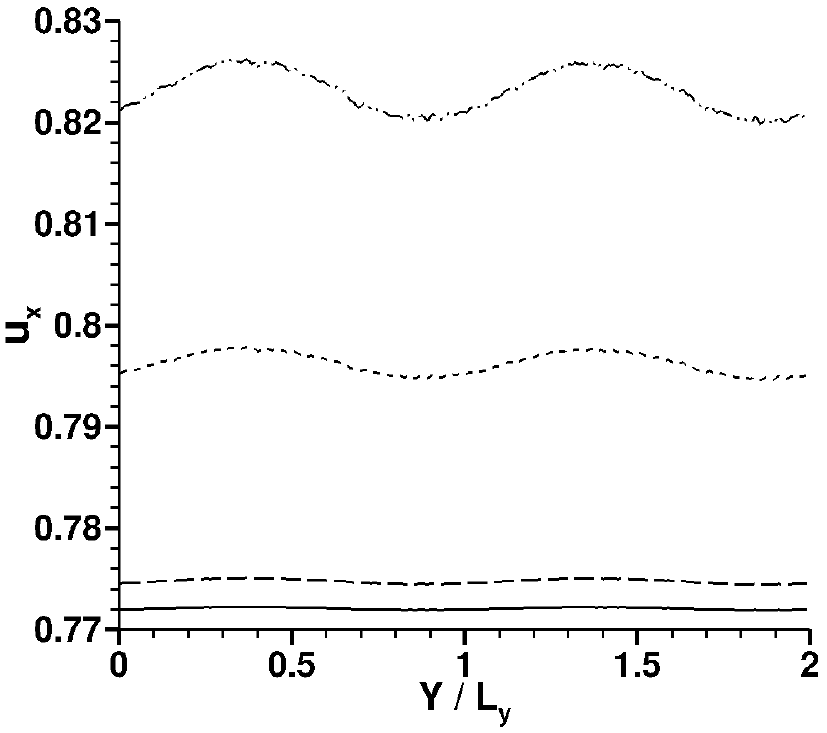}}}\,
    \sidesubfloat[]{\label{f:BowShockCorrugation}{\includegraphics[width=0.46\textwidth]{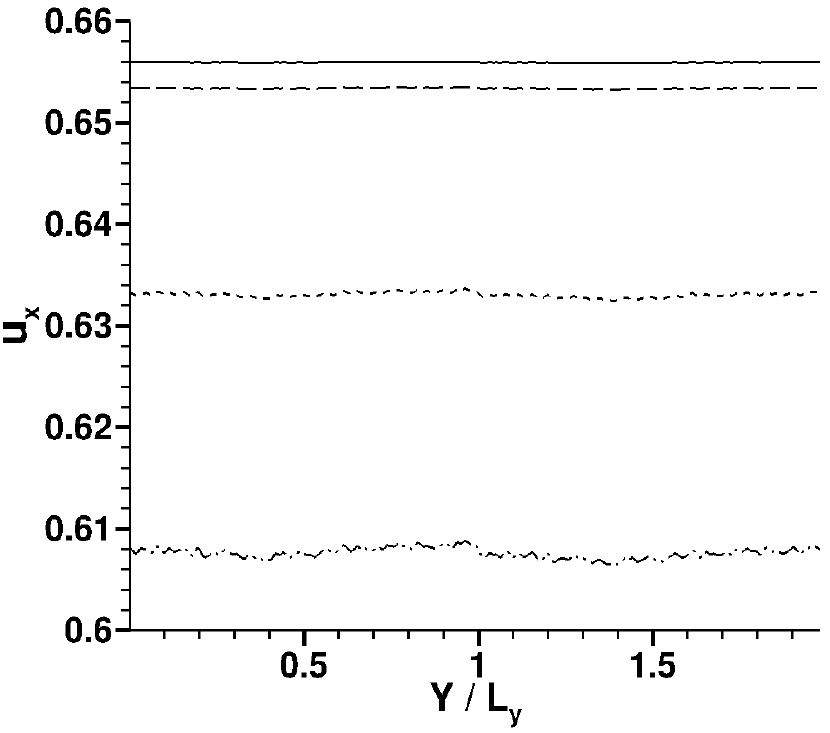}}}\hfill
    \sidesubfloat[]{\label{f:ws_5em1}{\includegraphics[width=0.46\textwidth]{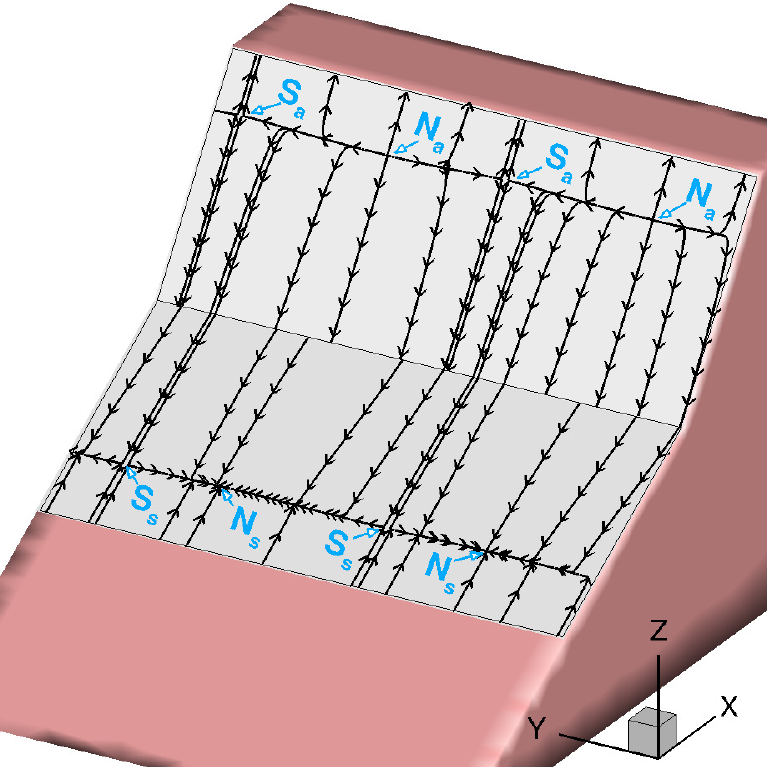}}}\,
    \sidesubfloat[]{\label{f:ws_5em1}{\includegraphics[width=0.46\textwidth]{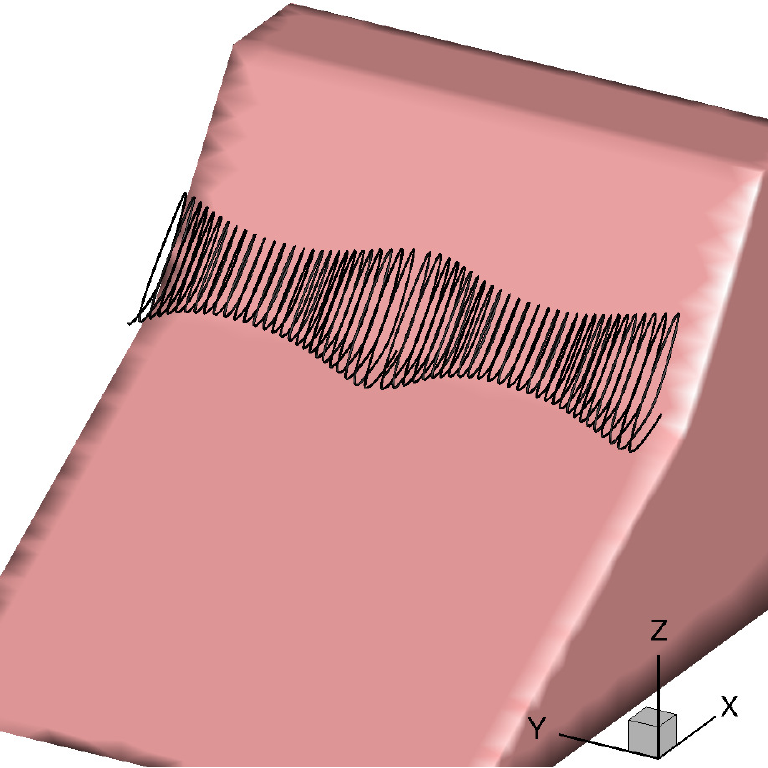}}}\,
    \caption{Streamwise velocity, $u_x$, obtained from the superposition of normalized base and perturbed flow field showing corrugations of (\textit{a}) the separation shock at $H_y/L_y$=0.41 on the $S$-plane and (\textit{b}) the detached shock at $H_u/L_y$=0.65 on the $R$-plane. These locations are inside the shock layer as seen from figures~\ref{f:uy_R} and~\ref{f:uy_S}. Corresponding legends: \protect (\full) $\epsilon=0.005$, (\longdashed) $\epsilon=0.01$, (\dashed) $\epsilon=0.05$, (\dashdoubledot) $\epsilon=0.1$.\\
(\textit{c}) Wall streamlines and (\textit{d}) volume lines inside the separation bubble for $\epsilon=0.1$.}
\label{f:topology2}
\end{figure}
When the perturbation velocity field is normalized by maximum perturbation velocity component inside the shock, the linear coupling of shock and the separation bubble is taken into account.
Figure~\ref{f:topology2} shows the features of the superposed flow field.
The spanwise corrugations of the separation and detached shocks in flow fields composed with four increasing amplitudes of linear perturbations are seen in figures~\ref{f:SepShockCorrugation} and~\ref{f:BowShockCorrugation}, respectively.
The three-dimensionality of the separation shock is more than the detached shock and becomes prominent for the largest amplitude of $\epsilon=0.1$.
The wall-streamlines in figure~\ref{f:ws_5em1} for $\epsilon=0.1$ reveal alternate node and saddle points on the separation and reattachment lines, where the node and saddle points on the two lines are not aligned.
This topology is similar to that in figure~\ref{f:ws_1em2} for amplitude $\epsilon=0.005$, when the effect of shock was not taken into account.
Similarly, the recirculation streamline inside the separation bubble shows a low degree of spanwise modulation, where the axis of rotation is the $Y$-axis for the largest amplitude of $\epsilon=0.1$ in figure~\ref{f:ws_5em1}, similar to figure~\ref{f:ss_1em2} for $\epsilon=0.005$ (without coupling).
Therefore, the coupled analysis indicates that the deviation from two-dimensionality in the shock structure dominates the deviation in the separation bubble.
As a result, the study of three-dimensionality in the topology of an LSB cannot be done by ignoring their coupling with shock structures.

\section{Conclusion}\label{sec:Conclusion}
The 3-D LSB induced by a laminar SBLI on a spanwise-periodic, Mach 7 hypersonic flow of nitrogen over a $30^{\circ}-55^{\circ}$ double wedge was simulated using the massively parallel SUGAR DSMC solver using billions of computational particles and collision cells on an adaptively refined octree grid.
The fully resolved kinetic solution resulted in accurate modelling of the internal structure of shocks, surface rarefaction effects, thermal nonequilibrium, and time-accurate evolution of 3-D, self-excited perturbations. {This is the first simulation that analyzes the linear instability of a 2-D base flow to  self-excited, small-amplitude, spanwise-homogeneous perturbations in the  low Reynolds number regime.}
\vspace{\baselineskip}

In line with the findings of \citet{tumuklu2018POF2} of Mach 16 flows over axisymmetric double cone and \citet{tumuklu2018PhysRevF} of a 2-D, Mach 7 flow over the double wedge, the 3-D LSB was found to be strongly coupled with the separation and detached shocks.
The presence of linear instability led to the formation of spanwise periodic flow structures in 3-D perturbations of macroscopic flow parameters not only inside the LSB, but also in the internal structure of the separation shock.
The spanwise periodicity length of the structures at these two zones was found to be the same and their amplitude was found to grow with an average, linear temporal growth rate of 5.0~kHz $\pm$ 0.16\%.
We obtained a larger value of $\num{0.0057}$ for the nondimensional growth rate compared to that of \citet{sidharth2018onset} for double wedges with lower angles, which is qualitatively consistent.
\vspace{\baselineskip}

The boundary-layer profiles in the 2-D base flow were compared with those obtained from the 3-D flow with perturbations at $T$=90.5 at two spanwise locations 
corresponding to the peak and trough of the spanwise sinusoidal mode.
The comparison these profiles upstream and downstream as well as at the point of separation revealed that the linear instability originates 
 in the interaction region of the separation shock with the LSB.
The difference between the peak and trough of wall-tangential velocities revealed that the amplitude of perturbations increases inside the  recirculation zone from the separation to the reattachment point.
All boundary-layer profiles exhibited nonzero wall-tangential velocities at the wall in the  Knudsen layer region.  
The profiles inside the separation zone also showed the presence of two GIPs, one between the wall and  shear layer and the other between the shear layer and  supersonic flow outside the bubble.
\vspace{\baselineskip}

The onset of linear instability at $T=50$ was followed by the low-frequency unsteadiness of the triple point $T_2$ at $T=70$.
The oscillation frequency corresponds to a Strouhal number of $St \sim 0.02$, consistent with the existing literature on turbulent SBLI, but in contrast with the 3-D, finite-span double wedge simulation of \citet{reinert2020simulations} at a factor of eight times higher density which did not reveal such unsteadiness. 
To resolve these predictions, the slow linear growth and long time-scale of low-frequency unsteadiness ($\sim 0.57$~ms) suggests that experimental test times must be significantly long to capture these effects.
In addition, the long-time ($T > 100$) spatio-temporal evolution of the flow at the triple point $T_2$ revealed for the first time the presence of spanwise corrugation as well as sinusoidal oscillations in time.
\vspace{\baselineskip}

Finally, the topology signature in the wall-streamlines of the 3-D flow constructed by superposition of the 2-D base flow and 3-D linear perturbations was analyzed with and without accounting for the coupling between the shocks and the LSB.
For a given amplitude of perturbations, significant differences were observed in the topology with versus without coupling.
The analysis with coupling also revealed an increase in the corrugation of the separation and detached shocks with increase in amplitude of 3D perturbations.
These findings further emphasize that, at these conditions, the 3-D changes to the topology of an LSB cannot be studied without taking into account the coupling with the shock structure.
\vspace{\baselineskip}

\noindent{\bf Acknowledgements\bf{.}} 
The authors acknowledge the Texas Advanced Computing Center (TACC) at the University of Texas at Austin for providing high performance computing resources on Frontera supercomputer under the Leadership Resource Allocation (LRAC) award CTS20001 of 200k SUs that have contributed to the research results reported within this paper.
This work also used the Stampede2 supercomputing resources of 400k SUs provided by the Extreme Science and Engineering Discovery Environment (XSEDE) TACC through allocation TG-PHY160006.
A part of the simulation was also carried out on Blue Waters supercomputer under projects ILL-BAWV and ILL-BBBK.
The Blue Waters sustained-petascale computing project is supported by the National Science Foundation (awards OCI-0725070 and ACI-1238993) the State of Illinois, and as of December, 2019, the National Geospatial-Intelligence Agency. Blue Waters is a joint effort of the University of Illinois at Urbana-Champaign and its National Center for Supercomputing Applications.
In addition, the authors thank Dr. Ozgur Tumuklu for providing the 2-D steady flow solution.
\\

\noindent{\bf Funding\bf{.}} The research conducted in this paper is supported by the Office of Naval Research under the grant No. N000141202195 titled, “Multi-scale modelling of unsteady shock-boundary layer hypersonic flow instabilities” with Dr. Eric Marineau as the program officer.\\

\noindent{\bf Declaration of Interests\bf{.}} The authors report no conflict of interest. \\

\noindent{\bf  Author ORCID\bf{.}} Authors may include the ORCID identifers as follows.  S. Sawant, https://orcid.org/0000-0002-2931-9299; D. Levin, https://orcid.org/0000-0002-6109-283X; V. Theofilis, https://orcid.org/0000-0002-7720-3434.

\newpage
\appendix
\section{}\label{A:MCCS}
In a typical DSMC simulation, the collision pairs selected using the MFS or the no time counter (NTC) scheme are allowed to collide with probability,
\begin{equation} 
P_c=\frac{\sigma_Tc_r}{(\sigma_T c_r)_{max}}
\end{equation}
where $\sigma_T=\pi d^2$ is the total cross-section, $d$ is the molecular diameter, and $c_r$ is the relative speed.
The maximum collision cross-section, $(\sigma_T c_r)_{max}$, is stored for each collision cell and is estimated at the beginning of the simulation to {\it a reasonably large value}.
Bird estimates this number as [Sec. 11.1 Bird 1994],
\begin{equation} 
(\sigma_T c_r)_{max} = (\pi d_r^2) 300 \sqrt{T_{tr}/300}
\label{SigmaCrGuess}
\end{equation}
where $d_r$ is the reference molecular diameter.
As the simulation progresses, the parameter is updated if a larger value is encountered in a collision cell.
However, a problem occurs at an AMR step, where the old $C$-mesh is deleted, and a new one is constructed.
For the newly created collision cells, an estimate of $(\sigma_T c_r)_{max}$ is required.
If the parameter value is arbitrarily guessed based on equation~\ref{SigmaCrGuess}, then the instantaneous temporal signals of macroscopic parameters exhibit kinks at the timesteps when the AMR step is performed.
Although these kinks decay in approximately 3 to 4~$\mu$s, they can spuriously reveal a dominant frequency equal to the inverse of the time period between two AMR steps.
To avoid the corruption of instantaneous signals with such artificial perturbations, at an AMR step, each root cell stores the smallest value of $(\sigma_T c_r)_{max}$ among all of its collision cells before deleting the $C$-mesh.
After a new $C$-mesh is formed, the value stored in the root is assigned as the lowest estimated guess to all collision cells in a given root.
Those newly formed collision cells, for which the actual value of $(\sigma_T c_r)_{max}$ must be larger than that assigned as an estimate, quickly update to this value within the next 0.2~$\mu$s.
This strategy avoids the kinks in the instantaneous residual.

\section{}\label{A:PODCompare}
This appendix shows the use of the POD method~\citep{luchtenburg2009introduction} to remove the statistical noise in instantaneous perturbation macroscopic flow parameter fields obtained from DSMC.
The use of the POD method to reduce statistical noise in a stochastic simulation can be found in a number of resources~\citep{grinberg2012proper,tumuklu2018PhysRevF}.
This method performs the singular value decomposition (SVD) of the input data matrix $\mathcal{D}$ formed from the solution of any given macroscopic flow parameter such that the number of rows and columns are equal to the number of total sampling cells $N_{c}$ in the DSMC domain and the instantaneous time snapshots $N_{s}$, respectively.
The SVD procedure results in the decomposition,
\begin{equation} 
\centering
\begin{split}
\mathcal{D} &= \phi \mathcal{S} \mathcal{T}\\
\end{split}
\label{Ansatz}
\end{equation}
where $\phi$ is the matrix of spatial modes having dimensions $N_c \times N_r$, $N_r$ is the user-specified rank of the reduced SVD approximation to $\mathcal{D}$, $\mathcal{S}$ is the square diagonal matrix of singular values having dimensions $N_r \times N_r$, and $\mathcal{T}$ is the matrix of temporal modes of dimensions $N_r \times N_s$.
The $i^{th}$ spatial and temporal modes are stored in the $i^{th}$ column of $\phi$ and row of $\mathcal{T}$, respectively.
The singular values in $\mathcal{S}$ are arranged in decreasing order, and their square corresponds to the amount of energy in the mode.
After the decomposition, a reduced-order, noise-filtered representation of $\mathcal{D}$ can be constructed by forming a new data matrix $\mathcal{D}_2$ from a user-specified number of ranks $N_{r2}$, which is smaller than $N_r$.
$N_{r2}$ is chosen such that the difference between any time snapshot of $\mathcal{D}_2$ and that of $\mathcal{D}$ is within statistical noise.
\begin{figure}[H]
    \centering
       \sidesubfloat[]{\label{f:modalenergy}{\includegraphics[width=0.46\textwidth]{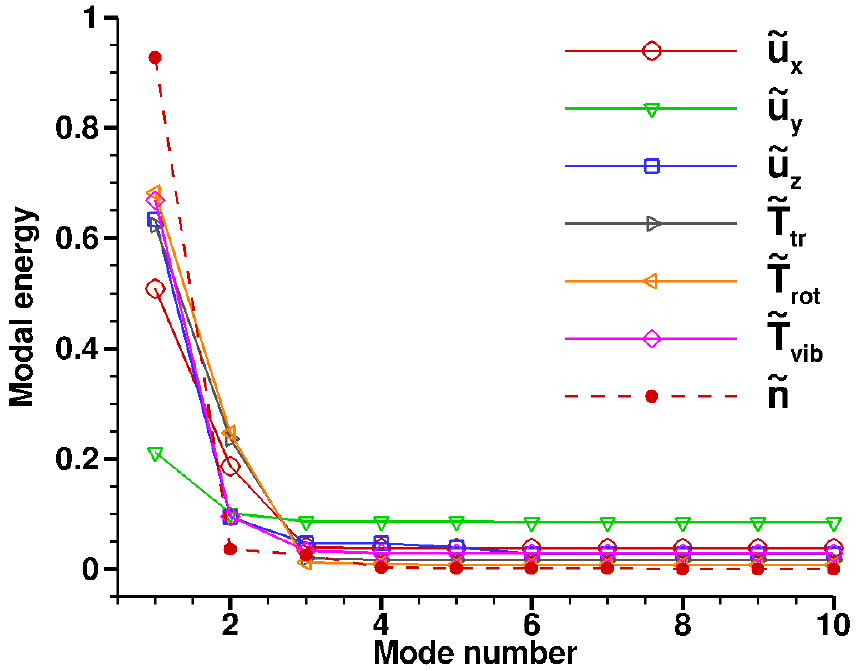}}}\hfill
       \sidesubfloat[]{\label{f:uy_PODCompare}{\includegraphics[width=0.46\textwidth]{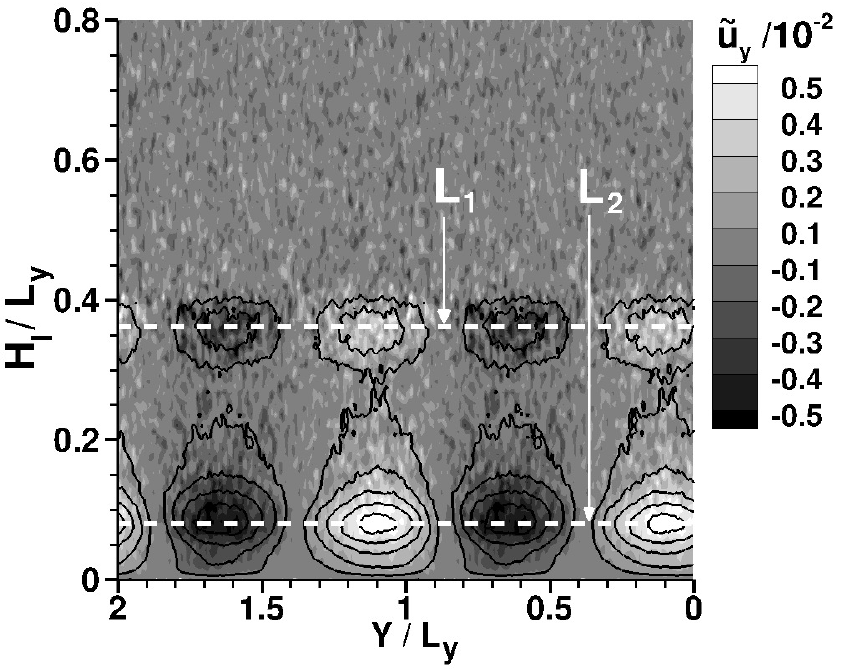}}}\hfill
       \sidesubfloat[]{\label{f:uy_1DCompare}{\includegraphics[width=0.46\textwidth]{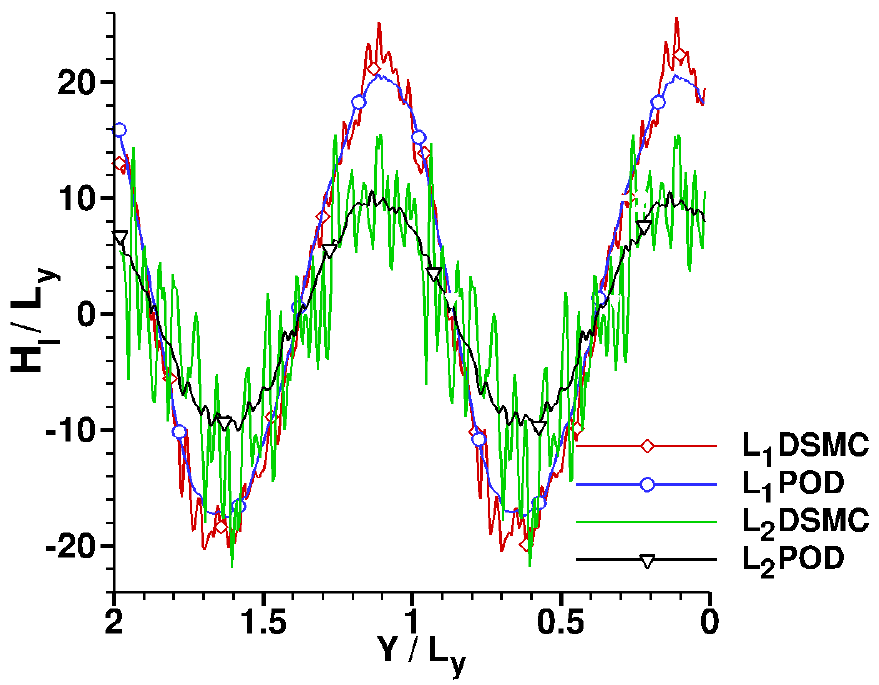}}}\hfill
    \caption{(\textit{a}) Modal energy in perturbation macroscopic flow parameters based on singular values obtained from proper orthogonal decomposition. (\textit{b}) Contours of unfiltered (raw DSMC data) $\tilde{u}_y$ normalized by $u_{x,1}$ at $T$=90.5 on a plane defined along wall-normal direction $S$ as in figure~\ref{f:uy_R}. Overlaid on it the contour lines of noise-filtered reconstruction of $\tilde{u}_y$ from the first two proper orthogonal modes. (\textit{c}) Comparison of unfiltered (DSMC) and filtered (POD) $\tilde{u}_y$ along lines $L_1$ and $L_2$ denoted in (\textit{b}).}
\label{f:PODCompare}
\end{figure}

For the double wedge solution, the data matrix for each macroscopic flow parameter was formed by the number of sampling cells, $N_c=\num{23.04e6}$ and number of time snapshots, $N_s=450$.
The instantaneous snapshots were collected from $T$=48.0312 to 90.9162, at an interval of 0.0953 flow time, which corresponds to the frequency of 1~MHz.
Initially, $N_r$=10 was chosen; however, $N_{r2}=2$ was sufficient as the modal energy of higher modes is less than 10\%, as shown in figure~\ref{f:modalenergy}.
The modal energy, $E_i$, of the $i^{th}$ mode is defined as,
\begin{equation} 
\centering
\begin{split}
E_i &= \frac{S_i^2}{\sum_{j=1}^{N_r}S_j^2}
\end{split}
\label{eq:ModalEnergy}
\end{equation}
where $S_i$ is the $i^{th}$ singular value.
The total modal energy of the first two modes of perturbation parameters other than $\tilde{u}_y$ is almost 70\%.
For $\tilde{u}_y$, this number is lower because the shock structure has little influence on its flowfield, and it is composed only of a slowly growing linear mode and statistical noise.
Note that the data matrix itself requires 77.24~GBs of run time memory, larger than the typical compute nodes of supercomputing clusters.
Therefore, the method was parallelized based on the Tall and Skinny QR factorization (TSQR) algorithm~\citep{sayadi2016parallel} to overcome storage requirements and speed up the SVD procedure.
Figure~\ref{f:uy_PODCompare} shows the original noise-contained DSMC solution of perturbation spanwise velocity at $T$=90.5 on the $S$-plane wall-normal to the lower wedge along with the noise-filtered contour lines of the solution reconstructed using POD.
The figure also shows two horizontal dashed lines $L_1$ and $L_2$ along which the DSMC data is extracted and compared in figure~\ref{f:uy_1DCompare}.
The POD-reconstructed data exhibits the same spatial spanwise variation but contains very low statistical noise compared to the DSMC solution.

\bibliographystyle{jfm}
\bibliography{References}

\begin{thebibliography}{109}
\expandafter\ifx\csname natexlab\endcsname\relax\def\natexlab#1{#1}\fi
\def\au#1{#1} \def\ed#1{#1} \def\yr#1{#1}\def\at#1{#1}\def\jt#1{\textit{#1}}
  \def\bt#1{#1}\def\bvol#1{\textbf{#1}} \def\vol#1{#1} \def\pg#1{#1}
  \def\publ#1{#1}\def\arxiv#1{#1}\def\org#1{#1}\def\st#1{\textit{#1}}

\bibitem[Anderson(2003)]{andersonModern}
{\sc \au{Anderson, John~D}} \yr{2003} {\em Modern compressible flow, with
  historical perspective\/}, 3rd edn.  \publ{Tata McGraw-Hill}.

\bibitem[Babinsky \& Harvey(2011)]{babinsky_harvey_2011}
{\sc \au{Babinsky, Holger} \& \au{Harvey, John~K.}} \yr{2011} {\em \it{Shock
  Wave--Boundary-Layer Interactions}\/}.  \publ{Cambridge University Press}.

\bibitem[Balakumar {\em et~al.\/}(2005)Balakumar, Zhao \&
  Atkins]{balakumar2005stability}
{\sc \au{Balakumar, Ponnampalam}, \au{Zhao, Hongwu} \& \au{Atkins, Harold}}
  \yr{2005}  \at{Stability of hypersonic boundary layers over a compression
  corner}.  \jt{AIAA journal}  \bvol{43}~(4),  \pg{760--767}.

\bibitem[Bird(1970)]{bird1970aspects}
{\sc \au{Bird, G.~A.}} \yr{1970}  \at{Aspects of the structure of strong shock
  waves}.  \jt{The Physics of Fluids}  \bvol{13}~(5),  \pg{1172--1177}.

\bibitem[Bird(1994)]{bird:94mgd}
{\sc \au{Bird, G.~A.}} \yr{1994} {\em Molecular Gas Dynamics and the Direct
  Simulation of Gas Flows\/}, 2nd edn.  \publ{Clarendon Press}.

\bibitem[Bird(1998)]{bird1998recent}
{\sc \au{Bird, G.~A.}} \yr{1998}  \at{Recent advances and current challenges
  for {DSMC}}.  \jt{Computers \& Mathematics with Applications}
  \bvol{35}~(1-2),  \pg{1--14}.

\bibitem[Boin {\em et~al.\/}(2006)Boin, Robinet, Corre \& Deniau]{boin20063d}
{\sc \au{Boin, J-Ph}, \au{Robinet, J~Ch}, \au{Corre, Ch} \& \au{Deniau, H}}
  \yr{2006}  \at{3{D} steady and unsteady bifurcations in a shock-wave/laminar
  boundary layer interaction: a numerical study}.  \jt{Theoretical and
  Computational Fluid Dynamics}  \bvol{20}~(3),  \pg{163--180}.

\bibitem[Borgnakke \& Larsen(1975)]{borgnakke:75BLContinuous}
{\sc \au{Borgnakke, Claus} \& \au{Larsen, Poul~S.}} \yr{1975}  \at{Statistical
  collision model for {M}onte {C}arlo simulation of polyatomic gas mixture}.
  \jt{Journal of Computational Physics}  \bvol{18}~(4),  \pg{405--420}.

\bibitem[Bruno(2019)]{bruno2019direct}
{\sc \au{Bruno, Domenico}} \yr{2019}  \at{Direct {S}imulation {M}onte {C}arlo
  simulation of thermal fluctuations in gases}.  \jt{Physics of Fluids}
  \bvol{31}~(4),  \pg{047105}.

\bibitem[Cassel {\em et~al.\/}(1995)Cassel, Ruban \&
  Walker]{cassel_ruban_walker_1995}
{\sc \au{Cassel, K.~W.}, \au{Ruban, A.~I.} \& \au{Walker, J. D.~A.}} \yr{1995}
  \at{An instability in supersonic boundary-layer flow over a compression
  ramp}.  \jt{Journal of Fluid Mechanics}  \bvol{300},  \pg{265–285}.

\bibitem[Cercignani {\em et~al.\/}(1999)Cercignani, Frezzotti \&
  Grosfils]{cercignani1999structure}
{\sc \au{Cercignani, Carlo}, \au{Frezzotti, Aldo} \& \au{Grosfils, Patrick}}
  \yr{1999}  \at{The structure of an infinitely strong shock wave}.
  \jt{Physics of fluids}  \bvol{11}~(9),  \pg{2757--2764}.

\bibitem[Chambre \& Schaaf(1961)]{SchaafAndChambre}
{\sc \au{Chambre, Paul~A.} \& \au{Schaaf, Samuel~A.}} \yr{1961} {\em Flow of
  Rarefied Gases\/}.  \publ{Princeton University Press}.

\bibitem[Chapman {\em et~al.\/}(1958)Chapman, Kuehn \&
  Larson]{chapman1958investigation}
{\sc \au{Chapman, Dean~R}, \au{Kuehn, Donald~M} \& \au{Larson, Howard~K}}
  \yr{1958}  \bt{Investigation of separated flows in supersonic and subsonic
  streams with emphasis on the effect of transition}. {\em Tech. Rep.\/} 1356.
  \org{NACA}.

\bibitem[Chuvakhov {\em et~al.\/}(2017)Chuvakhov, Borovoy, Egorov, Radchenko,
  Olivier \& Roghelia]{chuvakhov2017effect}
{\sc \au{Chuvakhov, PV}, \au{Borovoy, V~Ya}, \au{Egorov, IV}, \au{Radchenko,
  VN}, \au{Olivier, H} \& \au{Roghelia, A}} \yr{2017}  \at{Effect of small
  bluntness on formation of {G}{\"o}rtler vortices in a supersonic compression
  corner flow}.  \jt{Journal of Applied Mechanics and Technical Physics}
  \bvol{58}~(6),  \pg{975--989}.

\bibitem[Clemens \& Narayanaswamy(2014)]{clemens2014low}
{\sc \au{Clemens, Noel~T} \& \au{Narayanaswamy, Venkateswaran}} \yr{2014}
  \at{Low-frequency unsteadiness of shock wave/turbulent boundary layer
  interactions}.  \jt{Annual Review of Fluid Mechanics}  \bvol{46},
  \pg{469--492}.

\bibitem[Cowley \& Hall(1990)]{cowley_hall_1990}
{\sc \au{Cowley, Stephen} \& \au{Hall, Philip}} \yr{1990}  \at{On the
  instability of hypersonic flow past a wedge}.  \jt{Journal of Fluid
  Mechanics}  \bvol{214},  \pg{17–42}.

\bibitem[Crouch {\em et~al.\/}(2007)Crouch, Garbaruk \&
  Magidov]{crouch2007predicting}
{\sc \au{Crouch, JD}, \au{Garbaruk, A} \& \au{Magidov, D}} \yr{2007}
  \at{Predicting the onset of flow unsteadiness based on global instability}.
  \jt{Journal of Computational Physics}  \bvol{224}~(2),  \pg{924--940}.

\bibitem[Czarnecki \& Mueller(1950)]{czarnecki1950investigation}
{\sc \au{Czarnecki, KR} \& \au{Mueller, James~N}} \yr{1950}  \bt{Investigation
  at mach number 1.62 of the pressure distribution over a rectangular wing with
  symmetrical circular-arc section and 30-percent-chord trailing-edge flap}.
  {\em Tech. Rep.\/} RM L9JO5.  \org{NACA}.

\bibitem[Dallmann(1983)]{dallmann1983topological}
{\sc \au{Dallmann, Uve}} \yr{1983} Topological structures of three-dimensional
  vortex flow separation.  \bt{In {\em AIAA 16th Fluid and Plasmadynamics
  Conference\/}}. AIAA-83-1735.

\bibitem[Dallmann(1985)]{dallmann1985structural}
{\sc \au{Dallmann, U}} \yr{1985}  \at{Structural stability of three-dimensional
  vortex flows}.  \bt{In {\em Nonlinear Dynamics of Transcritical Flows\/}},
  \pg{pp. 81--102}.  \publ{Springer}.

\bibitem[Druguet {\em et~al.\/}(2005)Druguet, Candler \&
  Nompelis]{druguet2005effects}
{\sc \au{Druguet, Marie-Claude}, \au{Candler, Graham~V} \& \au{Nompelis,
  Ioannis}} \yr{2005}  \at{Effects of numerics on navier-stokes computations of
  hypersonic double-cone flows}.  \jt{AIAA journal}  \bvol{43}~(3),
  \pg{616--623}.

\bibitem[Durna \& Celik(2020)]{durna2020effects}
{\sc \au{Durna, AS} \& \au{Celik, Bayram}} \yr{2020}  \at{Effects of
  double-wedge aft angle on hypersonic laminar flows}.  \jt{AIAA Journal}
  \bvol{58}~(4),  \pg{1689--1703}.

\bibitem[Durna {\em et~al.\/}(2016)Durna, El~Hajj Ali~Barada \&
  Celik]{durna2016shock}
{\sc \au{Durna, Ahmet~Selim}, \au{El~Hajj Ali~Barada, Mohamad} \& \au{Celik,
  Bayram}} \yr{2016}  \at{Shock interaction mechanisms on a double wedge at
  {M}ach 7}.  \jt{Physics of Fluids}  \bvol{28}~(9),  \pg{096101}.

\bibitem[Dussauge {\em et~al.\/}(2006)Dussauge, Dupont \&
  Debi{\`e}ve]{dussauge2006unsteadiness}
{\sc \au{Dussauge, Jean-Paul}, \au{Dupont, Pierre} \& \au{Debi{\`e}ve,
  Jean-Francois}} \yr{2006}  \at{Unsteadiness in shock wave boundary layer
  interactions with separation}.  \jt{Aerospace Science and Technology}
  \bvol{10}~(2),  \pg{85--91}.

\bibitem[Dwivedi {\em et~al.\/}(2019)Dwivedi, Sidharth, Nichols, Candler \&
  Jovanović]{dwivedi_2019}
{\sc \au{Dwivedi, Anubhav}, \au{Sidharth, G.~S.}, \au{Nichols, Joseph~W.},
  \au{Candler, Graham~V.} \& \au{Jovanović, Mihailo~R.}} \yr{2019}
  \at{Reattachment streaks in hypersonic compression ramp flow: an
  input–output analysis}.  \jt{Journal of Fluid Mechanics}  \bvol{880},
  \pg{113–135}.

\bibitem[Edney(1968)]{edney1968effects}
{\sc \au{Edney, Barry~E}} \yr{1968}  \at{Effects of shock impingement on the
  heat transfer around blunt bodies.}  \jt{AIAA Journal}  \bvol{6}~(1),
  \pg{15--21}.

\bibitem[Egorov {\em et~al.\/}(2011)Egorov, Neiland \&
  Shredchenko]{egorov2011three}
{\sc \au{Egorov, Ivan}, \au{Neiland, Vladimir} \& \au{Shredchenko, Vladimir}}
  \yr{2011} Three-dimensional flow structures at supersonic flow over the
  compression ramp.  \bt{In {\em 49th AIAA Aerospace Sciences Meeting\/}}. AIAA
  2011-730.

\bibitem[Elfstrom(1971)]{elifstromThesis}
{\sc \au{Elfstrom, GM}} \yr{1971}  \at{Turbulent separation in hypersonic
  flow}. PhD thesis, University of London,
  \url{https://spiral.imperial.ac.uk/bitstream/10044/1/16361/2/Elfstrom-GM-1971-PhD-Thesis.pdf}.

\bibitem[Elfstrom(1972)]{elfstrom1972turbulent}
{\sc \au{Elfstrom, GM}} \yr{1972}  \at{Turbulent hypersonic flow at a
  wedge-compression corner}.  \jt{Journal of fluid Mechanics}  \bvol{53}~(1),
  \pg{113--127}.

\bibitem[Fletcher {\em et~al.\/}(2004)Fletcher, Ruban \&
  Walker]{fletcher_ruban_walker_2004}
{\sc \au{Fletcher, A. J.~P.}, \au{Ruban, A.~I.} \& \au{Walker, J. D.~A.}}
  \yr{2004}  \at{Instabilities in supersonic compression ramp flow}.
  \jt{Journal of Fluid Mechanics}  \bvol{517},  \pg{309–330}.

\bibitem[Frontera supercomputer(2019)]{Frontera}
Frontera supercomputer \yr{2019} {\em System hardware and software
  overview,\url{https://www.tacc.utexas.edu/systems/frontera}\/}.

\bibitem[Gai \& Khraibut(2019)]{gai2019hypersonic}
{\sc \au{Gai, Sudhir~L} \& \au{Khraibut, Amna}} \yr{2019}  \at{Hypersonic
  compression corner flow with large separated regions}.  \jt{Journal of Fluid
  Mechanics}  \bvol{877},  \pg{471--494}.

\bibitem[Gaitonde(2015)]{gaitonde2015progress}
{\sc \au{Gaitonde, Datta~V}} \yr{2015}  \at{Progress in shock wave/boundary
  layer interactions}.  \jt{Progress in Aerospace Sciences}  \bvol{72},
  \pg{80--99}.

\bibitem[Gallis {\em et~al.\/}(2016)Gallis, Koehler, Torczynski \&
  Plimpton]{gallis2016direct}
{\sc \au{Gallis, Michail~A}, \au{Koehler, TP}, \au{Torczynski, John~R} \&
  \au{Plimpton, Steven~J}} \yr{2016}  \at{{D}irect {S}imulation {M}onte {C}arlo
  investigation of the {R}ayleigh-{T}aylor instability}.  \jt{Physical Review
  Fluids}  \bvol{1}~(4),  \pg{043403}.

\bibitem[Gallis {\em et~al.\/}(2015)Gallis, Koehler, Torczynski \&
  Plimpton]{gallis2015direct}
{\sc \au{Gallis, Michail~A}, \au{Koehler, Timothy~P}, \au{Torczynski, John~R}
  \& \au{Plimpton, Steven~J}} \yr{2015}  \at{{D}irect {S}imulation {M}onte
  {C}arlo investigation of the {R}ichtmyer-{M}eshkov instability}.  \jt{Physics
  of Fluids}  \bvol{27}~(8),  \pg{084105}.

\bibitem[Garcia(1986)]{garcia1986nonequilibrium}
{\sc \au{Garcia, Alejandro~L}} \yr{1986}  \at{Nonequilibrium fluctuations
  studied by a rarefied-gas simulation}.  \jt{Physical Review A}
  \bvol{34}~(2),  \pg{1454}.

\bibitem[Gimelshein {\em et~al.\/}(2002)Gimelshein, Gimelshein \&
  Levin]{gimelshein2002vibrational}
{\sc \au{Gimelshein, NE}, \au{Gimelshein, SF} \& \au{Levin, DA}} \yr{2002}
  \at{Vibrational relaxation rates in the {D}irect {S}imulation {M}onte {C}arlo
  method}.  \jt{Physics of Fluids}  \bvol{14}~(12),  \pg{4452--4455}.

\bibitem[Grilli {\em et~al.\/}(2012)Grilli, Schmid, Hickel \&
  Adams]{grilli_schmid_hickel_adams_2012}
{\sc \au{Grilli, Muzio}, \au{Schmid, Peter~J.}, \au{Hickel, Stefan} \&
  \au{Adams, Nikolaus~A.}} \yr{2012}  \at{Analysis of unsteady behaviour in
  shockwave turbulent boundary layer interaction}.  \jt{Journal of Fluid
  Mechanics}  \bvol{700},  \pg{16–28}.

\bibitem[Grinberg(2012)]{grinberg2012proper}
{\sc \au{Grinberg, Leopold}} \yr{2012}  \at{Proper orthogonal decomposition of
  atomistic flow simulations}.  \jt{Journal of Computational Physics}
  \bvol{231}~(16),  \pg{5542--5556}.

\bibitem[Hadjiconstantinou {\em et~al.\/}(2003)Hadjiconstantinou, Garcia,
  Bazant \& He]{hadjiconstantinou2003statistical}
{\sc \au{Hadjiconstantinou, Nicolas~G}, \au{Garcia, Alejandro~L}, \au{Bazant,
  Martin~Z} \& \au{He, Gang}} \yr{2003}  \at{Statistical error in particle
  simulations of hydrodynamic phenomena}.  \jt{Journal of computational
  physics}  \bvol{187}~(1),  \pg{274--297}.

\bibitem[Hankey~Jr \& Holden(1975)]{hankey1975two}
{\sc \au{Hankey~Jr, WL} \& \au{Holden, Michael~S}} \yr{1975}
  \bt{Two-dimensional shock wave-boundary layer interactions in high speed
  flows}. {\em Tech. Rep.\/} AGARDograph No. 203.  \org{AGARD}.

\bibitem[Hashimoto(2009)]{hashimoto2009experimental}
{\sc \au{Hashimoto, Tokitada}} \yr{2009}  \at{Experimental investigation of
  hypersonic flow induced separation over double wedges}.  \jt{Journal of
  Thermal Science}  \bvol{18}~(3),  \pg{220--225}.

\bibitem[Holden(1963)]{holdenThesis}
{\sc \au{Holden, Michael~S.}} \yr{1963}  \at{Heat transfer in separated flow}.
  PhD thesis, University of London,
  \url{https://spiral.imperial.ac.uk/bitstream/10044/1/16813/2/Holden-MS-1964-PhD-Thesis.pdf}.

\bibitem[Holden(1978)]{holdenRamp1978}
{\sc \au{Holden, Michael~S.}} \yr{1978} A study of flow separation in regions
  of shock wave-boundary layer interaction in hypersonic flow.  \bt{In {\em
  AIAA 11th Fluid and Plasma Dynamics Conference\/}}. AIAA 1978-1169.

\bibitem[Ivanov \& Rogasinsky(1988)]{ivanov1988analysis}
{\sc \au{Ivanov, M.~S.} \& \au{Rogasinsky, S.~V.}} \yr{1988}  \at{Analysis of
  numerical techniques of the direct simulation {M}onte {C}arlo method in the
  rarefied gas dynamics}.  \jt{Russian Journal of numerical analysis and
  mathematical modelling}  \bvol{3}~(6),  \pg{453--466}.

\bibitem[Kadau {\em et~al.\/}(2010)Kadau, Barber, Germann, Holian \&
  Alder]{kadau2010atomistic}
{\sc \au{Kadau, Kai}, \au{Barber, John~L}, \au{Germann, Timothy~C}, \au{Holian,
  Brad~L} \& \au{Alder, Berni~J}} \yr{2010}  \at{Atomistic methods in fluid
  simulation}.  \jt{Philosophical Transactions of the Royal Society A:
  Mathematical, Physical and Engineering Sciences}  \bvol{368}~(1916),
  \pg{1547--1560}.

\bibitem[Kadau {\em et~al.\/}(2004)Kadau, Germann, Hadjiconstantinou, Lomdahl,
  Dimonte, Holian \& Alder]{kadau2004nanohydrodynamics}
{\sc \au{Kadau, Kai}, \au{Germann, Timothy~C}, \au{Hadjiconstantinou,
  Nicolas~G}, \au{Lomdahl, Peter~S}, \au{Dimonte, Guy}, \au{Holian, Brad~Lee}
  \& \au{Alder, Berni~J}} \yr{2004}  \at{Nanohydrodynamics simulations: an
  atomistic view of the {R}ayleigh--{T}aylor instability}.  \jt{Proceedings of
  the National Academy of Sciences}  \bvol{101}~(16),  \pg{5851--5855}.

\bibitem[Knight {\em et~al.\/}(2017)Knight, Chazot, Austin, Badr, Candler,
  Celik, de~Rosa, Donelli, Komives, Lani {\em et~al.\/}]{knight2017assessment}
{\sc \au{Knight, Doyle}, \au{Chazot, Olivier}, \au{Austin, Joanna}, \au{Badr,
  Mohammad~Ali}, \au{Candler, Graham}, \au{Celik, Bayram}, \au{de~Rosa,
  Donato}, \au{Donelli, Raffaele}, \au{Komives, Jeffrey}, \au{Lani, Andrea} \&
  \au{others}} \yr{2017}  \at{Assessment of predictive capabilities for
  aerodynamic heating in hypersonic flow}.  \jt{Progress in Aerospace Sciences}
   \bvol{90},  \pg{39--53}.

\bibitem[Knisely \& Austin(2016)]{knisely2016geometry}
{\sc \au{Knisely, Andrew~M} \& \au{Austin, Joanna~M}} \yr{2016} Geometry and
  test-time effects on hypervelocity shock-boundary layer interaction.  \bt{In
  {\em 54th AIAA Aerospace Sciences Meeting\/}}. AIAA 2016-1979.

\bibitem[Kogan(1969)]{koganRGDSpringer}
{\sc \au{Kogan, Maurice~N.}} \yr{1969} {\em Rarefied Gas Dynamics\/}, 1st edn.
  \publ{Springer US}.

\bibitem[Korolev {\em et~al.\/}(2002)Korolev, Gajjar \&
  Ruban]{korolev_gajjar_ruban_2002}
{\sc \au{Korolev, G.~L.}, \au{Gajjar, J. S.~B.} \& \au{Ruban, A.~I.}} \yr{2002}
   \at{Once again on the supersonic flow separation near a corner}.
  \jt{Journal of Fluid Mechanics}  \bvol{463},  \pg{173–199}.

\bibitem[Landau \& Lifshitz(1980)]{landauLifshitz}
{\sc \au{Landau, LD} \& \au{Lifshitz, EM}} \yr{1980} {\em Statistical Physics:
  Part 1\/}, ,  \vol{vol.~5}.  \publ{Pergamon Press}.

\bibitem[Liepmann {\em et~al.\/}(1951)Liepmann, Roshko \&
  Dhawan]{liepmann1951reflection}
{\sc \au{Liepmann, Hans~Wolfgang}, \au{Roshko, Anatol} \& \au{Dhawan, Satish}}
  \yr{1951}  \bt{On reflection of shock waves from boundary layers}. {\em Tech.
  Rep.\/} NACA TN 2334.  \org{California Institute of Technology, Pasadena,
  CA}.

\bibitem[Lighthill(1953)]{lighthill1953boundary2}
{\sc \au{Lighthill, Michael~James}} \yr{1953}  \at{On boundary layers and
  upstream influence ii. supersonic flows without separation}.  \jt{Proceedings
  of the Royal Society of London. Series A. Mathematical and Physical Sciences}
   \bvol{217}~(1131),  \pg{478--507}.

\bibitem[Lighthill(1963)]{lighthill}
{\sc \au{Lighthill, M.~J.}} \yr{1963}  \at{Attachment and separation in
  three-dimensional flow}.  \bt{In {\em Laminar Boundary Layers, Section II
  2.6, Rosenhead, L. ed.\/}},  \pg{pp. 72--82}.  \publ{Oxford University
  Press}.

\bibitem[Lighthill \& Newman(1953)]{lighthill1953boundary1}
{\sc \au{Lighthill, Michael~James} \& \au{Newman, Maxwell Herman~Alexander}}
  \yr{1953}  \at{On boundary layers and upstream influence. i. a comparison
  between subsonic and supersonic flows}.  \jt{Proceedings of the Royal Society
  of London. Series A. Mathematical and Physical Sciences}  \bvol{217}~(1130),
  \pg{344--357}.

\bibitem[Lighthill(2000)]{lighthill2000upstream}
{\sc \au{Lighthill, Sir~James}} \yr{2000}  \at{Upstream influence in boundary
  layers 45 years ago}.  \jt{Philosophical Transactions of the Royal Society of
  London. Series A: Mathematical, Physical and Engineering Sciences}
  \bvol{358}~(1777),  \pg{3047--3061}.

\bibitem[LMFIT(Version 1.0.1)]{LMFIT}
LMFIT \yr{Version 1.0.1} {\em Non-linear Least-Squares Minimization and
  Curve-Fitting for Python,\url{https://lmfit.github.io/lmfit-py/}\/}.

\bibitem[Luchtenburg {\em et~al.\/}(2009)Luchtenburg, Noack \&
  Schlegel]{luchtenburg2009introduction}
{\sc \au{Luchtenburg, DM}, \au{Noack, BR} \& \au{Schlegel, M}} \yr{2009}
  \at{An introduction to the pod galerkin method for fluid flows with
  analytical examples and matlab source codes}.  \jt{Berlin Institute of
  Technology MB1, Muller-Breslau-Strabe}  \bvol{11}.

\bibitem[{Lumpkin III} {\em et~al.\/}(1991){Lumpkin III}, Haas \&
  Boyd]{lumpkin:91Zcorrection}
{\sc \au{{Lumpkin III}, Forrest~E.}, \au{Haas, Brian~L.} \& \au{Boyd, Iain~D.}}
  \yr{1991}  \at{Resolution of differences between collision number definitions
  in particle and continuum simulations}.  \jt{Physics of Fluids A: Fluid
  Dynamics}  \bvol{3}~(9),  \pg{2282--2284}.

\bibitem[Lusher \& Sandham(2020)]{lusher_sandham_2020}
{\sc \au{Lusher, David~J.} \& \au{Sandham, Neil~D.}} \yr{2020}  \at{The effect
  of flow confinement on laminar shock-wave/boundary-layer interactions}.
  \jt{Journal of Fluid Mechanics}  \bvol{897},  \pg{A18}.

\bibitem[Millikan \& White(1963)]{millikan1963systematics}
{\sc \au{Millikan, Roger~C.} \& \au{White, Donald~R.}} \yr{1963}
  \at{Systematics of vibrational relaxation}.  \jt{The Journal of Chemical
  Physics}  \bvol{39}~(12),  \pg{3209--3213}.

\bibitem[Moss(2001)]{moss2001dsmc}
{\sc \au{Moss, James}} \yr{2001} Dsmc computations for regions of shock/shock
  and shock/boundary layer interaction.  \bt{In {\em 39th Aerospace Sciences
  Meeting and Exhibit\/}},  \pg{p. 1027}.

\bibitem[Moss \& Bird(2005)]{moss2005direct}
{\sc \au{Moss, James~N} \& \au{Bird, Graeme~A}} \yr{2005}  \at{Direct
  simulation monte carlo simulations of hypersonic flows with shock
  interactions}.  \jt{AIAA journal}  \bvol{43}~(12),  \pg{2565--2573}.

\bibitem[Needham(1965{\natexlab{{\em a\/}}})]{needham1965heat}
{\sc \au{Needham, David~A}} \yr{1965{\natexlab{{\em a\/}}}}  \at{A
  heat-transfer criterion for the detection of incipient separation in
  hypersonic flow}.  \jt{AIAA Journal}  \bvol{3}~(4),  \pg{781--783}.

\bibitem[Needham(1965{\natexlab{{\em b\/}}})]{needhamThesis}
{\sc \au{Needham, David~A}} \yr{1965{\natexlab{{\em b\/}}}}  \at{Laminar
  separation in hypersonic flow}. PhD thesis, University of London,
  \url{https://spiral.imperial.ac.uk/bitstream/10044/1/11846/2/Needham-DA-1965-PhD-Thesis.pdf}.

\bibitem[Neiland(2008)]{neyland2008asymptotic}
{\sc \au{Neiland, Vladimir}} \yr{2008} {\em Asymptotic theory of supersonic
  viscous gas flows\/}.  \publ{Butterworth-Heinemann}.

\bibitem[Park(1984)]{park1984problems}
{\sc \au{Park, Chul}} \yr{1984} Problems of rate chemistry in the flight
  regimes of aeroassisted orbital transfer vehicles.  \bt{In {\em 19th
  Thermophysics Conference\/}},  \pg{p. 1730}.

\bibitem[Parker(1959)]{parker:59rotational}
{\sc \au{Parker, J.~G.}} \yr{1959}  \at{Rotational and vibrational relaxation
  in diatomic gases}.  \jt{The Physics of Fluids}  \bvol{2}~(4),
  \pg{449--462}.

\bibitem[Pasquariello {\em et~al.\/}(2017)Pasquariello, Hickel \&
  Adams]{pasquariello2017unsteady}
{\sc \au{Pasquariello, Vito}, \au{Hickel, Stefan} \& \au{Adams, Nikolaus~A}}
  \yr{2017}  \at{Unsteady effects of strong shock-wave/boundary-layer
  interaction at high reynolds number}.  \jt{J. Fluid Mech}  \bvol{823}~(617),
  \pg{014602--19}.

\bibitem[Perry \& Chong(1987)]{perryAndChong}
{\sc \au{Perry, A~E} \& \au{Chong, M~S}} \yr{1987}  \at{A description of
  eddying motions and flow patterns using critical-point concepts}.  \jt{Annual
  Review of Fluid Mechanics}  \bvol{19}~(1),  \pg{125--155},  \arxiv{arXiv:
  https://doi.org/10.1146/annurev.fl.19.010187.001013}.

\bibitem[Perry \& Hornung(1984{\natexlab{{\em a\/}}})]{hornungAndPerry}
{\sc \au{Perry, A.~E.} \& \au{Hornung, H.~G.}} \yr{1984{\natexlab{{\em a\/}}}}
  Some aspects of three-dimensional separation. part i. streamsurface
  bifurcations.  \bt{In {\em Z. Flugwiss. Weltraumforsch\/}}, ,  \vol{vol.~8},
  \pg{pp. 77--87}.

\bibitem[Perry \& Hornung(1984{\natexlab{{\em b\/}}})]{perryAndHornung}
{\sc \au{Perry, A.~E.} \& \au{Hornung, H.~G.}} \yr{1984{\natexlab{{\em b\/}}}}
  Some aspects of three-dimensional separation. part ii. vortex skeletons.
  \bt{In {\em Z. Flugwiss. Weltraumforsch\/}}, ,  \vol{vol.~8},  \pg{pp.
  155--160}.

\bibitem[Piponniau {\em et~al.\/}(2009)Piponniau, Dussauge, Debi{\`e}ve \&
  Dupont]{piponniau2009simple}
{\sc \au{Piponniau, S{\'e}bastien}, \au{Dussauge, Jean-Paul}, \au{Debi{\`e}ve,
  Jean-Fran{\c{c}}ois} \& \au{Dupont, Pierre}} \yr{2009}  \at{A simple model
  for low-frequency unsteadiness in shock-induced separation}.  \jt{Journal of
  Fluid Mechanics}  \bvol{629},  \pg{87--108}.

\bibitem[Pirozzoli \& Grasso(2006)]{pirozzoli2006direct}
{\sc \au{Pirozzoli, Sergio} \& \au{Grasso, Francesco}} \yr{2006}  \at{Direct
  numerical simulation of impinging shock wave/turbulent boundary layer
  interaction at {M}= 2.25}.  \jt{Physics of Fluids}  \bvol{18}~(6),
  \pg{065113}.

\bibitem[Priebe \& Mart{\'\i}n(2012)]{priebe_martin_2012}
{\sc \au{Priebe, Stephan} \& \au{Mart{\'\i}n, M.~Pino}} \yr{2012}
  \at{Low-frequency unsteadiness in shock wave–turbulent boundary layer
  interaction}.  \jt{Journal of Fluid Mechanics}  \bvol{699},  \pg{1–49}.

\bibitem[Priebe {\em et~al.\/}(2016)Priebe, Tu, Rowley \&
  Mart{\'\i}n]{priebe2016low}
{\sc \au{Priebe, Stephan}, \au{Tu, Jonathan~H}, \au{Rowley, Clarence~W} \&
  \au{Mart{\'\i}n, M~Pino}} \yr{2016}  \at{Low-frequency dynamics in a
  shock-induced separated flow}.  \jt{Journal of Fluid Mechanics}  \bvol{807},
  \pg{441--477}.

\bibitem[Reinert {\em et~al.\/}(2020)Reinert, Candler \&
  Komives]{reinert2020simulations}
{\sc \au{Reinert, John~D}, \au{Candler, Graham~V} \& \au{Komives, Jeffrey~R}}
  \yr{2020}  \at{Simulations of unsteady three-dimensional hypersonic
  double-wedge flow experiments}.  \jt{AIAA Journal}  \bvol{58}~(9),
  \pg{4055--4067}.

\bibitem[Rizzetta {\em et~al.\/}(1978)Rizzetta, Burggraf \&
  Jenson]{rizzetta_burggraf_jenson_1978}
{\sc \au{Rizzetta, D.~P.}, \au{Burggraf, O.~R.} \& \au{Jenson, Richard}}
  \yr{1978}  \at{Triple-deck solutions for viscous supersonic and hypersonic
  flow past corners}.  \jt{Journal of Fluid Mechanics}  \bvol{89}~(3),
  \pg{535–552}.

\bibitem[Robinet(2007)]{robinet_2007}
{\sc \au{Robinet, J.-CH.}} \yr{2007}  \at{Bifurcations in
  shock-wave/laminar-boundary-layer interaction: global instability approach}.
  \jt{Journal of Fluid Mechanics}  \bvol{579},  \pg{85–112}.

\bibitem[Rodríguez \& Theofilis(2010)]{rodriguez_theofilis_2010}
{\sc \au{Rodríguez, D.} \& \au{Theofilis, V.}} \yr{2010}  \at{Structural
  changes of laminar separation bubbles induced by global linear instability}.
  \jt{Journal of Fluid Mechanics}  \bvol{655},  \pg{280–305}.

\bibitem[Roghelia {\em et~al.\/}(2017)Roghelia, Olivier, Egorov \&
  Chuvakhov]{roghelia2017experimental}
{\sc \au{Roghelia, Amit}, \au{Olivier, Herbert}, \au{Egorov, Ivan} \&
  \au{Chuvakhov, Pavel}} \yr{2017}  \at{Experimental investigation of
  {G}{\"o}rtler vortices in hypersonic ramp flows}.  \jt{Experiments in Fluids}
   \bvol{58}~(10),  \pg{139}.

\bibitem[Rudy {\em et~al.\/}(1991)Rudy, Thomas, Kumar, Gnoffo \&
  Chakravarthy]{rudy1991computation}
{\sc \au{Rudy, David~H}, \au{Thomas, James~L}, \au{Kumar, Ajay}, \au{Gnoffo,
  Peter~A} \& \au{Chakravarthy, Sukumar~R}} \yr{1991}  \at{Computation of
  laminar hypersonic compression-corner flows}.  \jt{AIAA journal}
  \bvol{29}~(7),  \pg{1108--1113}.

\bibitem[Sansica {\em et~al.\/}(2016)Sansica, Sandham \&
  Hu]{sansica_sandham_hu_2016}
{\sc \au{Sansica, Andrea}, \au{Sandham, Neil~D.} \& \au{Hu, Zhiwei}} \yr{2016}
  \at{Instability and low-frequency unsteadiness in a shock-induced laminar
  separation bubble}.  \jt{Journal of Fluid Mechanics}  \bvol{798},
  \pg{5–26}.

\bibitem[Sawant {\em et~al.\/}(2020)Sawant, Levin \&
  Theofilis]{sawant2020Kinetic}
{\sc \au{Sawant, Saurabh~S.}, \au{Levin, Deborah~A.} \& \au{Theofilis,
  Vassilios}} \yr{2020} A kinetic approach to studying low-frequency molecular
  fluctuations in a one-dimensional shock,  \arxiv{arXiv: 2012.14593}.

\bibitem[Sawant {\em et~al.\/}(2018)Sawant, Tumuklu, Jambunathan \&
  Levin]{sawant2018application}
{\sc \au{Sawant, Saurabh~S}, \au{Tumuklu, Ozgur}, \au{Jambunathan, Revathi} \&
  \au{Levin, Deborah~A}} \yr{2018}  \at{Application of adaptively refined
  unstructured grids in dsmc to shock wave simulations}.  \jt{Computers \&
  Fluids}  \bvol{170},  \pg{197--212}.

\bibitem[Sawant {\em et~al.\/}(2019)Sawant, Tumuklu, Theofilis \&
  Levin]{sawantIUTAM2019}
{\sc \au{Sawant, Saurabh~S.}, \au{Tumuklu, Ozgur}, \au{Theofilis, Vassilis} \&
  \au{Levin, Deborah~A.}} \yr{2019} Linear instability of shock-dominated
  laminar hypersonic separated flows.  \bt{In {\em The IUTAM Transition 2019
  Proceedings\/}}. Springer, (accepted),  \arxiv{arXiv: 2101.03688}.

\bibitem[Sayadi \& Schmid(2016)]{sayadi2016parallel}
{\sc \au{Sayadi, Taraneh} \& \au{Schmid, Peter~J}} \yr{2016}  \at{Parallel
  data-driven decomposition algorithm for large-scale datasets: with
  application to transitional boundary layers}.  \jt{Theoretical and
  Computational Fluid Dynamics}  \bvol{30}~(5),  \pg{415--428}.

\bibitem[Schneider(2004)]{schneider2004hypersonic}
{\sc \au{Schneider, Steven~P}} \yr{2004}  \at{Hypersonic laminar--turbulent
  transition on circular cones and scramjet forebodies}.  \jt{Progress in
  Aerospace Sciences}  \bvol{40}~(1-2),  \pg{1--50}.

\bibitem[Schrijer {\em et~al.\/}(2009)Schrijer, Caljouw, Scarano \&
  Van~Oudheusden]{schrijer2009three}
{\sc \au{Schrijer, F. F.~J.}, \au{Caljouw, R}, \au{Scarano, F} \&
  \au{Van~Oudheusden, B.~W.}} \yr{2009}  \at{Three dimensional experimental
  investigation of a hypersonic double-ramp flow}.  \bt{In {\em Shock
  Waves\/}},  \pg{pp. 719--724}.  \publ{Springer}.

\bibitem[Schrijer {\em et~al.\/}(2006)Schrijer, Van~Oudheusden, Dierksheide \&
  Scarano]{schrijer2006quantitative}
{\sc \au{Schrijer, F. F.~J.}, \au{Van~Oudheusden, B.~W.}, \au{Dierksheide, U}
  \& \au{Scarano, F}} \yr{2006} Quantitative visualization of a hypersonic
  double-ramp flow using {PIV} and schlieren.  \bt{In {\em 12th International
  Symposium on Flow Visualization\/}}. German Aerospace Center (DLR).

\bibitem[Sidharth {\em et~al.\/}(2018)Sidharth, Dwivedi, Candler \&
  Nichols]{sidharth2018onset}
{\sc \au{Sidharth, GS}, \au{Dwivedi, Anubhav}, \au{Candler, Graham~V} \&
  \au{Nichols, Joseph~W}} \yr{2018}  \at{Onset of three-dimensionality in
  supersonic flow over a slender double wedge}.  \jt{Physical Review Fluids}
  \bvol{3}~(9),  \pg{093901}.

\bibitem[Simeonides \& Haase(1995)]{simeonides1995experimental}
{\sc \au{Simeonides, G} \& \au{Haase, W}} \yr{1995}  \at{Experimental and
  computational investigations of hypersonic flow about compression ramps}.
  \jt{Journal of Fluid Mechanics}  \bvol{283},  \pg{17--42}.

\bibitem[Smith(1986)]{smith1986steady}
{\sc \au{Smith, F.~T.}} \yr{1986}  \at{Steady and unsteady boundary-layer
  separation}.  \jt{Annual {R}eview of {F}luid {M}echanics}  \bvol{18}~(1),
  \pg{197--220}.

\bibitem[Smith \& Khorrami(1991)]{smith_khorrami_1991}
{\sc \au{Smith, F.~T.} \& \au{Khorrami, A.~Farid}} \yr{1991}  \at{The
  interactive breakdown in supersonic ramp flow}.  \jt{Journal of Fluid
  Mechanics}  \bvol{224},  \pg{197–215}.

\bibitem[Stefanov {\em et~al.\/}(2002{\natexlab{{\em a\/}}})Stefanov, Roussinov
  \& Cercignani]{stefanov_part1}
{\sc \au{Stefanov, S}, \au{Roussinov, V} \& \au{Cercignani, C}}
  \yr{2002{\natexlab{{\em a\/}}}}  \at{Rayleigh--{B}{\'e}nard flow of a
  rarefied gas and its attractors. {I}. {C}onvection regime}.  \jt{Physics of
  Fluids}  \bvol{14}~(7),  \pg{2255--2269}.

\bibitem[Stefanov {\em et~al.\/}(2002{\natexlab{{\em b\/}}})Stefanov, Roussinov
  \& Cercignani]{stefanov_part2}
{\sc \au{Stefanov, S}, \au{Roussinov, V} \& \au{Cercignani, C}}
  \yr{2002{\natexlab{{\em b\/}}}}  \at{Rayleigh--{B}{\'e}nard flow of a
  rarefied gas and its attractors. {II}. {C}haotic and periodic convective
  regimes}.  \jt{Physics of Fluids}  \bvol{14}~(7),  \pg{2270--2288}.

\bibitem[Stefanov {\em et~al.\/}(2007)Stefanov, Roussinov \&
  Cercignani]{stefanov_part3}
{\sc \au{Stefanov, S}, \au{Roussinov, V} \& \au{Cercignani, C}} \yr{2007}
  \at{Rayleigh--{B}{\'e}nard flow of a rarefied gas and its attractors. {III}.
  {T}hree-dimensional computer simulations}.  \jt{Physics of Fluids}
  \bvol{19}~(12),  \pg{124101}.

\bibitem[Stewartson(1964)]{stewartson1964theory}
{\sc \au{Stewartson, Keith}} \yr{1964} {\em The Theory of Laminar Boundary
  Layers in Compressible Fluids\/}.  \publ{Clarendon Press Oxford}.

\bibitem[Stewartson \& Williams(1969)]{stewartson1969self}
{\sc \au{Stewartson, Keith} \& \au{Williams, PG}} \yr{1969}  \at{Self-induced
  separation}.  \jt{Proceedings of the Royal Society of London. A. Mathematical
  and Physical Sciences}  \bvol{312}~(1509),  \pg{181--206}.

\bibitem[Swantek \& Austin(2015)]{swantek2015flowfield}
{\sc \au{Swantek, AB} \& \au{Austin, JM}} \yr{2015}  \at{Flowfield
  establishment in hypervelocity shock-wave/boundary-layer interactions}.
  \jt{AIAA Journal}  \bvol{53}~(2),  \pg{311--320}.

\bibitem[Tecplot-360(2020 R1)]{Tecplot}
Tecplot-360 \yr{2020 R1} {\em
  \url{https://www.tecplot.com/products/tecplot-360/}\/}.

\bibitem[Theofilis {\em et~al.\/}(2000)Theofilis, Hein \&
  Dallmann]{theofilis2000origins}
{\sc \au{Theofilis, Vassilios}, \au{Hein, Stefan} \& \au{Dallmann, Uwe}}
  \yr{2000}  \at{On the origins of unsteadiness and three-dimensionality in a
  laminar separation bubble}.  \jt{Philosophical Transactions of the Royal
  Society of London. Series A: Mathematical, Physical and Engineering Sciences}
   \bvol{358}~(1777),  \pg{3229--3246}.

\bibitem[Tobak \& Peake(1982)]{tobak1982topology}
{\sc \au{Tobak, Murray} \& \au{Peake, David~J}} \yr{1982}  \at{Topology of
  three-dimensional separated flows}.  \jt{Annual review of fluid mechanics}
  \bvol{14}~(1),  \pg{61--85}.

\bibitem[Touber \& Sandham(2009)]{touber2009large}
{\sc \au{Touber, Emile} \& \au{Sandham, Neil~D}} \yr{2009}  \at{Large-eddy
  simulation of low-frequency unsteadiness in a turbulent shock-induced
  separation bubble}.  \jt{Theoretical and Computational Fluid Dynamics}
  \bvol{23}~(2),  \pg{79--107}.

\bibitem[Tumuklu {\em et~al.\/}(2018{\natexlab{{\em a\/}}})Tumuklu, Levin \&
  Theofilis]{tumuklu2018POF1}
{\sc \au{Tumuklu, Ozgur}, \au{Levin, Deborah~A.} \& \au{Theofilis, Vassilis}}
  \yr{2018{\natexlab{{\em a\/}}}}  \at{Investigation of unsteady, hypersonic,
  laminar separated flows over a double cone geometry using a kinetic
  approach}.  \jt{Physics of Fluids}  \bvol{30}~(4),  \pg{046103}.

\bibitem[Tumuklu {\em et~al.\/}(2019)Tumuklu, Levin \&
  Theofilis]{tumuklu2018PhysRevF}
{\sc \au{Tumuklu, Ozgur}, \au{Levin, Deborah~A} \& \au{Theofilis, Vassilis}}
  \yr{2019}  \at{Modal analysis with proper orthogonal decomposition of
  hypersonic separated flows over a double wedge}.  \jt{Physical Review Fluids}
   \bvol{4}~(3),  \pg{033403}.

\bibitem[Tumuklu {\em et~al.\/}(2018{\natexlab{{\em b\/}}})Tumuklu, Theofilis
  \& Levin]{tumuklu2018POF2}
{\sc \au{Tumuklu, Ozgur}, \au{Theofilis, Vassilis} \& \au{Levin, Deborah~A}}
  \yr{2018{\natexlab{{\em b\/}}}}  \at{On the unsteadiness of shock--laminar
  boundary layer interactions of hypersonic flows over a double cone}.
  \jt{Physics of Fluids}  \bvol{30}~(10),  \pg{106111}.

\bibitem[Vincenti \& Kruger(1965)]{vincenti1965introduction}
{\sc \au{Vincenti, Walter~Guido} \& \au{Kruger, Charles~H}} \yr{1965} {\em
  Introduction to Physical Gas Dynamics\/}.  \publ{Wiley, New York}.

\end{thebibliography}

\newpage
\end{document}